\documentclass[twocolumn,showpacs,floatfix,superscriptaddress,amsmath,amssymb,prx]{revtex4}
\usepackage{mathrsfs}
\usepackage{txfonts}
\usepackage{amssymb}
\usepackage{graphicx}
\usepackage{hyperref}
\usepackage{ulem}
 \usepackage{overpic}
 \usepackage{psfrag}
\usepackage{tabularx}
\usepackage{array}
\usepackage{placeins}
\usepackage{subfigure}
\usepackage[toc,page]{appendix}
\makeatletter
\newcommand{\be}{\begin{equation}}
\newcommand{\ee}{\end{equation}}

\newcommand{\PreserveBackslash}[1]{\let\temp=\\#1\let\\=\temp}

\begin{document}

\title{Are degenerate groundstates induced by spontaneous symmetry breakings in quantum phase transitions?}
\author{Mei He}
\affiliation{Centre for Modern physics, Chongqing University,
Chongqing 400044, China}
\affiliation{Department of Physics, Chongqing University,
Chongqing 400044, China}
\author{Qian-Qian Shi}
\affiliation{Centre for Modern physics, Chongqing University,
Chongqing 400044, China}
\affiliation{College of Materials Science
and Engineering, Chongqing University, Chongqing 400044, China}
\author{Sam Young Cho}
\altaffiliation{sycho@cqu.edu.cn}
\affiliation{Centre for Modern physics, Chongqing University,
Chongqing 400044, China}
\affiliation{Department of Physics, Chongqing University,
Chongqing 400044, China}

\begin{abstract}
 Recently, emergent symmetry is one of fast-growing intriguing issues in many-body systems.
 Its roles and consequential physics have not been well understood in quantum phase transitions.
 Emergent symmetry of degenerate groundstates is discussed in possible connection to spontaneous symmetry breaking within
 the Landau theory.
 For a clear discussion, a quantum spin-$1/2$ plaquette chain system
 is shown to have rich emergent symmetry phenomena in its groundstates.
 A covering symmetry group over all emergent symmetries responsible for degenerate groundstates
 in the plaquette chain system
 is found to correspond to a largest common symmetry group of constituent Hamiltonians
 describing the plaquette system.
 Consequently, this result suggests that,
 as a guiding symmetry principle in quantum phase transitions,
 {\it degenerate groundstates are induced by a spontaneous breaking of
 symmetries belonging to a largest common symmetry group of continent Hamiltonians
 describing a given system  but can have more symmetries than the largest common symmetry}.

\end{abstract}
%\pacs{}
%64.70.Tg Quantum phase transition
%73.22.Gk Broken symmetry phase
%75.10.Pq Spin chain models

\maketitle

\section{Introduction}
 Symmetry plays an important and indispensable role in understanding physics.
 Revealing many properties of nature are allowed, in fact, owing
 to understanding a mechanism of symmetry breaking
 \cite{Landau,Anderson,anderson2,Laughlin,nuclear,classical}.
 As is known, symmetry of nature can be broken in two different ways.
 One is explicit symmetry breaking, e.g., the isotopic symmetry of the nuclear force~\cite{nuclear}.
 The other is spontaneous symmetry breaking, which plays a more profound role than explicit symmetry breaking.
 As a modern terminology,
 spontaneous symmetry breaking, though appeared first in Baker and Glashow's paper~\cite{baker},
  is considered as an emergent phenomenon
 because once it happens to a system, an underlying physics of the system is not reducible to some sort of sum of behaviors of its parts,
 and not predictable~\cite{anderson2,Laughlin}.
 Spontaneous symmetry breaking phenomena have been observed ubiquitously in macroscopic systems
 including not only classical systems~\cite{classical} but also quantum systems such as quantum Ising systems, and also
 theoretically introduced quantum systems and experimentally prepared systems that can be described by an
 effective Hamiltonian, e.g., spontaneous particle-hole symmetry breaking in the $\nu=5/2$ fractional quantum Hall effect~\cite{hall}.
 Quantum phase transitions can then be a prototypical example for spontaneous symmetry breaking in many-body systems.
 The Landau theory for spontaneous symmetry breaking has guided how to understand quantum phase transitions
 \cite{Landau,Anderson,SSB,Coleman}.

 In the Landau theory,
 a consequence of spontaneous symmetry breaking is a degeneracy of groundstates in broken-symmetry phase.
 Degenerate groundstates from spontaneous symmetry breaking is normally believed
 to have a less symmetry than their Hamiltonian for a given system parameter ~\cite{SSB,Anderson}.
 More precisely, if a given system Hamiltonian $H$ is invariant under an unitary transformation $U$, i.e., $U H U^\dagger = H$,
 and the unitary transformation $U$ is related to an element of the Hamiltonian symmetry group $G$,
 there are two possible groundstates, i.e., $|\psi_{gs}\rangle$ and $ U |\psi_{gs}\rangle$, that satisfies
 $H |\psi_{gs} \rangle =E_{gs} |\psi_{gs} \rangle$ and
 $H U |\psi_{gs} \rangle =E_{gs} U |\psi_{gs} \rangle$, respectively.
 When the two groundstates become $|\psi_{gs}\rangle \neq U |\psi_{gs}\rangle$,
 the system undergoes a spontaneous breaking of which symmetry consists of a subgroup $g$ of the Hamiltonian symmetry group $G$
 in association with the unitary transformation $U$.
 Then the two degenerate groundstates seem to have a lower symmetry than the Hamiltonian.
 However, in general, groundstates can have a symmetry which does not belong to their Hamiltonian symmetry.
 For instance, the two-fold degenerate dimerized states can have a $\mathrm{U}(1)$ symmetry
 induced by a local projector on the singlet state of the bond
 at the Majumdar-Ghosh point in the frustrated antiferromagnetic Heisenberg chain,
 so called, the $J_1$-$J_2$ model ~\cite{emergent}.
 Such a symmetry, which is absent in Hamiltonian for a fixed parameter, is called {\it emergent symmetry}
 \cite{emergent,emergent2,Schmaliah,Batista,Young,Liu,Zhang,Silvi,Chen,Armoni}.
 However, the two dimerized states are not connected by
 a unitary transformation relevant to the emergent $\mathrm{U}(1)$ symmetry.
 The actual broken-symmetry giving rise to the dimer phase
 is the one-site translational symmetry
 because the two dimerized states are not one-site translational invariant
 but one of them becomes the other under the one-site translational transformation.
 The broken-symmetry for the two degenerate groundstates
 belongs to the symmetry of the frustrated antiferromagnetic Heisenberg chain
 Hamiltonian.
 In contrast to the normal belief, then,
 for a spontaneous symmetry breaking,
 induced degenerate groundstates do not have a broken-symmetry belonging to
 Hamiltonian symmetry but can have more symmetries not belonging to
 Hamiltonian symmetry.
 However,
 such an emergent symmetry phenomenon in
 groundstates can allow a more crucial question, i.e.,
 are degenerate groundstates induced by the spontaneous symmetry breaking in the Landau
 theory?
 Supposed that there are two degenerate groundstates
 $\left|\psi_1\right\rangle$ and $\left|\psi_2\right\rangle$, i.e.,
 $\left|\psi_1\right\rangle \neq \left|\psi_2\right\rangle$,
 in a system Hamiltonian $H$ for a {\it fixed system parameter},
 where
 the two groundstates satisfy
 $H|\psi_{1/2}\rangle=E_{gs}|\psi_{1/2}\rangle$.  %with a groundstate energy $E_{gs}$ of $H$.
 The two groundstates can be related by a unitary transformation $U$, i.e.,
 $\left|\psi_1\right\rangle = U \left|\psi_2\right\rangle$.
 The unitary transformation $U$ can be applied to see
 how the system Hamiltonian can be transformed by $U$, i.e.,
 $UHU^{\dagger}$.
 In general, due to occurring an emergent symmetry, there are two possible situations, i.e.,
 (i) the Hamiltonian $H$ is invariant under the unitary transformation, i.e.,
 $UHU^{\dagger}=H$,
 or (ii) the Hamiltonian $H$ is not invariant under the unitary transformation, i.e.,
 $UHU^{\dagger} \neq H$.
 Straightforwardly, the case (i) is equivalent to what the Landau theory normally states
 about the spontaneous symmetry breaking.
 For the case (ii), however,
 the two degenerate groundstates might not be understood
 within the spontaneous symmetry breaking in the Landau theory.
 One may then first ask whether such a situation in (ii) can occur in any quantum system.
 In fact, the answer to this question is {\it yes}, i.e.,
 the model system in Eq. (\ref{Ham})
 will be shown to have
 such a two-fold degenerate groundstates, e.g., in the anti-ferromagnetic
 plaquette (AFP)
 phase ($h\neq0$) and the staggered bond (SB) phase.
 Hence, it is natural to ask
 whether an emergent symmetry can be introduced to understand
 why the degenerate groundstates occur in association with the unitary transformation
 $U$.
 Further, can
 such degenerate groundstates be understood within
 a spontaneous symmetry breaking and Hamiltonian symmetry?

 For clear discussions on an emergent symmetry of groundstates in quantum phase transitions,
 we will numerically investigate the spin-$1/2$ infinite plaquette chain in Eq. (\ref{Ham}).
 The infinite matrix product state (iMPS) representation  \cite{mps,Ostlund,vidal}
 is employed for wavefunctions and
 the infinite time-evolving block decimation (iTEBD) method \cite{frank}
 is used to get a groundstate wavefunction
(See Fig. \ref{model}).
 To detect degenerate groundstates for a given parameter,
 we have used the quantum fidelity
 \cite{fida,fidc,deg}
 defined as a overlap measurement between calculated groundstates and an arbitrary reference state.
 From the singular behaviors of the quantum fidelities, also, the phase boundaries are determined.
 To see symmetry of groundstates properly,
 we obtain an explicit form of groundstates
 from careful analysis on local magnetizations and two-site spin correlations in each phase.
 In order to understand a relation between a degenerate groundstate
 and Hamiltonian symmetry,
 we investigate groundstates in whole parameter range of the model Hamiltonian.
 Seven phases with two-fold degenerate groundstates and two phases with a single groundstate are clarified.
 The single groundstates in the two phases are found to have emergent symmetries.
 Furthermore, we find that a two-fold degenerate groundstates
 could not be understood within the spontaneous symmetry breaking picture
 in Landau theory.
 Based on the common properties of the emergent symmetries in our model,
 we discuss and suggest
 a possible extension of the spontaneous symmetry braking picture
 in association with a largest common symmetry group of the constituent
 Hamiltonians.
 Also, the spin structure factors for the degenerate groundstates with an emergent symmetry in each phase
 are shown to have experimentally distinguishable peak structures.

% $-----------------------$

\section{Quantum spin-1/2 plaquette chain}

%%%%%%%%%%%%%%%%%%%%%%%%%%%%%%%%%%%%%%%%%%%%%%%%%%%%%%%%%%%%%%%%%%%%%%%%%%%%
\begin{figure}
\begin{center}
 \begin{overpic}[width=0.4\textwidth]{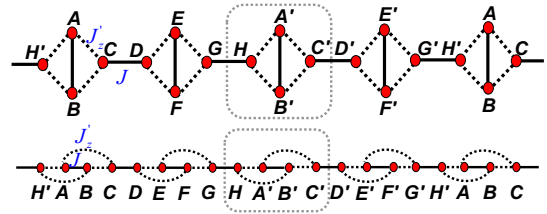}
 \end{overpic}
 \end{center}
\caption{(color online) Top: Spin-$1/2$ plaquette chain in the quasi-one-dimensional
 lattice. The dotted box indicates a plaquette lattice where the four spins interact one another with
 the Ising interaction strength $J'_z$ denoted by the dashed lines. The thick solid lines indicate
 the Heisenberg interactions $J$ between two spins.
 Bottom: One-dimensional spin chain mapped from the quasi-one-dimensional lattice
 for the infinite matrix product state (iMPS) representation.
 The labels from $A$-$H'$ indicate a $16$-site unit cell in the iMPS representation.
 The labels in the one-dimensional lattice correspond to ones in the quasi-one-dimensional lattice.
 }
\label{model}
\end{figure}
%%%%%%%%%%%%%%%%%%%%%%%%%%%%%%%%%%%%%%%%%%%%%%%%%%%

 In order to investigate a relation between spontaneous symmetry breakings and degenerate ground states,
 we consider a quantum spin-1/2 plaquette chain in a magnetic field in Fig. \ref{model}.
 The chain Hamiltonian with the Heisenberg intradimer and the Ising interdimer interactions can be written as
\begin{equation}
 H =  H_{dimer} +H_{plaq} + H_{field},
 \label{Ham}
\end{equation}
 where
\begin{subequations}
\begin{eqnarray}
H_{dimer} &=& J\sum_{i}({\bf S}_{i,u}\cdot {\bf S}_{i,d}+{\bf S}_{i,r}\cdot {\bf S}_{i+1,l}),\\
H_{plaq} &=& J'_z\sum_{i}(S_{i,l}^{z}+S_{i,r}^z)(S_{i,u}^{z}+S_{i,d}^{z}),\\
H_{field} &=& -h\sum_{i}(S_{i,u}^{z}+S_{i,d}^{z}+S_{i,l}^{z}+S_{i,r}^{z}).
\end{eqnarray}
\end{subequations}
  Here. $S_{i,u},S_{i,d},S_{i,l},$ and $S_{i,r}$ represent the spins in the $i$th plaquette, where
  $u$, $d$, $l$, and $r$ denote their corresponding lattice sites, i.e., respectively, at the up, down,
  left, and right sites of the plaquette (See Fig.~\ref{model}).  $H_{dimer}$ and $H_{plaq}$ describe
  the interactions on dimers and plaquettes with
  their corresponding interaction strengths $J$ and $J'_z$ respectively.
  In $H_{field}$, $h$ is the strength of external magnetic field along the $z$ direction.

  For $J \ll J'_z$ and $h=0$, for instance, the Hamiltonian becomes $H \approx H_{plaq}$.
  Because $H_{plaq}$ consists of four-site Ising plaquettes with a periodic boundary condition
  and possesses a global $Z_2\otimes\mathrm{U}(1)$ symmetry,
  the system can have degenerate ground states due to a spontaneous $Z_2$ symmetry breaking in each plaquette.
  Also, for $J \gg J'_z$ and $h=0$,
  the Hamiltonian becomes the dimer Hamiltonian $H \approx H_{dimer}$ that has a global
  $\mathrm{SU}(2)$-rotational symmetry and then the dimerized spins are in a singlet state for $J > 0$.
  Thus the system can have a single ground state that has the same $\mathrm{SU}(2)$ symmetry
  with the dimer Hamiltonian.
  If the applied magnetic field is relatively very strong, i.e., $h\gg J$, $J'_z$,
  the system Hamiltonian has a global $\mathrm{U}(1)$-rotational symmetry
  and the system may be in a fully polarized ground state with the same $\mathrm{U}(1)$ symmetry.
  Then, if the interaction parameters varies
  the system may undergo other types of spontaneous symmetry breakings because
  the above examples show that groundstates can have a different symmetry for different system parameters.
  In this study, for whole system parameter range,
  we will numerically investigate all degenerate ground states and their relations to
  a spontaneous symmetry breaking in the dimer-plaquette chain.

  For our numerical study, we use the iMPS algorithm~\cite{vidal}.
  In order to employ the iMPS representation for wavefunctions,
  one needs to map the dimer-plaquette chain
  to a one-dimensional chain. In the mapped one-dimensional chain in Fig. \ref{model},
 the mapped Hamiltonian is
 four-site translational invariant. To ensure a lattice symmetry breaking in the one-dimensional lattice,
 our numerical simulations are performed in the $4$-site, $8$-site, and $16$-site iMPS wavefunctions.
 In fact, we have found that the $8$-site and $16$-site unit cells in the iMPS representations
 give a same result with a negligible numerical accuracy difference.
 Then, we will discuss our results based on the $16$-site iMPS representation.

\section{Fidelity per lattice site and degenerate ground states}
 For given parameters in an initial iMPS wavefunction, once the energy of the system becomes saturated during
 the iTEBD procedure, one can obtain a groundstate wavefunction.
 If a different choice of initial states leads to a different groundstate with a same saturated energy,
 the system has a degenerate groundstate.
 When the system has more than one groundstates,
 degenerate groundstates can be distinguished by using the quantum fidelity per lattice site (FLS)
 with an arbitrary reference state in Ref. \onlinecite{deg}.
 The FLS $d(|\psi\rangle,|\phi\rangle)$~\cite{fida,deg} is defined as
\begin{equation}
 \ln{d(|\psi\rangle,|\phi\rangle)}=\lim_{L\rightarrow \infty}{\frac{\ln{F(|\psi\rangle,|\phi\rangle)}}{L}},
\end{equation}
 where $L$ is the system size.
 Here, the fidelity $F(|\psi\rangle,|\phi\rangle)$ between a reference state $|\phi\rangle$ and an iMPS ground
 state $|\psi\rangle$ is defined as $F(|\psi\rangle,|\phi\rangle)=|\langle\psi|\phi\rangle|$,
 where $|\psi\rangle \in \{ |\psi_n^{gs}\rangle  \}$ with $n=1, \cdots, N$
 when the system has $N$-fold degenerate ground states  $\{ |\psi^{gs}_{n}\rangle \}$.
 As a scaling parameter~\cite{fida} in the thermodynamic limit, the FLS $d$ is well defined with the
 characteristic properties, i.e.,
(i) normalization $d(|\psi\rangle,|\phi\rangle\equiv|\psi\rangle)= 1$ with $F(|\psi\rangle,|\phi\rangle\equiv|\psi\rangle)= 1$,
(ii) symmetry $d(|\psi\rangle,|\phi\rangle)= d(|\phi\rangle,|\psi\rangle)$ with $F(|\psi\rangle,|\phi\rangle)= F(|\phi\rangle,|\psi\rangle)$, and
(iii) its range $0 \leq d(|\psi\rangle,|\phi\rangle)\leq 1$ with $F(|\psi\rangle,|\phi\rangle)\in{\{0,1\}}$.
 Actually, the FLS corresponds to a projection of degenerate groundstates into a chosen reference state.
 Then, a number of different values of FLS indicates the degeneracy of the groundstates.

 In our numerical calculation, for a given parameter,
 we have used many different initial states randomly chosen numerically
 to determine whether the system has a degenerate groundstate with the FLS.
 If only one value of FLS is detected, the system has a single groundstate, while if $N$ different values
 of FLS are detected, the system has $N$-fold degenerate groundstates.
 However, if a chosen reference state is one of degenerate groundstates,
 the FLS cannot distinguish all of
 degenerate groundstates \cite{deg}.
 If a chosen reference state is not one of degenerate groundstates,
 the FLS distinguishes all degenerate groundstates for a given system parameter.
 Thus,
 to distinguish all different degenerate groundstates properly,
 we have randomly chosen several reference states numerically in our FLS calculation.
 Also, we have calculated groundstates with different truncation dimensions $\chi=4$, $8$, $16$ and $32$.
 The different truncation dimensions have been found to give a same numerical result.
 We present our numerical results for the truncation dimension $\chi=32$.

%%%%%%%%%%%%%%%%%%%%%%%%%Fig. 2%%%%%%%%%%%%%%%%%%%%%%%%%%%%%%%%%%
\begin{figure}
\begin{center}
 \begin{overpic}[width=0.4\textwidth]{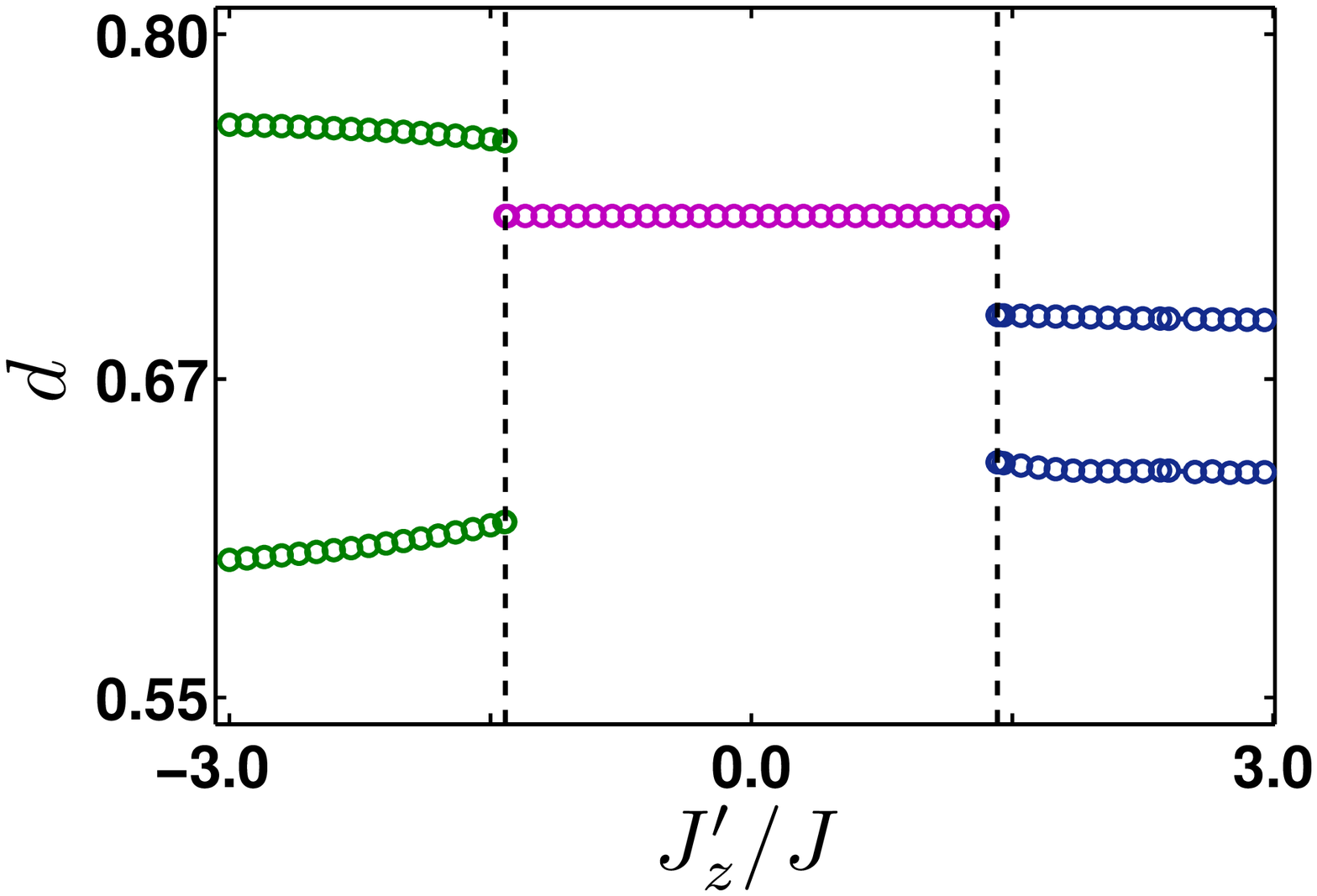}
            \put(1,60){(a)}
\end{overpic}
 \begin{overpic}[width=0.4\textwidth]{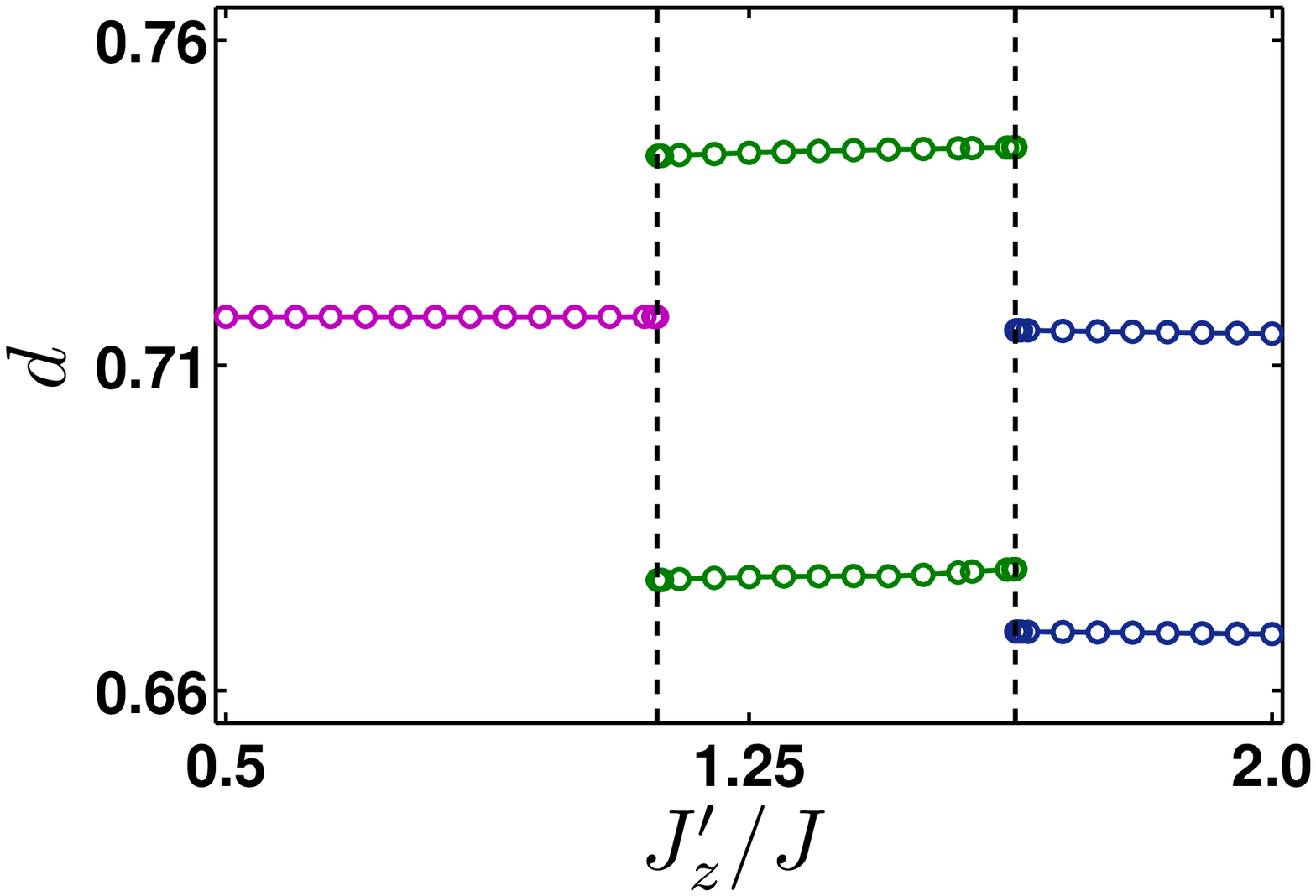}
            \put(1,60){(b)}
\end{overpic}
 \begin{overpic}[width=0.4\textwidth]{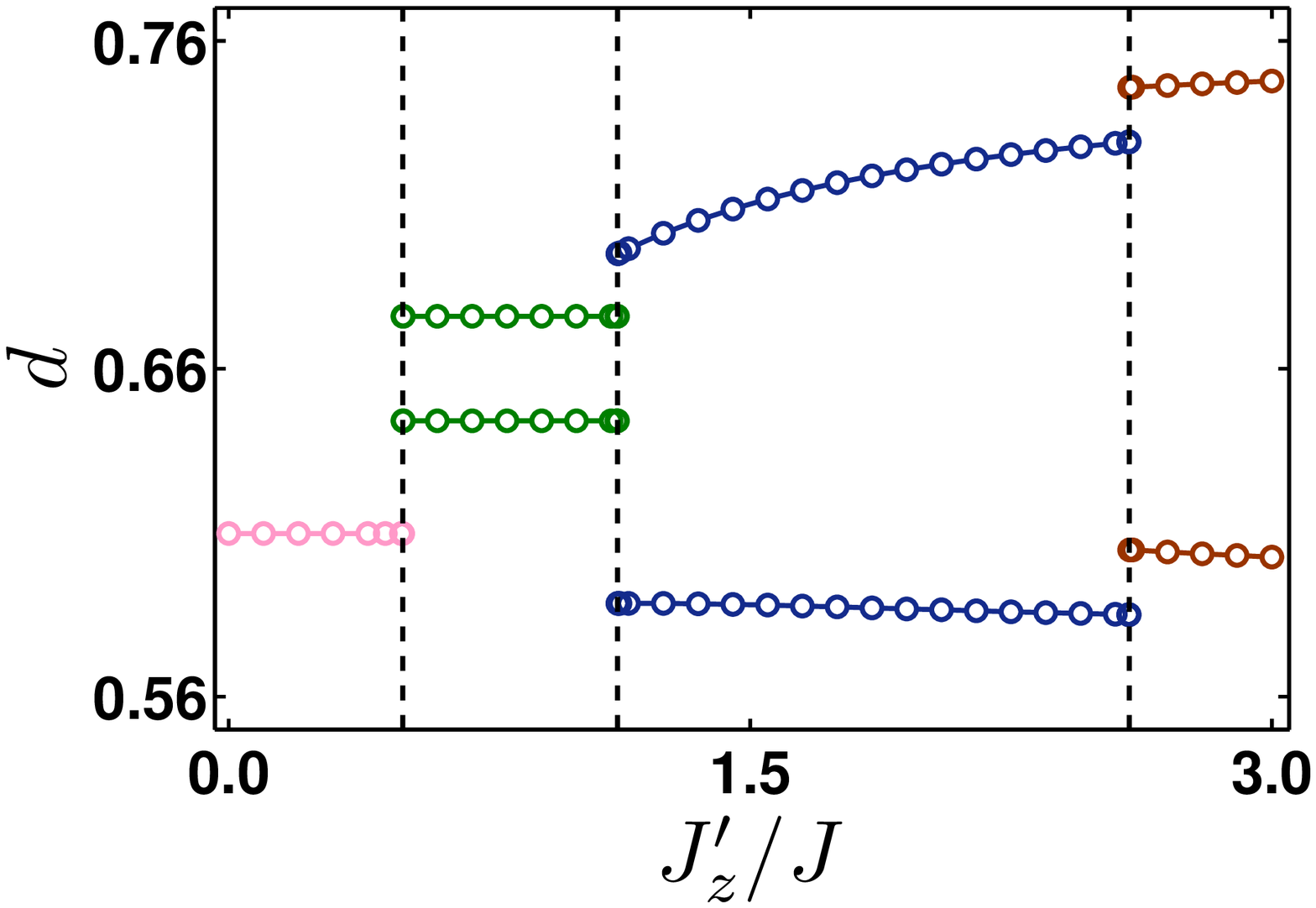}
            \put(1,60){(c)}
 \end{overpic}

 \end{center}
\caption{(color online) Fidelity per lattice
 site for (a) $h=0$, (b) $h=0.5J$ and (c) $h=1.5J$.
 In some ranges of $J'_z/J$, FLS has one or two values.
 A single value of FLS indicates a single groundstate,
 while two different values of FLS indicate doubly degenerate groundstates.
 Also, the discontinuous FLSs at some values of $J'_z/J$
 indicate occurring of phase transitions at those values.}
\label{bif}
\end{figure}
%%%%%%%%%%%%%%%%%%%%%%%%%%%%%%%%%%%%%%%%%%%%%%%%%%%%%%%%%%%%%%%%%%%%%%

 In order to show clearly how to determine a phase boundary from a characteristic behavior
 of FLS with degenerate groundstates, as examples,
 we display FLSs as a function of $J'_z/J$ for (a) $h=0$, (b) $h=0.5J$ and (c) $h=1.5J$ in Fig. \ref{bif}.
% because all possible phases in our model appear in the chosen parameter ranges.
%
%
 In Fig. \ref{bif}(a), we plot FLS as a function of $J'_z/J$ for $h=0$ with $-3J<J'_z<3J$.
 For $J'_z<-1.414J$, FLS has two values.
 At $J'_z=-1.414J$, the FLSs jump into a single value of FLS.
 The observed single value of FLS for $-1.414J < J'_z < 1.414J$ means the system has a single groundstate.
 Normally, a single groundstate implies that a spontaneous symmetry breaking may not happen to the system.
 The further incensement $J'_z/J$ gives rise to two values of FLS for $J'_z>1.414J$.
 For $J'_z < -1.414J$ and $J'_z > 1.414J$,
 the two values of FLS means a two-fold degenerate groundstates
 indicating that possibly a spontaneous symmetry breaking occurs.
 Note that
 the jumps of FLS occurs at $J'_z=-1.414J$ and $J'_z=1.414J$.
 Such discontinuous behaviors of FLS indicate discontinues phase transitions~\cite{fida,deg}.
 In the parameter range, then, there are two phase transitions at $J'_z=-1.414J$ and $J'_z=1.414J$.

 Similarly, in Fig. \ref{bif}(b), we plot FLS as a function of $J'_z/J$ for $h=0.5J$ with $0.5J<J'_z<2J$.
 As the interaction $J'_z/J$ increases from $J'_z=0.5J$, FLS has one value until $J'_z=1.118J$.
 The further incensement $J'_z/J$ gives rise to two values of FLS.
 At $J'_z=1.632J$, the two values of FLS jump into another two values of FLS.
 Then, one can see that FLS has a single value for $0.5J < J'_z < 1.118J$ and two values for $J'_z > 1.118J$.
 Note that at $J'_z=1.632J$ there occurs a phase transition between the two phases,
 both of which have two-fold degenerate groundstates.

 In Fig.~\ref{bif}(c), we display FLS as a function of $J'_z/J$ for $h=1.5J$ with $0<J'_z<3J$.
 When $J'_z/J$ varies from $0$ to $0.5$,
 there is only one value of FLS corresponding to a single ground state.
 With a jump at $J'_z=0.5J$,
 the FLS is split into two values for $J'_z>0.5J$, which means a two-fold degenerate groundstate.
 At $J'_z=1.118J$ and $J'_z=2.59J$, as the $J'_z/J$ increases further,
 two values of FLS jump into another two values.
 Then, the three jumps at $J'_z=0.5J$, $J'_z=1.118J$, and $J'_z=2.59J$
 respectively indicate discontinuous phase transitions \cite{deg,deg2d}
 separating four different phases.
%

%%%%%%%%%%%%%%%%%%%%%%%%%%Fig. 3%%%%%%%%%%%%%%%%%%%%%%%%%%%%%%%%%%%%%%%%%%
\begin{figure}
\vskip 5mm
\begin{center}
 \begin{overpic}[width=0.4\textwidth]{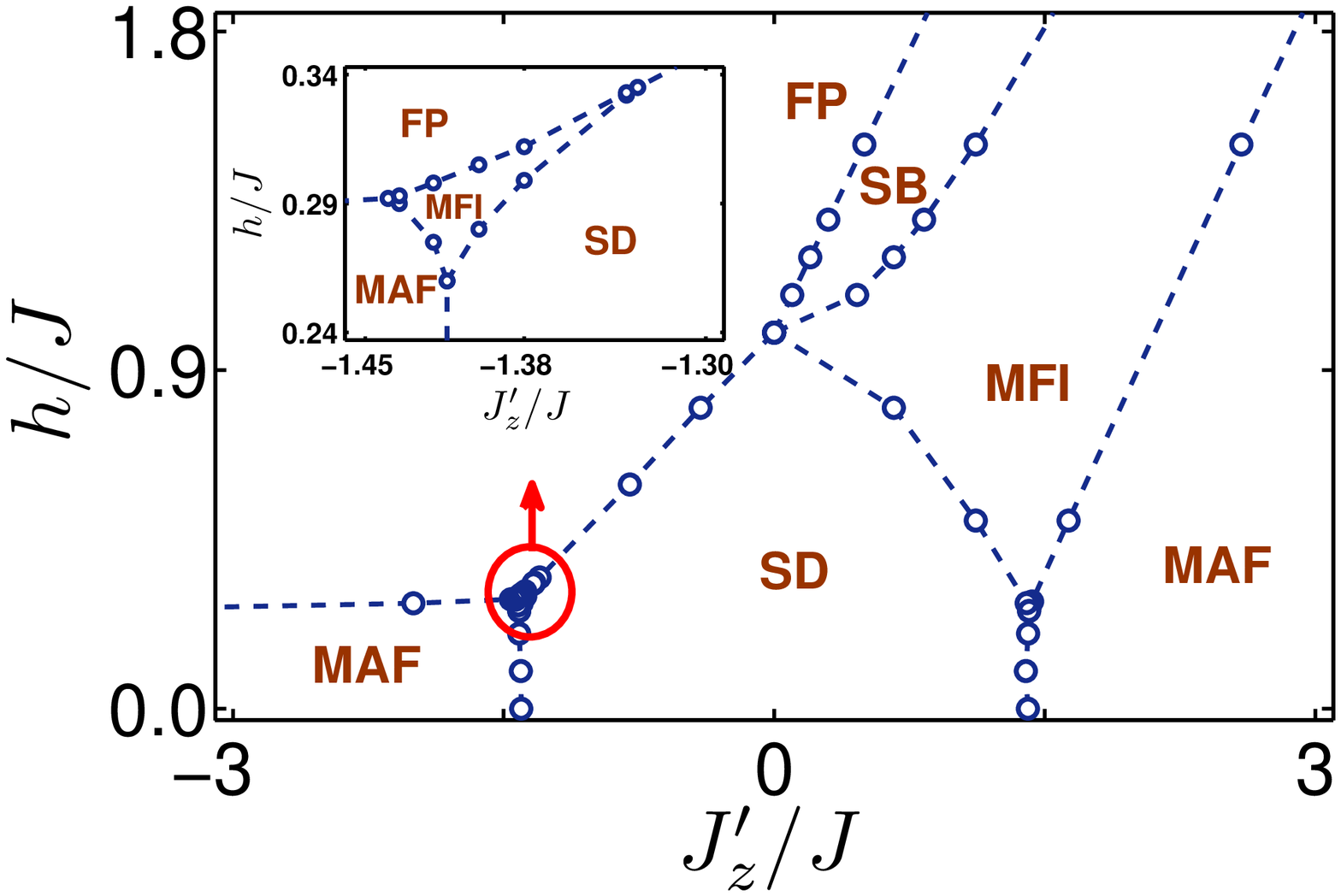}
                 \put(1,60){(a)}
\end{overpic}
 \begin{overpic}[width=0.4\textwidth]{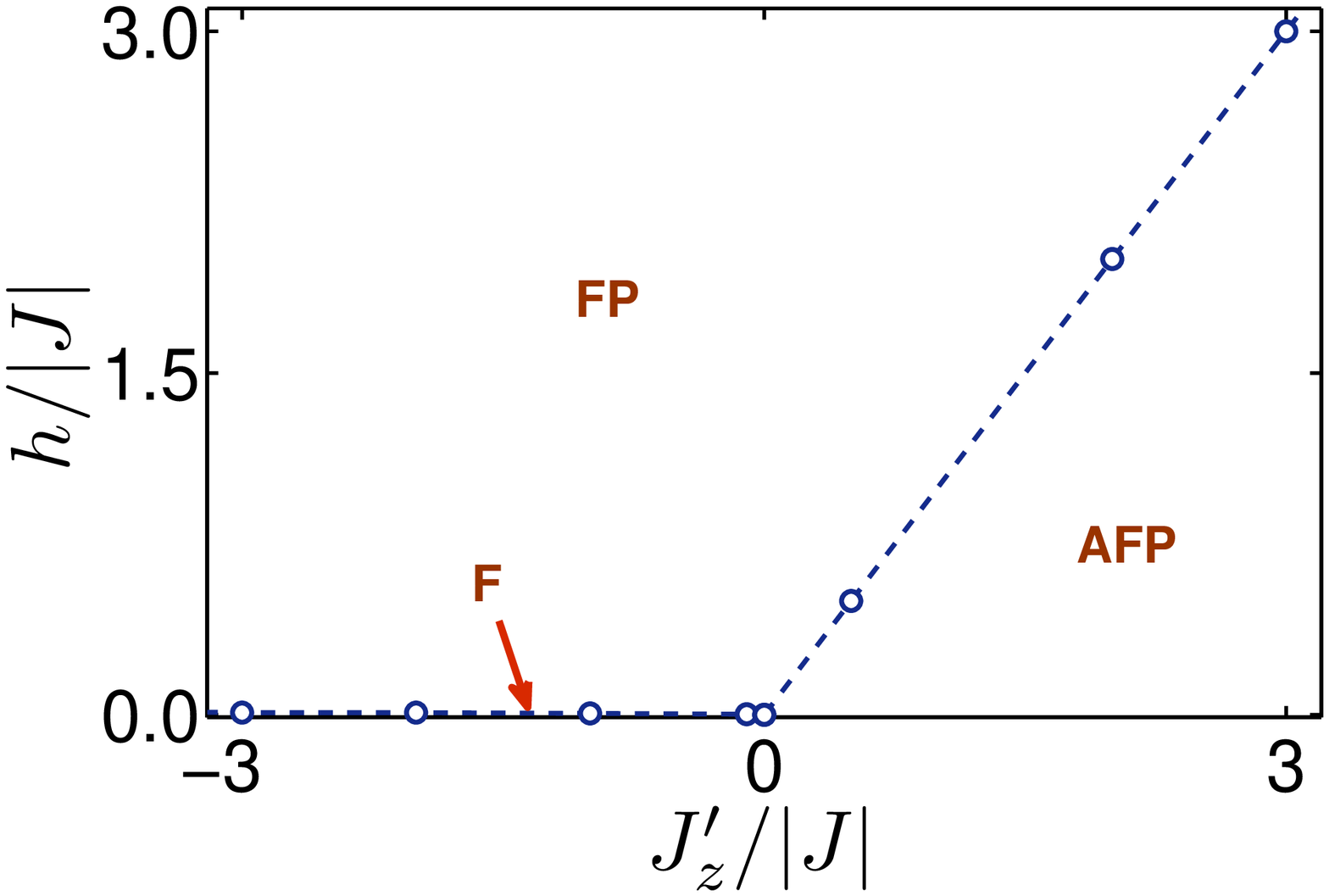}
                 \put(1,60){(b)}
\end{overpic}
 \end{center}
\caption{(color online)
 Groundstate phase diagrams for the spin-$1/2$ plaquette chain
 for (a) $J > 0$ and (b) $J < 0$ in the $J'_z$-$h$ plane.
 The phase boundaries are determined by the discontinuous behaviors
 of the fidelity per lattice site (FLS).
 For each phase, its characteristic property is discussed in the text
 in association with relevant symmetries of groundstates and the system Hamiltonian.
} \label{phase}
\end{figure}
%%%%%%%%%%%%%%%%%%%%%%%%%%%%%%%%%%%%%%%%%%%%%%%%%%%%%%%%%%%%%%%%%%%%%%

 Actually, by using the property of FLS,
 we have detected seven phases and their phase boundaries for the spin-$1/2$ plaquette
 model.
 Figure~\ref{phase} shows the phase boundaries between the nine different phases denoted as
 singlet-dimerized (SD), fully polarized (FP),
 staggered bond (SB), modulated anti-ferromagnetic plaquette (MAFP),
 modulated ferromagnetic plaquette (MFP),
 staggered anti-ferromagnetic plaquette (SAFP),
 staggered ferromagnetic plaquette (SFP), anti-ferromagnetic plaquette (SAF), and
 ferromagnetic (F) phases.
 The SD and FP phases have a single groundstate from a single value of FLS.
 While the other seven phases, i.e.,
 SB, MAFP, MFP, SAFP, SFP, AFP, and F phases, have a two-fold degenerate groundstate.
 Note that, without knowing whether any spontaneous symmetry breaking occurs or not,
 our FLS is shown to determine the phase boundaries and
 the degeneracy of groundstates for each phase.
 For the parameter range, i.e.,  $ J, J'_z, h \geq 0$,
 a transfer matrix method has been used to obtain a similar phase diagram
 and groundstates in Ref. \onlinecite{ES}.
 In order to discuss about a relation between spontaneous symmetry breakings and degenerate groundstates
 for the phases, one should know an explicit symmetry of groundstates.
 To do this in our study,
 we will investigate a local property of groundstate wavefunctions, i.e., local mangetizations
 and two-point spin correlations for each phase in the following sections.

\section{Groundstate wavefunctions and emergent symmetries}
 Symmetries of calculated iMPS groundstates can be understood
 from their local property and two-point spin correlations.
 Especially, local magnetizations can give us an information
 about spin rotational symmetries of groundstates.
 Also,
 if they have a periodic structure, lattice symmetries of them could also be understood.
 Based on such properties of local magnetizations and two-point spin correlations,
 one may discuss symmetries of groundstate wavefunctions.
 However, the best way to discuss symmetries of groundstates wavefunction
 requires knowing an explicit form of groundstate wavefunctions.
 In our model, we have found that
 for our whole parameter range, i.e., all nine phases, only $z$-components of local magnetizations
 are nonzero from all groundstates, i.e., $\langle S_x\rangle = 0 = \langle S_y \rangle$.
 Furthermore, two-point spin correlations defined as
 $C^{\alpha\alpha'}(|i-j|)=\langle S^\alpha_i S^{\alpha'}_j\rangle$
 with $\alpha,\alpha' \in \{ x, y, z \}$  are found to have a periodic structure.
 These facts allow us to extract an explicit form of groundstate wavefunctions from
 the two-point spin correlations and local magnetizations for each phase.
 We will discuss emergent symmetries of groundstates in each phase.

%%%%%%%%%%%%%%%%%%%%%%%%%%%%%%%fig. 4%%%%%%%%%%%%%%%%%%%%%%%%%
 \begin{figure}
 \begin{center}
  \begin{overpic}[width=0.4\textwidth]{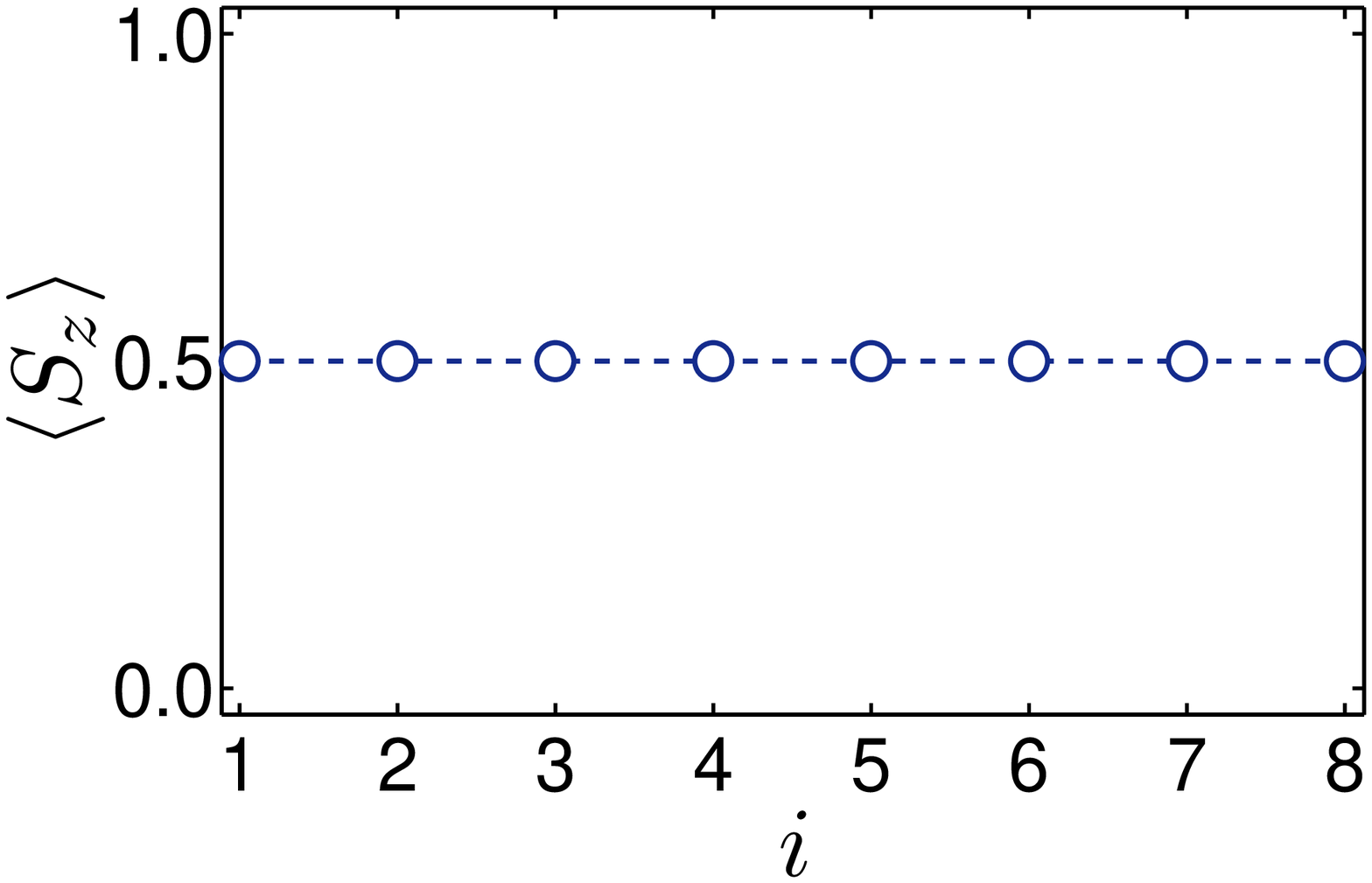}
            \put(1,60){(a)}
  \end{overpic}
  \begin{overpic}[width=0.4\textwidth]{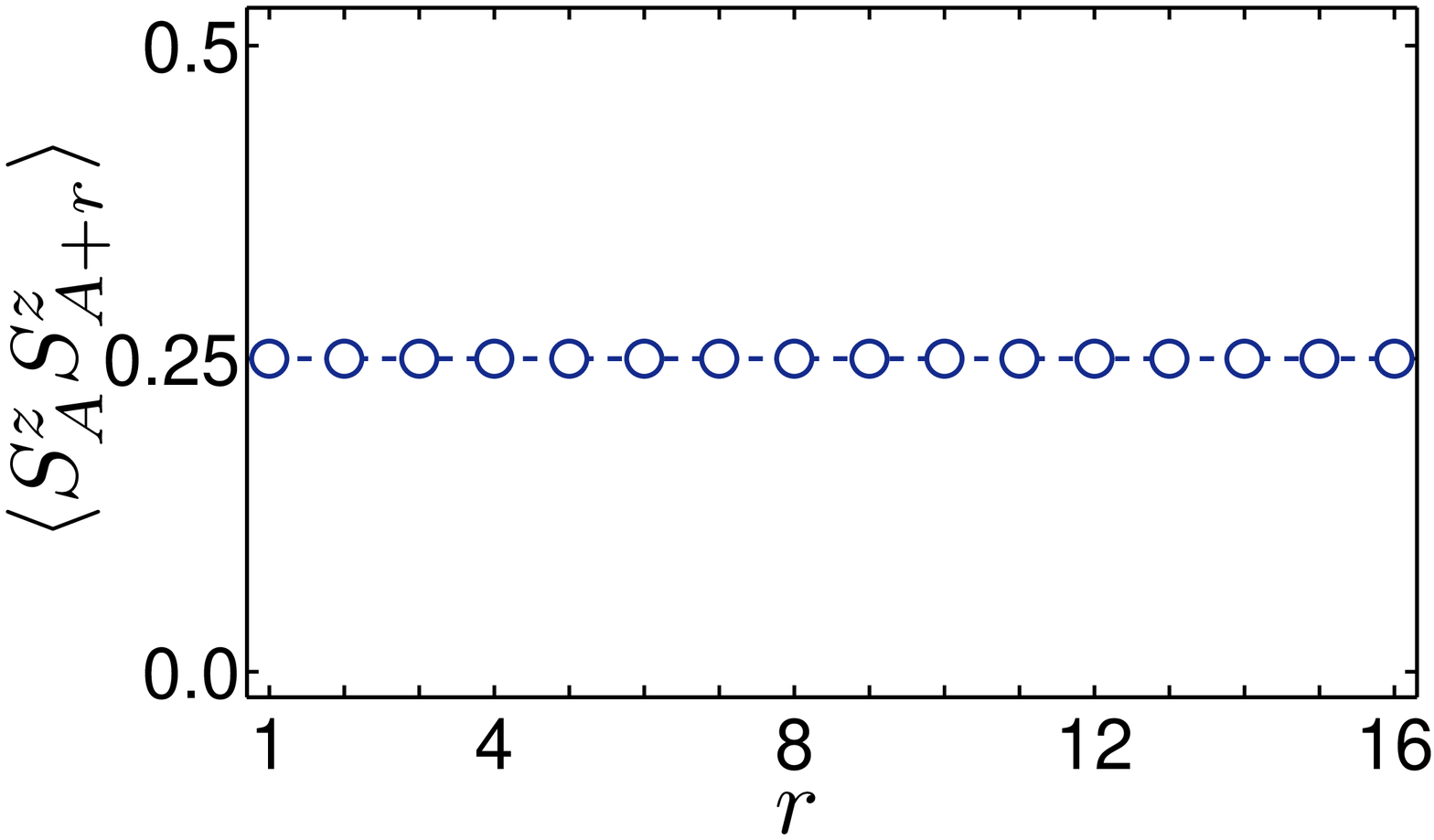}
            \put(1,54){(b)}
  \end{overpic}
  \begin{overpic}[width=0.35\textwidth]{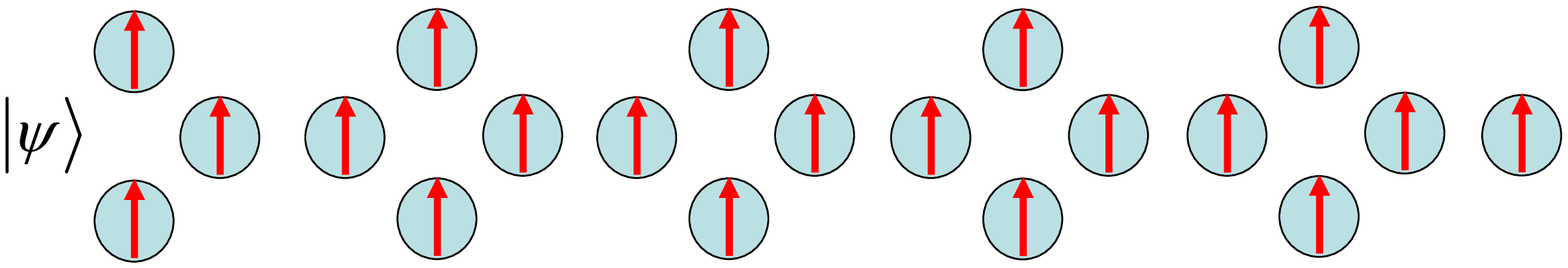}
            \put(-6,20){(c)}
  \end{overpic}
 \end{center}
\caption{(color online)
 (a) Local magnetization $\langle S_z\rangle$ at the lattice site $i$
 ($i = 1,2, \cdots$ correspond to $A$-$H$ in regular sequence in Fig. \ref{model}.)
 and (b) two-point spin correlations $\langle S^{z}_{A} S^{z}_{A + r}\rangle$
 as a function of lattice distance between two sites
 $A$ and $A + r$ for $J'_z=0.2J$ and $h=1.5J$.
 (c) Pictorial representation of the ground state with spin configuration in the FP phase.
 The circles in the diagram are spins at each site.
 A red arrow indicate a fully polarized magnetization. }
 \label{FP}
\end{figure}
%%%%%%%%%%%%%%%%%%%%%%%%%%%%%%%%%%%%%%%%%%%%%%%%%%%%%%%%%%%%%%%%%%%%%%%%%%%%%%%%%%%%%

\subsection{Fully polarized phase}
 For the FP phase, there is a single groundstate from the FLS calculation.
 In Fig. \ref{FP}(a), the local magnetizations at the lattice site $i$
 and (b) the two-spin correlations as a function of lattice distance $r$
 are plotted for $J'_z=0.2J$ and $h=1.5J$.
 The local magnetizations have their maximum value, i.e. $\langle S_z \rangle=1/2$ at all lattice sites.
 In Fig.~\ref{FP}(b), the two-point correlations satisfy
 $\langle S^z_i S^z_j\rangle =\langle S^z_i\rangle\langle S^z_j\rangle$
 for any pair of two spins.
 Actually, we have also observed that $\langle S^x_i S^x_j\rangle = 0 =\langle S^y_i S^y_j\rangle$ and
  $\langle S^{\alpha}_i S^{\alpha'\neq \alpha}_j\rangle = 0$ with $\alpha, \alpha' \in \{x,y,z\}$
  (not displayed).
 In general,
 if spin correlation between two sites $i$ and $j$ satisfies
\begin{subequations}
 \begin{eqnarray}
 \langle S^{\alpha}_i S^{\alpha}_j\rangle &=& \langle S^{\alpha}_i\rangle \langle S^{\alpha}_j\rangle,
 \label{eq:6a}
  \\
 \langle S^{\alpha}_i S^{\alpha'\neq \alpha}_j\rangle &=& 0,
 \label{eq:6b}
  \end{eqnarray}
 \end{subequations}
 the spin states for the two sites are a product state.
 Also, if any pair of two sites satisfies the conditions in Eqs.~(\ref{eq:6a}) and (\ref{eq:6b}),
 the state for the system is in a product state of the spin states of each site.
 For our FP groundstates,
 any two-point spin correlation satisfies the conditions in Eqs.~(\ref{eq:6a}) and (\ref{eq:6b}),
 which implies that the groundstate is a product state of the spin states of each lattice site.
 Consequently, in the original quasi-one-dimensional lattice,
 with the fully polarized magnetizations $\langle S^z_i \rangle = \langle S^z_j \rangle = 1/2$,
 the groundstate for the FP phase can be written as
 \begin{equation}
 \left|\psi\right\rangle
 =\prod_i \left|\uparrow_{i,u} \uparrow_{i,d} \uparrow_{i,l} \uparrow_{i,r} \right\rangle.
\label{FPW}
 \end{equation}
 In Fig.~\ref{FP} (c), the groundstate for the FP phase are presented pictorially.

 The groundstate in Eq.~(\ref{FPW}) has a $\mathrm{U}(1)$ rotational symmetry in the $x$-$y$ plane.
 Also, the Hamiltonian in Eq. (\ref{Ham}) has the $\mathrm{U}(1)$ rotational symmetry.
 Both the groundstate and the Hamiltonian are one-plaquette translational invariant.
 Hence, for the FP phase, no spontaneous symmetry  breaking occurs,
 which results in the single groundstate.
 As a result, this FP phase can be explained within the Landau's spontaneous symmetry breaking theory.
 By comparing with the symmetry of the Hamiltonian,
 however, one can notice emergent symmetries for the groundstate.
 For instance, for a lattice rotation of each plaquette,
 the groundstate is invariant but the Hamiltonian is not.
 For a vertical-to-horizontal site exchange
 in each plaquette, e.g, $(A,B) \leftrightarrow (C,D)$, also, the groundstate is invariant
 but the Hamiltonian is not.
 These imply that the lattice-rotation and the exchange symmetries are emergent for the groundstate.
 Consequently, the groundstate has such emergent symmetries not belonging to the Hamiltonian symmetry.

%%%%%%%%%%%%%%%%%%%%%%%%%%%%%%%%%%%%%fig. 5%%%%%%%%%%%%%%%%%%%%%%%%%%%%%%%%%%%%%%%%%%%
 \begin{figure}
 \begin{center}
  \begin{overpic}[width=0.4\textwidth]{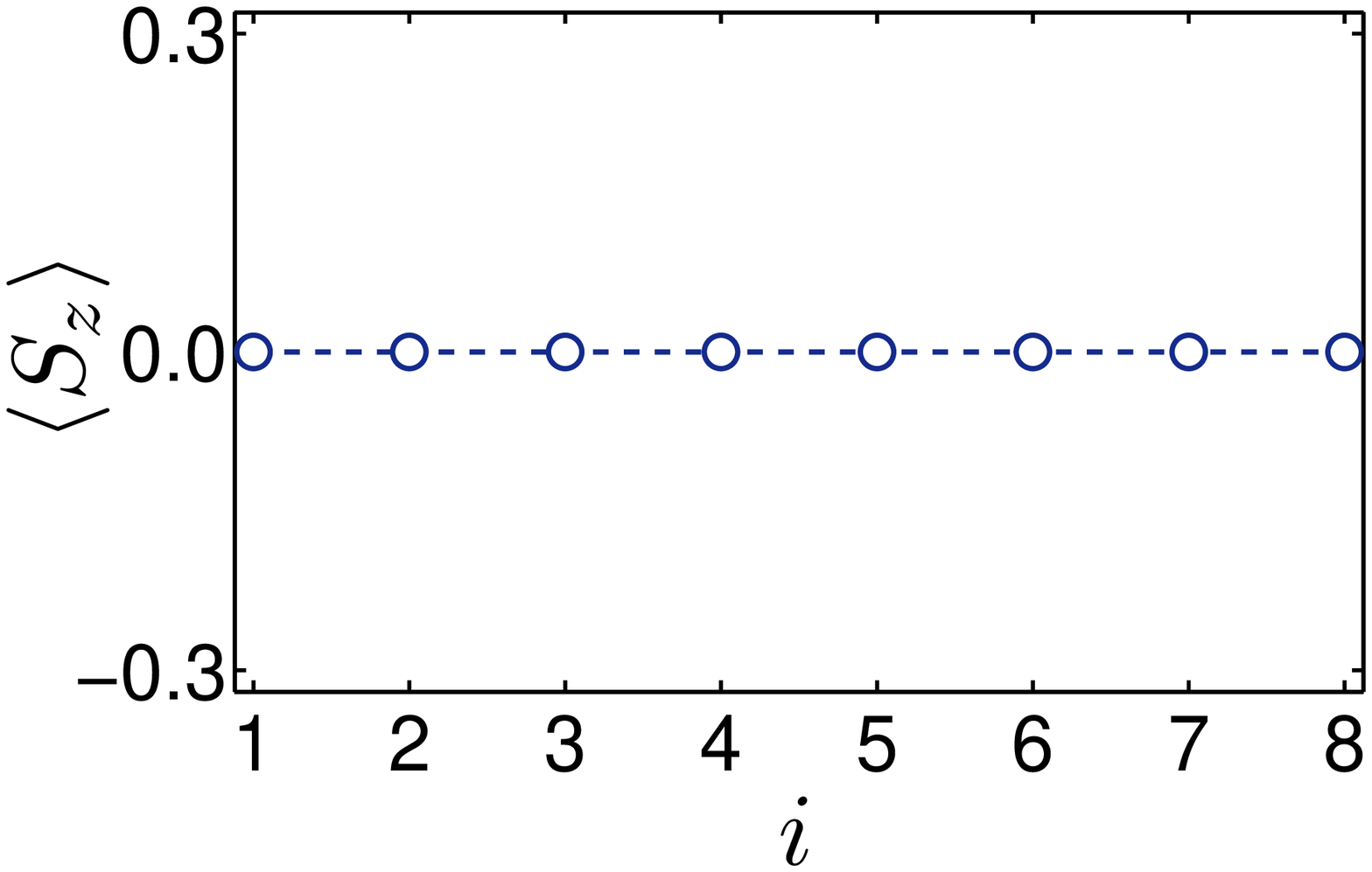}
            \put(1,58){(a)}
 \end{overpic}
  \begin{overpic}[width=0.4\textwidth]{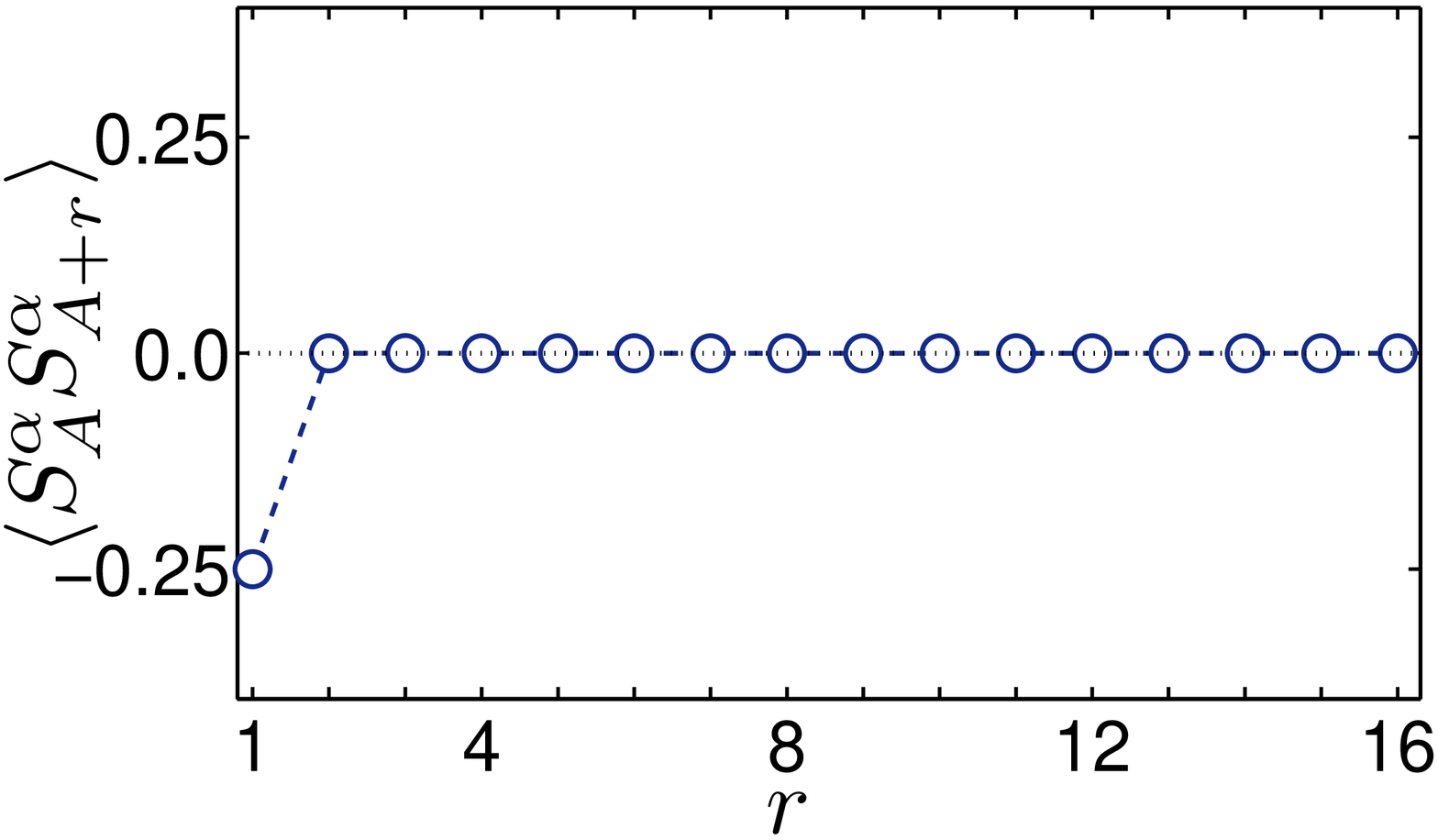}
            \put(1,55){(b)}
  \end{overpic}
  \begin{overpic}[width=0.4\textwidth]{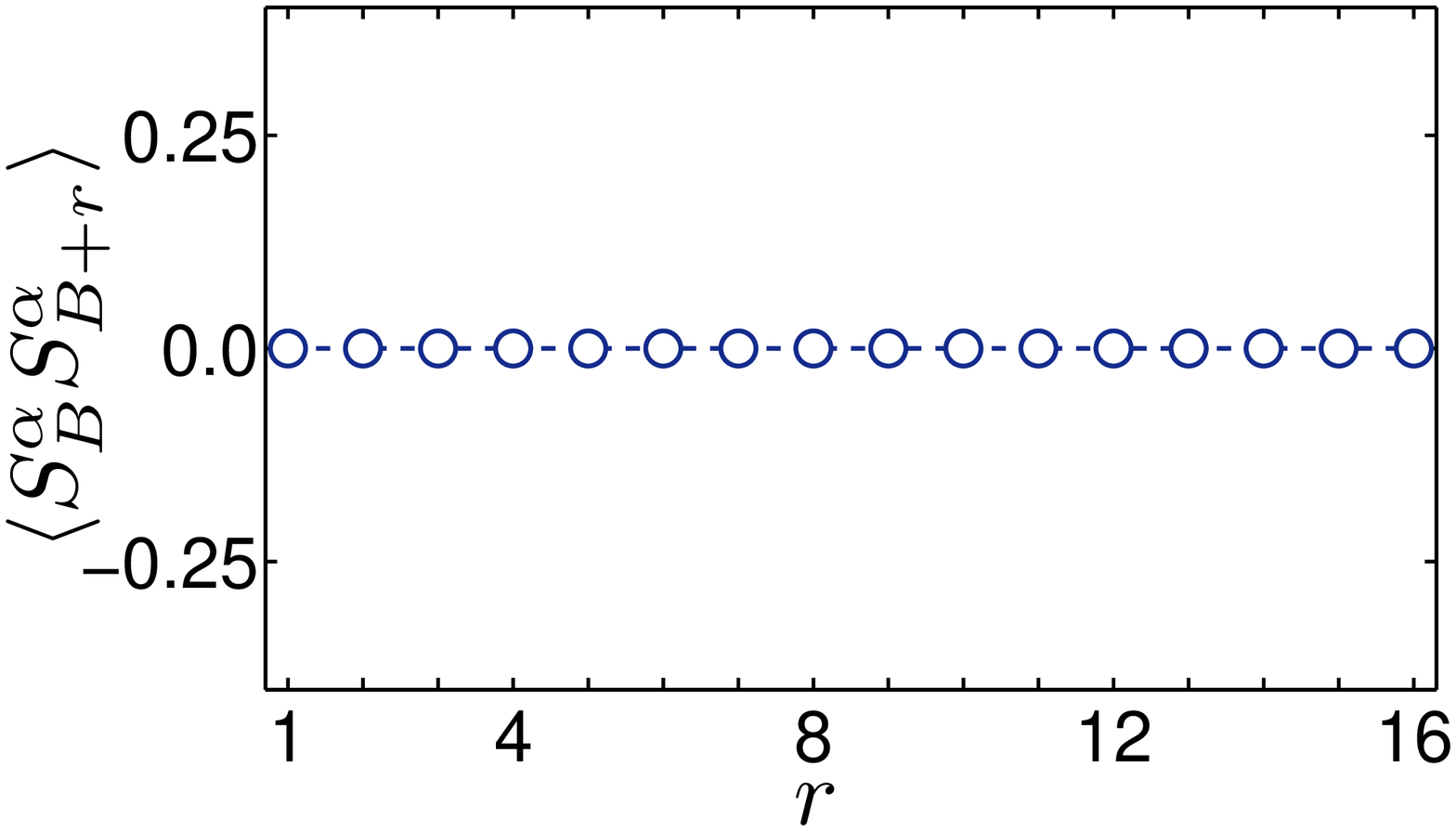}
            \put(1,52){(c)}
  \end{overpic}
  \begin{overpic}[width=0.35\textwidth]{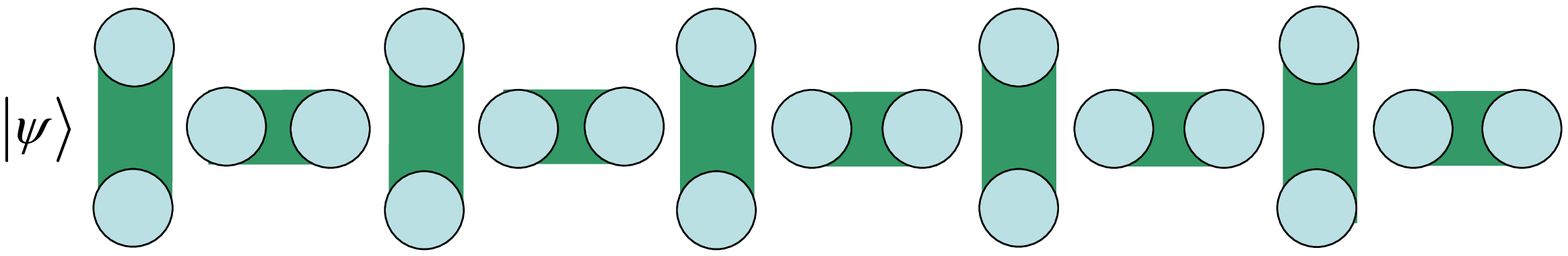}
            \put(-6,20){(d)}
  \end{overpic}
 \end{center}
\caption{
 (a) Local magnetization $\langle S_z\rangle$ at the lattice site $i$
 for $J'_z=0.6J$ and $h=0.5J$ in the SD phase.
 Two-point spin correlations (b) $\langle S^{\alpha}_{A} S^{\alpha}_{A + r}\rangle$
 and (c) $\langle S^{\alpha}_{B} S^{\alpha}_{B + r}\rangle$ with $\alpha=x,y,z$
 as a function of lattice distance between two sites
 $A/B$ and $A/B + r$ for $J'_z=0.6J$ and $h=0.5J$.
 For the singlet dimerized phase, two spins $AB$ are dimerized and in a singlet pair state.
 Note that for the sites $C$ and $D$, the two-point spin correlations
 $\langle S^{\alpha}_{C/D} S^{\alpha}_{C/D + r}\rangle
  = \langle S^{\alpha}_{A/B} S^{\alpha}_{A/B + r}\rangle$.
 (d) Pictorial representation of the ground state with spin configuration.
 The circles in the diagram are spins at each site.
 A thick green line connecting two spins represents a singlet state.}
 \label{SDcorrab}
\end{figure}
%%%%%%%%%%%%%%%%%%%%%%%%%%%%%%%%%%%%%%%%%%%%%%%%%%%%%%%%%%%%%%%%%%%%%%%%%%%%%%%%%%%%%

\subsection{Singlet dimerized phase}
 Similar to the FP phase, for the SD phase, a single groundstate is detected from the FLS calculation.
 In Figs. \ref{SDcorrab}(a)-(c),
 the local magnetizations at the lattice site $i$
 and the two-spin correlations as a function of lattice distance $r$
 are respectively plotted for $J'_z=0.6J$ and $h=0.5J$.
 For the two-point spin correlations,
 we observe that $\langle S^{\alpha}_i S^{\alpha}_j\rangle$ have a same behavior and
 $\langle S^{\alpha}_i S^{\alpha'\neq \alpha}_j\rangle = 0$.
 In Figs. ~\ref{SDcorrab}(b) and \ref{SDcorrab}(c), thus,
 the two-point spin correlations $\langle S^\alpha_i S^\alpha_{i+r}\rangle$
 show that except for the nearest
 two-spin correlation $\langle S^\alpha_A S^\alpha_{B}\rangle$
 and $\langle S^\alpha_C S^\alpha_{D}\rangle$,
 all other two-point spin correlations are zero.
 In general, if two-point spin correlations between sites $i$ and $j$ satisfies
\begin{subequations}
 \begin{eqnarray}
 \langle S^{\alpha}_i S^{\alpha}_{i+1} \rangle
 & \neq & \langle S^{\alpha}_i\rangle \langle S^{\alpha}_{i+1} \rangle,
 \label{eq:8a}
\\
 \langle S^{\alpha}_i S^{\alpha}_j \rangle
 &=& \langle S^{\alpha}_i\rangle \langle S^{\alpha}_j \rangle \mbox{~~for~~} j\neq i+1,
 \label{eq:8b}
\\
  \langle S^{\alpha}_i S^{\alpha'}_j \rangle
  &=& 0,
 \label{eq:8c}
 \end{eqnarray}
\end{subequations}
 the two spins for sites $i$ and $i+1$ are dimerized.
 Hence, the two sites $(A,B)$ and $(C,D)$ are dimerized.
 Also, the groundstate is a product state of the dimerized spin pairs, i.e.,
 the two sites $(A,B)$ and $(C,D)$.
 Furthermore,
 the local magnetizations are zero at all lattice sites, which
 implies that the two bases $\{ \left|\uparrow\right\rangle, \left|\downarrow\right\rangle\}$ at each site
 equally contribute for the local magnetizations.
 The locally dimerized spin pairs
 then are in an opposite spin state, i.e., for instance,
 $\left|\uparrow_A\right\rangle \left|\downarrow_{B}\right\rangle$
 or $\left|\downarrow_A\right\rangle \left|\uparrow_{B}\right\rangle$ because
 $\langle S^z_A S^z_{B}\rangle = -1/4$. This fact means that the two-spin state
 can be written by a linear combination of the two possible states, i.e.,
 $\left| \psi_{AB} \right\rangle = a
 \left|\uparrow_A\right\rangle \left|\downarrow_{B}\right\rangle
 + b \left|\downarrow_A\right\rangle \left|\uparrow_{B}\right\rangle$
 with $|a|^2+|b|^2=1$, where $a$ and $b$ are numerical coefficients.
 In this case, the zero magnetization at each site
 give a constraint condition, i.e., $|a|=|b|$.
 Such a condition $|a|=|b|$ allows only $|a|=|b|=1/\sqrt{2}$
 with the normalization condition $|a|^2+|b|^2=1$.
 In addition,
 from the property of $\langle \mathbf{S_A}\cdot \mathbf{S}_{B}\rangle = -3/4$
 with $\langle S^\alpha_A S^\alpha_{B}\rangle = -1/4$,
 one can determine the coefficients explicitly, i.e.,
 $a=-b=1/\sqrt{2}$.
 Then, the dimerized spin pairs are in a spin singlet state
 (denoted by a thick line between a pair of two spins in Fig.~\ref{SDcorrab}(d)),
 i.e.,  $\left| \psi_{AB} \right\rangle = \left(
 \left|\uparrow_A\right\rangle \left|\downarrow_{B}\right\rangle
 - \left|\downarrow_A\right\rangle \left|\uparrow_{B}\right\rangle \right)/\sqrt{2}$.
 Similar discussions
 should lead to a spin singlet state for other pairs of two spins $C$ and $D$.
Consequently, for the SD phase, the groundstate can be written as
\begin{equation}
 \left|\psi\right\rangle = \prod_i \frac{1}{2}
 \left( \left|\uparrow_{i,u}\downarrow_{i,d}\right\rangle-\left|\downarrow_{i,u}\uparrow_{i,d}\right \rangle \right)
 \left(\left |\uparrow_{i,r}\downarrow_{i+1,l}\right\rangle-\left|\downarrow_{i,r}\uparrow_{i+1,l}\right \rangle \right).
 \label{SDW}
\end{equation}

 Now, let's discuss a symmetry of the groundstate for the SD phase with $h \geq 0$.
 In the quasi-one dimensional plaquette lattice,
 the groundstate in Eq. (\ref{SDW}) for the SD phase is one-plaquette translational invariant.
 Also,
 the groundstate in the SD phase has a $\mathrm{SU}(2)$ spin-rotational symmetry for $h \geq 0$,
 while for $h=0$ ($J'_z\neq0$), due to the interdimer Ising interactions,
 the system Hamiltonian possesses a $Z_2$ spin-flip symmetry for the $z$-direction and
 a $\mathrm{U}(1)$-rotational symmetry along the $z$-axis, and
 for $h\neq0$, the original $\mathrm{Z}_2\otimes \mathrm{U}(1)$ symmetry of the Hamiltonian is broken into
 a $\mathrm{U}(1)$-rotational symmetry due to the magnetic field applied to the $z$-direction.
 Thus, interestingly, in both cases $h=0$ and $h\neq0$,
 the groundstate wavefunction possesses more symmetries than the Hamiltonian, i.e.,
 $\mathrm{Z}_2\otimes \mathrm{U}(1)\subset \mathrm{SU}(2)$ and $\mathrm{U}(1)\subset \mathrm{SU}(2)$, respectively.
 Actually, the $\mathrm{SU}(2)$-rotational symmetry of the groundstate
 could not be expected from the system Hamiltonian for the parameter range of the SD phase
 because the system Hamiltonian is not invariant under the $\mathrm{SU}(2)$ rotation.
 In addition, the groundstate has a vertical-to-horizontal site-exchange symmetry, i.e.,
 the groundstate is invariant for exchanging sites $(A,B) \leftrightarrow (C,D)$
 ($(E,F) \leftrightarrow (G,H)$) in each plaquette.
 The $\mathrm{SU}(2)$-rotational and the exchange symmetries are emergent for the groundstate.
 For the SD phase in the spin-$1/2$ plaquette chain,
 as a result, the single groundstate has the emergent $\mathrm{SU}(2)$-rotational and exchange symmetries.

%%%%%%%%%%%%%%%%%%%%%%%%%%%%%%%%%%%%%%%fig. 6%%%%%%%%%%%%%%%%%%%%%%%%%%%%%%%%%%%%%%%%%
 \begin{figure}
 \begin{center}
  \begin{overpic}[width=0.4\textwidth]{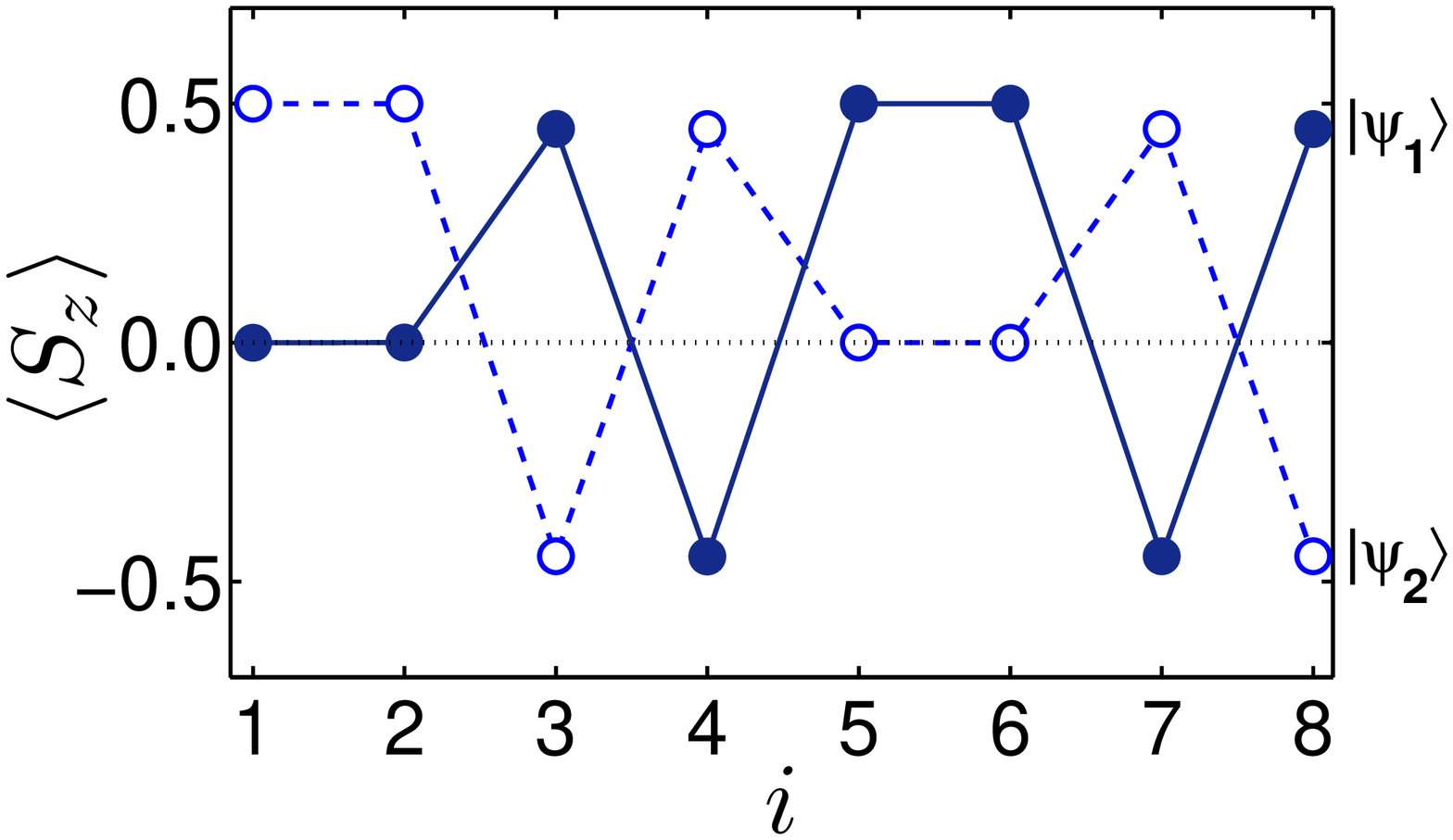}
   \put(1,55){(a)}
  \end{overpic}
%  \begin{center}
  \begin{overpic}[width=0.35\textwidth]{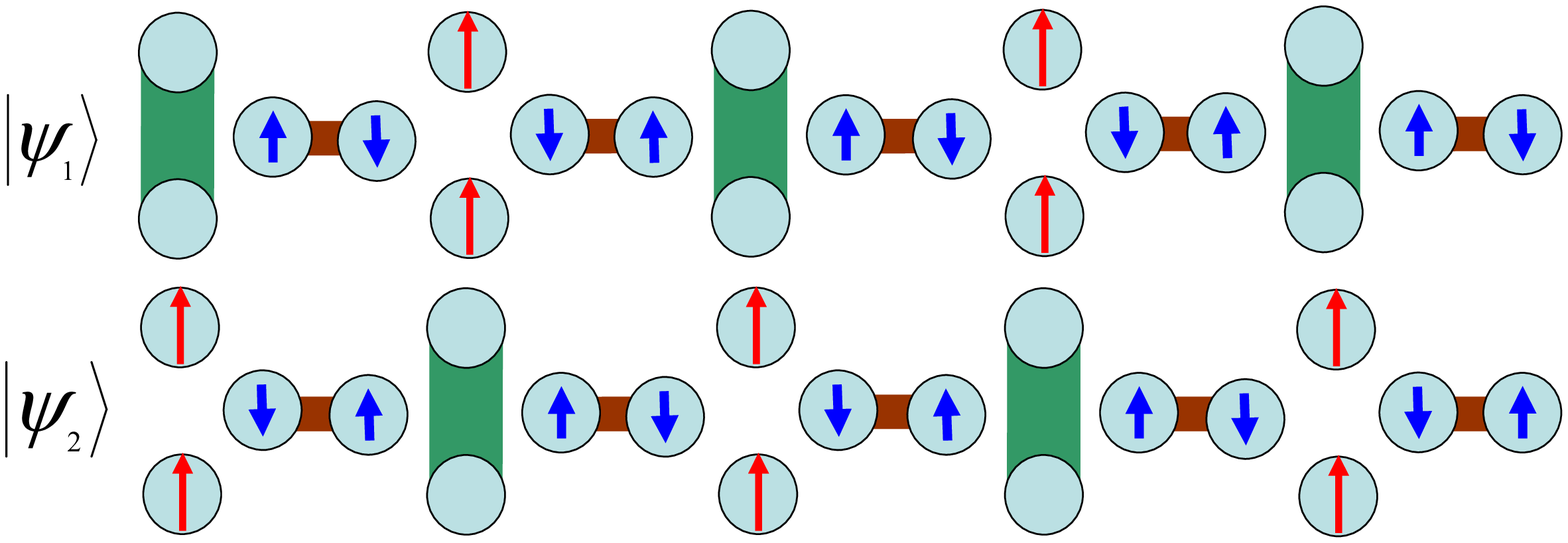}
   \put(-6,35){(b)}
  \end{overpic}
 \end{center}
\caption{(color online)
 (a) Local magnetization $\langle S_z\rangle$ at the lattice site $i$  for $J'_z=2J$ and $h=1.5J$
 in the MAFP phase.
 (b) Pictorial representation of the ground states with spin configuration.
 Red arrows indicate fully polarized spins.
 Two spins connected by a thick green line
 represents a singlet state, which means that the local magnetizations
 at the spin sites are zero.
 Two spins connected by a thinner brown line
 represents an entangled pair states, where blue arrows denote
 magnetizations at the spin sites.
 Note that each plaquette has an anti-ferromagnetic configuration
 of local magnetizations.
 }
 \label{MFImg}
\end{figure}
%%%%%%%%%%%%%%%%%%%%%%%%%%%%%%%%%%%%%%%%%%%%%%%%%%%%%%%%%%%%%%%%%%%%%%%%%%%%%%%%%%%%%

\subsection{Modulated anti-ferromagnetic plaquette phase}
 In contrast to the FP and SD phases,
 for the MAFP phase, a two-fold degenerate groundstate is detected.
 To discuss local magnetic properties of the states,
 let us first denote the two groundstate wavefunctions as $|\psi_n\rangle$
 with $n \in \{ 1, 2 \}$.
 For the system parameters $J'_z=2J$ and $h=1.5J$, from the two degenerate groundstates,
 the local magnetizations $\langle \psi_n |  S_z | \psi_n\rangle$
 are plotted at lattice sites in Fig.~\ref{MFImg}(a).
 For $|\psi_1\rangle$, the local magnetizations have a two-plaquette (eight-site) periodic structure where
 the first two sites, i.e., $A$ and $B$, have zero-magnetizations $\langle S_z \rangle =0$,
 the second two sites, i.e., $C$ and $D$,
 have an opposite value of magnetizations depending on $J'_z/J$, i.e., $\langle S_z^C\rangle=-\langle S_z^D\rangle >0$,
 the third two sites, i.e., $E$ and $F$, have an maximum magnetizations $\langle S_z \rangle = 1/2$,
 and the fourth two sites, i.e., $G$ and $H$,
 have an opposite value of magnetizations depending on $J'_z/J$, i.e., $\langle S_z^G\rangle=-\langle S_z^H\rangle <0$.
 For other values of system parameters,
 the characteristic behaviors of the local magnetizations are not changed
 and only the values of the local magnetizations at the sites $C$, $D$, $G$, and $H$
 are determined by $J'_z/J$.
 In this MAFP phase, also, the local magnetizations has a very characteristic property, i.e,
 an anti-ferromagnetic configuration in each plaquette.
 Note that for one-plaquette (four-site) shift,
 the local magentizations from $|\psi_2\rangle$ are equal to ones from $|\psi_1\rangle$.
 This implies that one groundstate becomes the other groundstate under one-plaquette (four-site) shift transformation.

 To obtain an explicit form of groundstates,
 we have discussed the detailed properties of two-site spin correlations in the Appendix~\ref{app:MAFP1}.
 From the discussions for $|\psi_1\rangle$,
 it is found that for the first two sites $i = A$ and $B$,
 the properties of the two-point spin correlations with the local zero magnetizations
 satisfy the dimerized conditions in Eqs. (\ref{eq:8a})-(\ref{eq:8c}) and
 the conditions for a singlet state discussed in the SD phase.
 The spin state for sites $A$ and $B$ can then be written as
 $|\psi_{AB}\rangle=\left(
 \left|\uparrow_A\right\rangle \left|\downarrow_{B}\right\rangle
 - \left|\downarrow_A\right\rangle \left|\uparrow_{B}\right\rangle \right)/\sqrt{2}$.
 For the third two sites $i = E$ and $F$ with the local maximum magnetizations,
 the two-point spin correlations between $E/F$ and any site $j$ in the system
 satisfy the conditions in Eqs.~(\ref{eq:6a}) and ~(\ref{eq:6b}), which
 imply that each of the two sites is in a fully polarized state, i.e.,
 $\left|\psi_{EF}\right\rangle=\left|\uparrow_E\right\rangle\left|\uparrow_F\right\rangle$.
 For the second two sites ($i = C$ and $D$) and the fourth two sites ($i = G$ and $H$),
 the properties of the two-point spin correlations satisfy the dimerized conditions
 in Eqs. (\ref{eq:8a})-(\ref{eq:8c}).
 However,
 $\langle S^{x/y}_{C/G} S^{x/y}_{C/G+1} \rangle \neq -1/4$ even though
 $\langle S^{z}_{C/G} S^{z}_{C/G+1} \rangle = -1/4$,
 which implies that the spin state for the dimerized two-sites
 is not in a singlet state.
 However, other local correlations and magentizations discussed in the Appendix~\ref{app:MAFP2}
 allow us to write the spin state of the two sites
 as a linear combinations of two possible spin states, i.e.,
 $\left|\psi_{i,i+1}\right\rangle = a_i \left|\uparrow_i\downarrow_{i+1}\right\rangle
  - b_i \left|\downarrow_{i}\uparrow_{i+1} \right\rangle $ with here
  $i= C$ and $G$, where $a_i$ and $b_i$ are numerical coefficients depending on
  $J'_z/J$ and satisfying
  $|a_i|^2+|b_i|^2=1$.
 Thus, the $\left|\psi_{GH}\right\rangle$ has a similar form with the $\left|\psi_{CD}\right\rangle$.
 Comparing with the local magnetizations and the two-point spin correlations in the second two sites
 and the fourth two sites,
 the relations between the coefficients are given as $a^2_C-|b_C|^2=-(a^2_G-|b_G|^2)$
 and $a_C|b_C|=a_G|b_G|$ with $|a_i|^2+|b_i|^2=1$, which leads to
 $a_C = |b_G|$ and $a_G = |b_C|$.
  For the second two sites ($i = C$ and $D$) and the fourth two sites ($i = G$ and $H$),
  the spin states can be written as
  $\left|\psi_{CD}\right\rangle = a\left|\uparrow_C\downarrow_D\right\rangle
  - |b| \left|\downarrow_C\uparrow_D \right\rangle$
 and
 $\left|\psi_{GH}\right\rangle = |b|\left|\uparrow_G\downarrow_H\right\rangle
  - a \left|\downarrow_G\uparrow_H \right\rangle$,
  where $a$ and $b$ are numerical coefficients depending on $J'_z/J$ with
  $|a|^2+|b|^2=1$.
 For $|\psi_2\rangle$,
 we have found similar properties of two-point spin correlations.
 Compared with the characteristic properties of local magnetizations
 and two-point spin correlations from $|\psi_1\rangle$,
 it is found that the properties of them from $|\psi_1\rangle$
 are equal to ones from $|\psi_2\rangle$
 for one-plaquette shift, which implies
 $|\psi_2\rangle$ are equal to $|\psi_1\rangle$ under one-plaquette shift operation.
 Consequently,
 the groundstates $\left|\psi_1\right\rangle$ and $\left|\psi_2\right\rangle$
 are given as
\begin{subequations}
 \begin{eqnarray}
 \left|\psi_1\right\rangle
 \!\!\!&=&\!\!\! \prod_i \left|\chi_{2i}\right\rangle
 \left|\phi_{2i,2i+1}\right\rangle \left|\uparrow_{2i+1,u}\right\rangle \left|\uparrow_{2i+1,d}\right\rangle
 \left|\varphi_{2i+1,2i+2}\right\rangle,
 \label{MFIW1}
\\
 \left|\psi_2\right\rangle
 \!\!\!&=&\!\!\! \prod_i\left|\uparrow_{2i,u}\right\rangle \left|\uparrow_{2i,d}\right\rangle
 \left|\varphi_{2i,2i+1}\right\rangle \left|\chi_{2i+1}\right\rangle
 \left|\phi_{2i+1,2i+2}\right\rangle,
 \label{MFIW2}
 \end{eqnarray}
\end{subequations}
 where
 $\left|\phi_{i,j}\right\rangle = a\left|\uparrow_{i,r}\downarrow_{j,l}\right\rangle
 - |b| \left|\downarrow_{i,r}\uparrow_{j,l} \right\rangle $,
 $\left|\varphi_{i,j}\right\rangle = |b|\left|\uparrow_{i,r}\downarrow_{j,l}\right\rangle
 - a \left|\downarrow_{i,r}\uparrow_{j,l} \right\rangle $,
 and
 $|\chi_{i}\rangle=\left(
 \left|\uparrow_{i,u}\right\rangle \left|\downarrow_{i,d}\right\rangle
 - \left|\downarrow_{i,u}\right\rangle \left|\uparrow_{i,d}\right\rangle \right)/\sqrt{2}$,
 Note that the two groundstates $|\psi_1\rangle$ and $|\psi_2\rangle$ are orthogonal to each other, i.e., $\langle\psi_1|\psi_2\rangle=0$.

% $---------------------$

  The two groundstates in Eqs.~(\ref{MFIW1}) and ~(\ref{MFIW2}) show
  the characteristic properties of their local symmetries as follows:
 (i) singlet states $|\chi\rangle$ are $\mathrm{SU}(2)$-rotational invariant,
 (ii) fully-polarized states are polarized along the $z$-direction
 and then they are $\mathrm{U}(1)$-rotational invariant on the $x$-$y$ plane, and
 (iii) two-spin states $|\phi\rangle$ and $|\varphi\rangle$ are
 $\mathrm{U}(1)$-rotational invariant on the $x$-$y$ plane because the
 $x$- and $y$-components of the local magnetizations are zero, i.e.,
 $\langle S_x\rangle=0=\langle S_y\rangle$.
 Hence, for the MAFP phase, the two degenerate groundstates in Eqs.~(\ref{MFIW1}) and ~(\ref{MFIW2})
 are globally $U(1)$-rotational invariant.
 Since the system Hamiltonian possesses the same $U(1)$-rotational symmetry with the two groundstates,
 for the MAFP phase, no global rotational symmetry is broken.
 Both of the two degenerate groundstates
 are two-plaquette translational invariant, while
 our system Hamiltonian is one-plaquette translational invariant.
 Consequently, for the MAFP phase, the plaquette-translational symmetry breaking occurs
 and result in the degenerate groundstates.
 It is shown that the two degenerate groundstates can be understood
 within the Landau's spontaneous symmetry breaking picture.
 However, each of the two degenerate groundstates have the local
 $\mathrm{SU}(2)$ symmetry that cannot be explained within the
 Hamiltonian symmetry. Hence, the groundstates has such an emergent
 symmetry in the MAFP phase.

%%%%%
%%%%%%%%%%%%%%%%%%%%%%%%%%%%%%%fig. 7%%%%%%%%%%%%%%%%%%%%%%%%%%%%%%%%%%%%%%%%%%%%%%%%%
 \begin{figure}
 \begin{center}
  \begin{overpic}[width=0.4\textwidth]{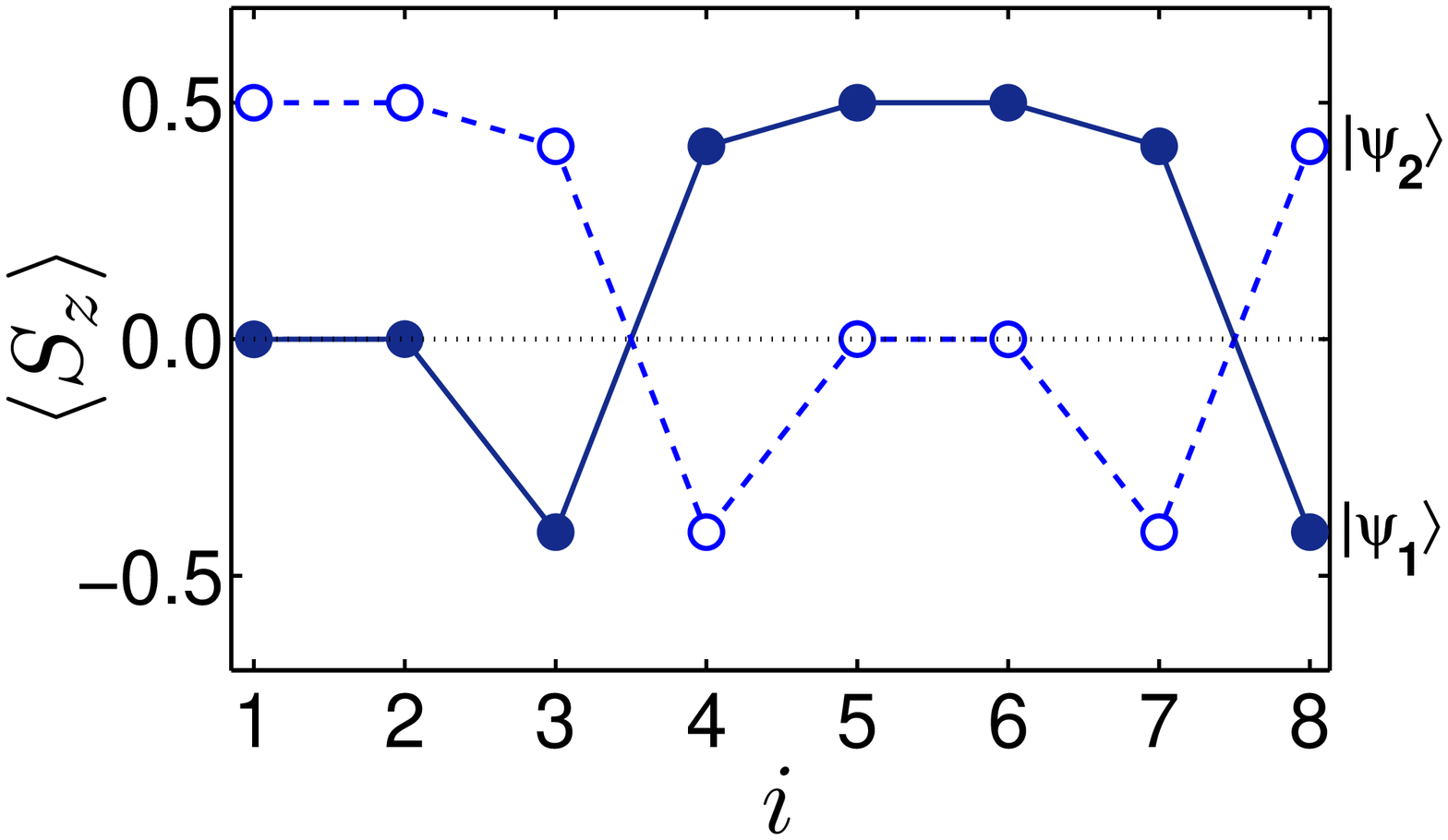}
   \put(1,58){(a)}
  \end{overpic}
%  \begin{center}
  \begin{overpic}[width=0.35\textwidth]{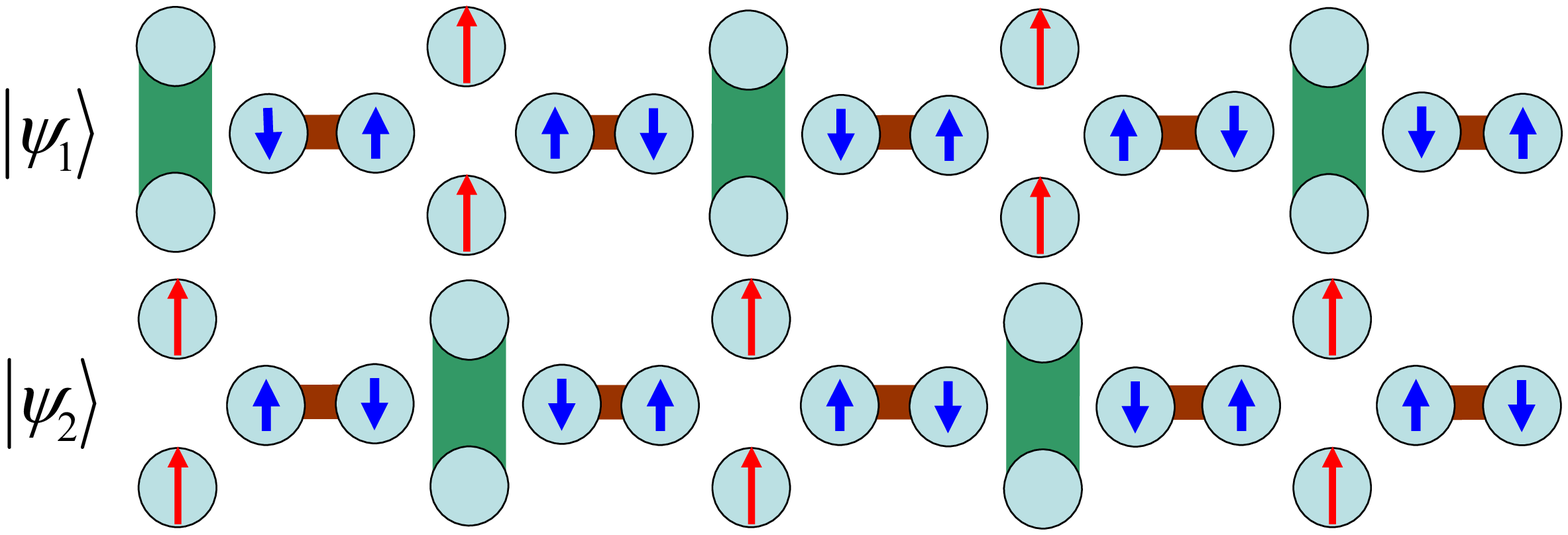}
   \put(-6,35){(b)}
  \end{overpic}
 \end{center}
\caption{(color online)
 (a) Local magnetization $\langle S_z\rangle$ at the lattice site $i$
 for $J'_z=-1.41J$ and $h=0.3$ in the MFP phase.
 (b) Pictorial representation of the groundstates with spin configuration.
  In contrast to the MAFP phase,
  note that each plaquette has a ferromagnetic configuration
  of local magnetizations in the MFP phase.
 }
 \label{MFPmg}
\end{figure}
%%%%%%%%%%%%%%%%%%%%%%%%%%%%%%%%%%%%%%%%%%%%%%%%%%%%%%%%%%%%%%%%%%%%%%%%%%%%%%%%%%%%%

%%%%%%%%%%%%%%%%%%%%%%%%%%%%%%%%%%%%%%%%%%%%%%%%%%%%%%%%%%%%%%%%%%%%%%%%%%%%%%%%%%%%%
\subsection{Modulated ferromagnetic-plaquette phase}
 The MFP phase is described by two-fold degenerate groundstates $|\psi_1\rangle$ and $|\psi_2\rangle$.
 Let us discuss first about local magnetizations.
 We plot the local magnetizations from the two degenerate groundstates
 for $J'_z=-1.41J$ and $h=0.3J$ in Fig.~\ref{MFPmg}.
 The local magnetizations from $|\psi_1\rangle$
 have a two-plaquette (eight-site) periodic structure where
 the first two sites, i.e., $A$ and $B$, have zero-magnetizations $\langle S^z \rangle =0$,
 the second two sites, i.e., $C$ and $D$,
 have
 $\langle S^z_C\rangle=-\langle S^z_D\rangle <0$,
 the third two sites, i.e., $E$ and $F$, have an maximum magnetizations $\langle S^z \rangle = 1/2$,
 and the fourth two sites, i.e., $G$ and $H$,
 have
 $\langle S^z_G\rangle=-\langle S^z_H\rangle >0$.
 For other values of system parameters,
 the characteristic behaviors of the local magnetizations are not changed
 and only the values of the local magnetizations at the sites $C$, $D$, $G$, and $H$
 are determined by $J'_z/J$.
 Compared with the local magnetizations from $|\psi_1\rangle$ for the MAFP phase in Fig. \ref{MFImg},
 one can notice that at the sites $D$ and $G$, the local magnetizations in the MFP phase
 have the opposite sign.
 This implies that
 the MFP phase has a distinguishable property of local magnetizations from the MAFP phase, i.e.,
 a ferromagnetic configuration in each plaquette.
 For $|\psi_2\rangle$, also,
 this difference can be easily conformed from the local magnetizations from the MAFP phase in Fig. \ref{MFImg} and the MFP phase in Fig. ~\ref{MFPmg}.
 Furthermore, note that for one-plaquette (four-site) shift,
 the local magentizations from $|\psi_2\rangle$ are equal to ones from $|\psi_1\rangle$.
 Similar to the MAFP phase,
 in the MFP phase,
 one groundstate becomes the other groundstate under one-plaquette shift transformation.

 Actually, from the calculations of two-point spin correlations for $|\psi_1\rangle$ in this MFP phase,
 we have found that
 for the first two sites $i = A$ and $B$ with the local zero magnetizations and
 the third two sites $i = E$ and $F$ with the local maximum magnetizations,
 the characteristic properties of the two-point spin correlations
 are the same with those in the MAFP phase.
 For the sites, their spin states have the forms:
 $|\psi_{AB}\rangle=\left(
 \left|\uparrow_A\right\rangle \left|\downarrow_{B}\right\rangle
 - \left|\downarrow_A\right\rangle \left|\uparrow_{B}\right\rangle \right)/\sqrt{2}$
 and
 $\left|\psi_{EF}\right\rangle=\left|\uparrow_E\right\rangle\left|\uparrow_F\right\rangle$.
 For the second two sites ($i = C$ and $D$) and the fourth two sites ($i = G$ and $H$),
 the properties of the two-point spin correlations have been found to be
 the same with those of the fourth two sites ($i = G$ and $H$) and
 the second two sites ($i = C$ and $D$) in the MAFP phase, respectively.
 This means that in the MFP phase,
 the spin states for the second and the fourth two sites are given as
 $\left|\psi_{CD}\right\rangle = |b|\left|\uparrow_C\downarrow_D\right\rangle
  - a \left|\downarrow_C\uparrow_D \right\rangle$
 and
 $\left|\psi_{GH}\right\rangle = a \left|\uparrow_G\downarrow_H\right\rangle
  - |b| \left|\downarrow_G\uparrow_H \right\rangle$,
  where $a$ and $b$ are numerical coefficients depending on $J'_z/J$ with
  $|a|^2+|b|^2=1$.
 Similar to the MAFP phase,
 in this MFP phase, we find that
 the physical properties from $|\psi_2\rangle$
 are equal to ones from $|\psi_1\rangle$
 for one-plaquette shift.
 This implies
 $|\psi_2\rangle$ are equal to $|\psi_1\rangle$ under one-plaquette shift operation.
 Consequently, for the MFP phase, we obtain
 the groundstates $\left|\psi_1\right\rangle$ and $\left|\psi_2\right\rangle$ as
\begin{subequations}
 \begin{eqnarray}
 \left|\psi_1\right\rangle
 \!\!\!&=&\!\!\! \prod_i \left|\chi_{2i}\right\rangle
 \left|\phi_{2i,2i+1}\right\rangle \left|\uparrow_{2i+1,u}\right\rangle \left|\uparrow_{2i+1,d}\right\rangle
 \left|\varphi_{2i+1,2i+2}\right\rangle,
 \label{MFPW1}
\\
 \left|\psi_2\right\rangle
 \!\!\!&=&\!\!\! \prod_i\left|\uparrow_{2i,u}\right\rangle \left|\uparrow_{2i,d}\right\rangle
 \left|\varphi_{2i,2i+1}\right\rangle \left|\chi_{2i+1}\right\rangle
 \left|\phi_{2i+1,2i+2}\right\rangle.
 \label{MFPW2}
 \end{eqnarray}
\end{subequations}
 These two groundstates $|\psi_1\rangle$ and $|\psi_2\rangle$ are orthogonal to each other,
 i.e., $\langle\psi_1|\psi_2\rangle=0$.
 One can also find that one-plaquette translational operation on $|\psi_{1/2}\rangle$
 leads to $|\psi_{2/1}\rangle$.
 Basically,
 the two degenerate groundstates in Eqs. (\ref{MFPW1}) and (\ref{MFPW2})
 have the same characteristic local symmetries with those in Eqs.~(\ref{MFIW1}) and ~(\ref{MFIW2})
 in the MAFP phase.
 Also, both of the two degenerate groundstates
 are two-plaquette translational invariant, while
 our system Hamiltonian is one-plaquette translational invariant.
 Thus, for the MFP phase, the two degenerate groundstates in Eqs.~(\ref{MFPW1}) and ~(\ref{MFPW2})
 can be understood within the Landau's spontaneous symmetry breaking picture, i.e.,
 one-plaquette translational symmetry breaking results in
 the reduced symmetry of them comparing to the Hamiltonian symmetry.
 Similar to the MAFP phase, also,
 each of the two degenerate groundstates have the emergent local
 $\mathrm{SU}(2)$ symmetry that cannot be explained within the
 Hamiltonian symmetry in the MFP phase.

\subsection{ Staggered bond phase}
\begin{figure}
 \begin{center}
  \begin{overpic}[width=0.4\textwidth]{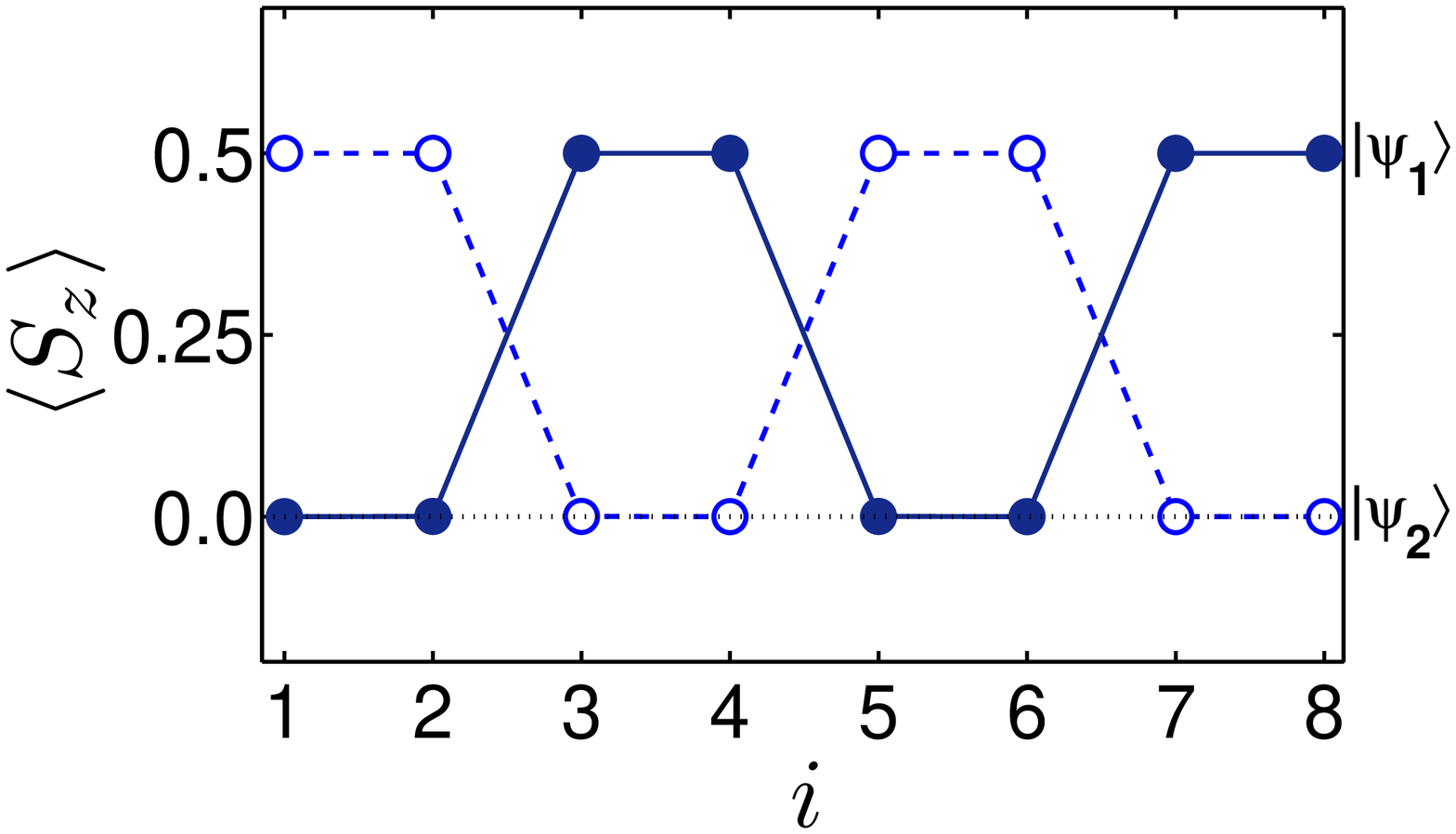}
   \put(1,58){(a)}
  \end{overpic}
  \begin{overpic}[width=0.35\textwidth]{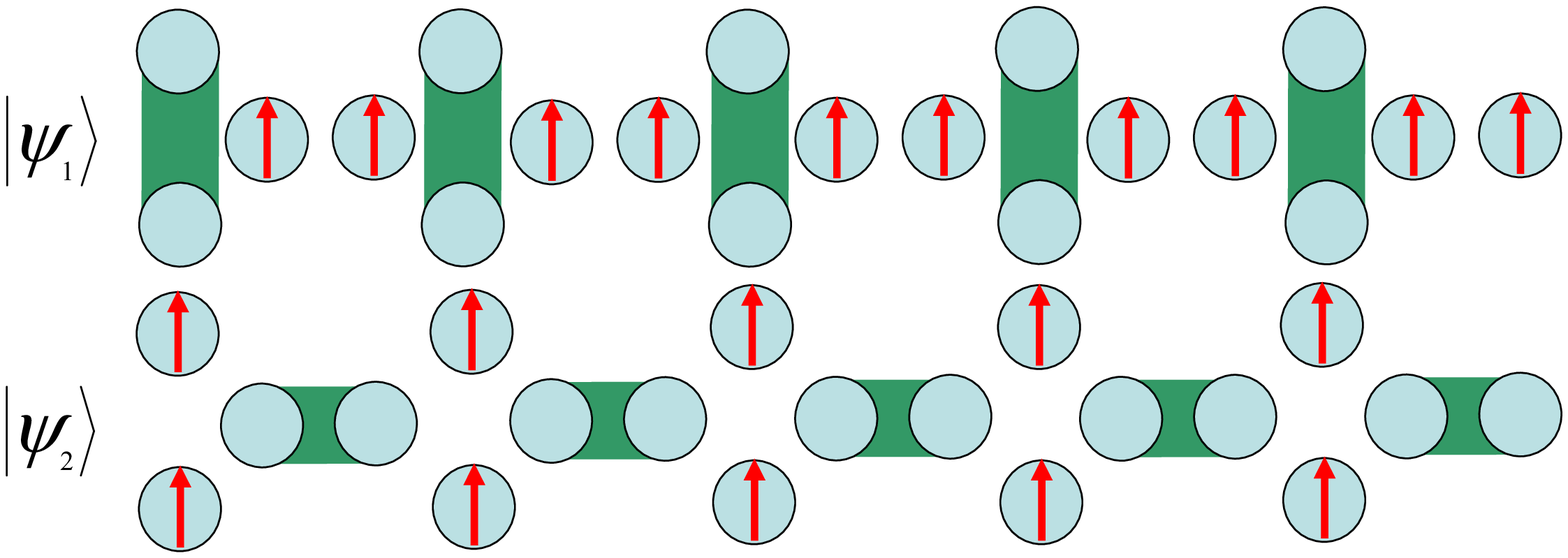}
   \put(-6,35){(b)}
  \end{overpic}
 \end{center}
\caption{(color online)
 (a) Local magnetization $\langle S_z\rangle$
 at the lattice site $i$ for $J'_z=0.8J$ and $h=1.5J$ in the SB phase.
 (b) Pictorial representation of the groundstates with spin configuration in the SB phase.
 }
 \label{SBmg}
\end{figure}

\begin{figure}
 \begin{center}
  \begin{overpic}[width=0.4\textwidth]{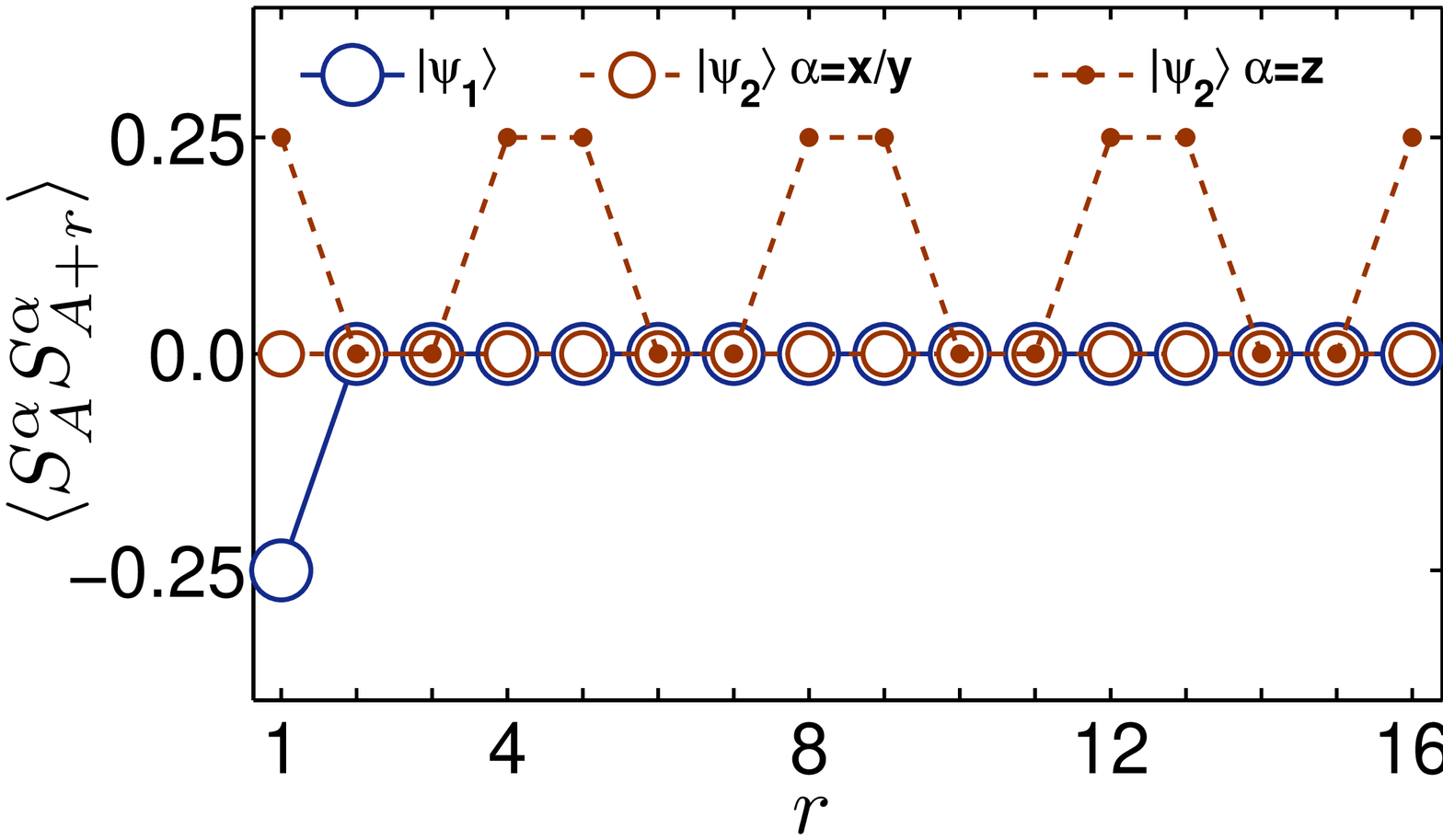}
            \put(1,52){(a)}
  \end{overpic}
  \begin{overpic}[width=0.4\textwidth]{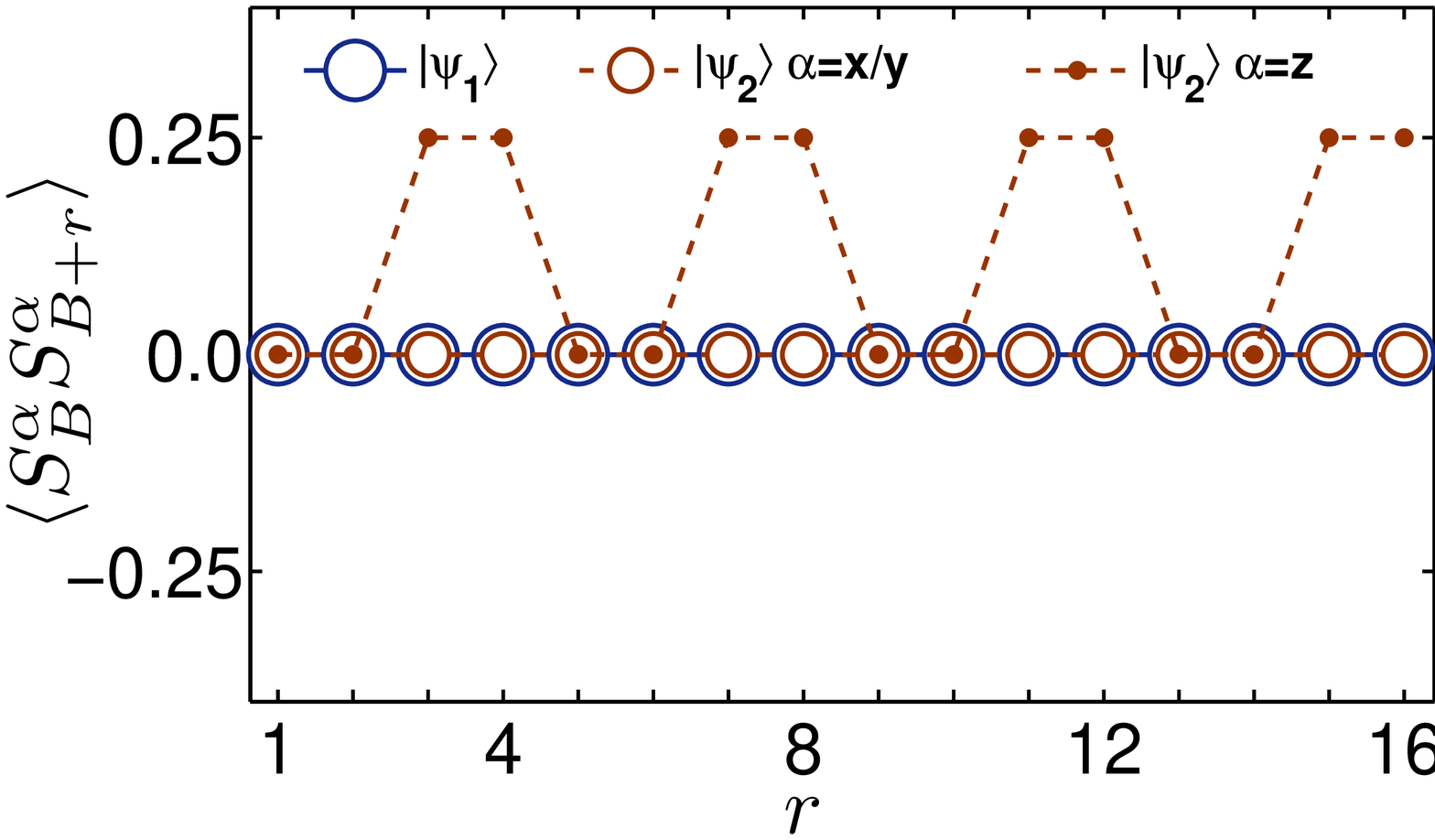}
            \put(1,52){(b)}
  \end{overpic}
  \begin{overpic}[width=0.4\textwidth]{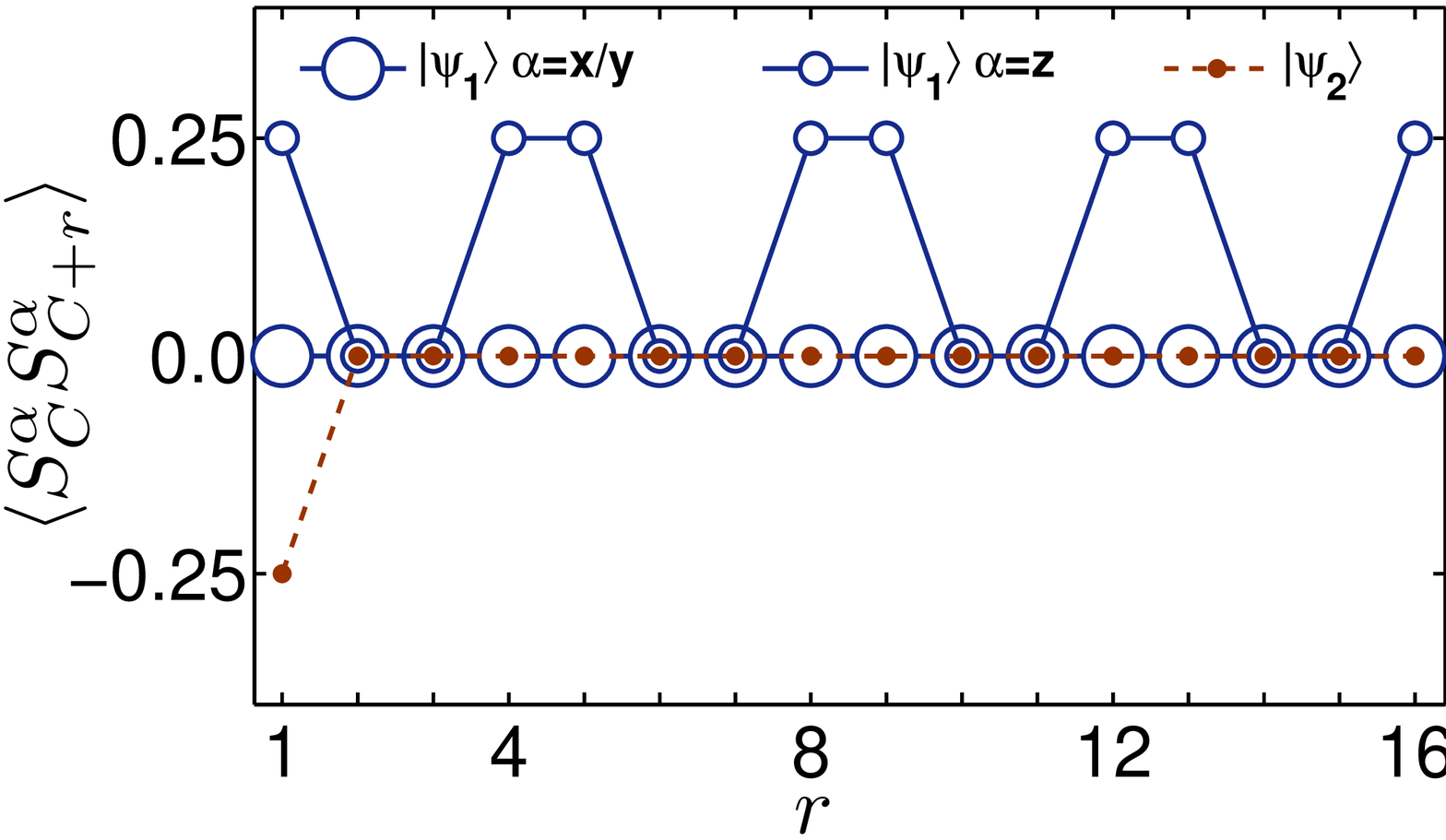}
            \put(1,52){(c)}
  \end{overpic}
  \begin{overpic}[width=0.4\textwidth]{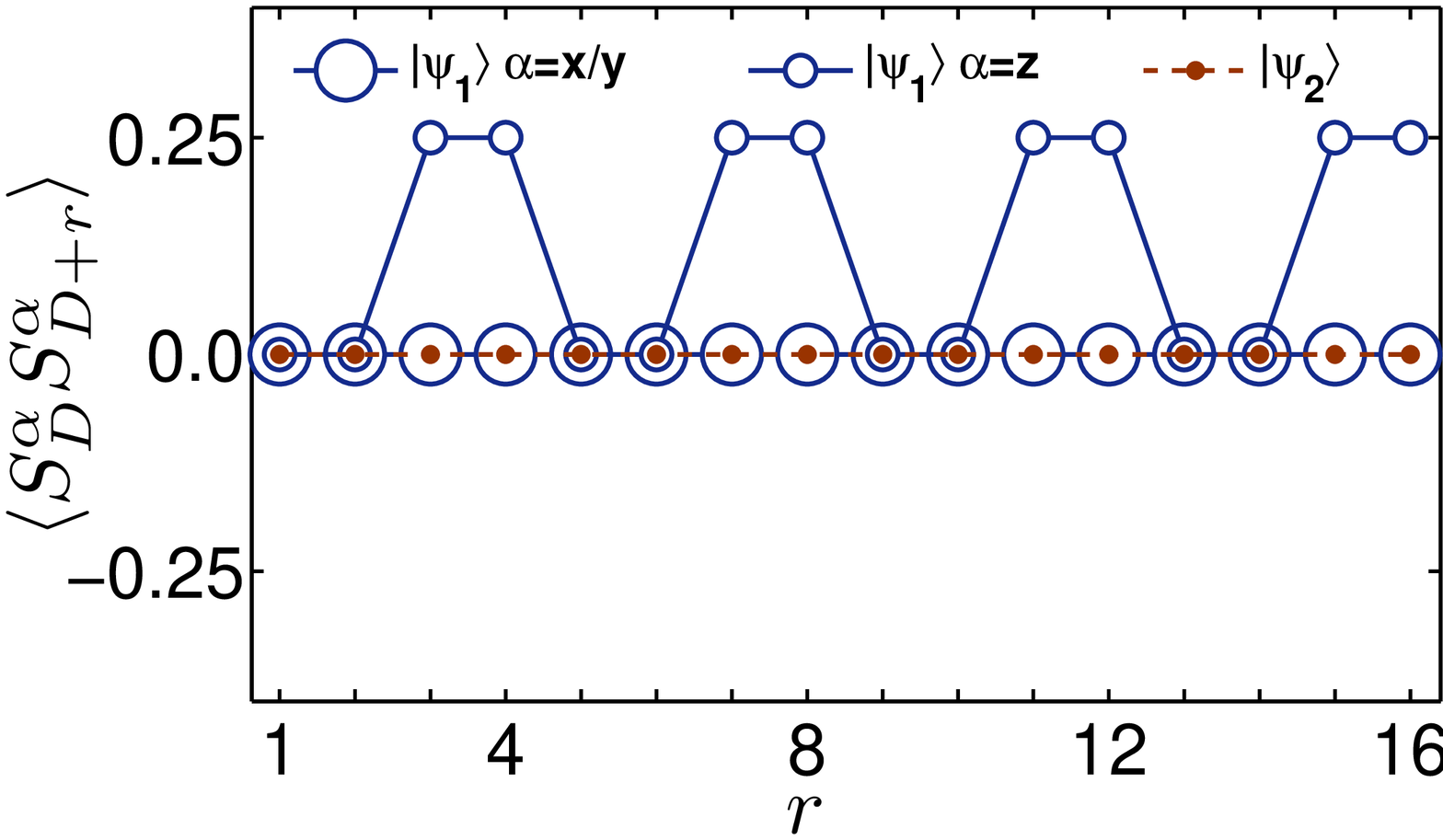}
            \put(1,52){(d)}
  \end{overpic} \end{center}
\caption{(color online)
 Two-point spin correlations $\langle S^{\alpha}_{i} S^{\alpha}_{i+ r}\rangle$ with
 $i = A, B, C, D$ and $\alpha=x,y,z$
 as a function of lattice distance $r$ between two sites
 $i$ and $i+ r$ for $J'_z=0.8J$ and $h=1.5J$ in the SB phase.}
 \label{SBcorrab}
\end{figure}

 In the SB phase, there are the two degenerate groundstates $|\psi_1\rangle$ and $|\psi_2\rangle$.
 From the two degenerate groundstates, for $J'_z=0.8J$ and $h=1.5J$,
 we plot
 the local magnetizations at lattice sites in Fig.~\ref{SBmg}(a) and
 the two-spin correlations
 as a function of lattice distance $r$ between site $i$ and $i+r$
 in Fig.~\ref{SBcorrab}.
 For $|\psi_1\rangle$ ($|\psi_2\rangle$) in Fig.~\ref{SBmg}(a),
 one can easily notice that
 the local magnetizations have one-plaquette periodic structure where
 the first (second) two sites have zero-magnetizations $\langle S_z \rangle =0$
 and the second (first) two sites
 have maximum magnetizations $\langle S_z \rangle = 1/2$.
 We have found that
 the local magnetizations are not changed for other values of system
 parameters.
 For the first two sites $i = A$ and $B$,
 Figs.~\ref{SBcorrab}(a) and \ref{SBcorrab}(b) show
 that the properties of two-point spin correlations for $|\psi_1\rangle$
 are summarized as
 (i)
 $\langle S^{\alpha}_A S^{\alpha}_{A+1} \rangle = -1/4$ and
 $\langle S^{\alpha}_A S^{\alpha}_{A+r} \rangle = 0$ for $r > 1 $,
 and (ii)
 $\langle S^{\alpha}_B S^{ \alpha}_{B+r} \rangle = 0$.
 Also,
 $\langle S^{\alpha}_{A/B}S^{\alpha'}_j\rangle = 0$ for $\alpha \neq \alpha'$
 (not displayed) have been observed.
 Since the local magnetizations are zero at the sites $i = A$ and $B$,
 the properties of the two-point spin correlations
 imply that the two sites are dimerized and satisfy the conditions for a singlet state
 discussed in the SD phase, i.e.,
 $|\psi_{AB}\rangle=\left(
 \left|\uparrow_A\right\rangle \left|\downarrow_{B}\right\rangle
 - \left|\downarrow_A\right\rangle \left|\uparrow_{B}\right\rangle \right)/\sqrt{2}$.
 Also,
 for the second two sites $i = C$ and $D$,
 Figs.~\ref{SBcorrab}(c) and \ref{SBcorrab}(d)
 show that the two-point spin correlations for $|\psi_1\rangle$
 have the characteristic properties, i.e.,
 (i)
 $\langle S^{z}_{C/D} S^{z}_{j} \rangle = 1/4=\langle S^{z}_{C/D} \rangle\langle S^{z}_{j}\rangle$
 for $j\in\{C,D\}$ and
 $\langle S^{z}_{C/D} S^{z}_{j} \rangle = 0$ for $j\neq C$ and $D$,
 and (ii)
 $\langle S^{x/y}_{C/D} S^{x/y}_{j} \rangle = 0$.
 We have noticed numerically that
 $\langle S^{\alpha}_{C/D}S^{\alpha'}_j\rangle = 0$ for $\alpha \neq \alpha'$
 (not displayed).
 Since the local magnetizations
 have their maximum value at the sites $i = C$ and
 $D$,
 the properties of the two-point spin correlations
 imply that each of the two sites
 satisfies the conditions for a fully polarized state
 discussed in the FP phase, i.e.,
 $|\psi_{CD}\rangle=\left|\uparrow_C\right\rangle \left|\uparrow_D\right\rangle$.
 Similar discussions can be made for the other groundstate, $\left|\psi_2\right\rangle$.
 As one can notice easily, the first (second) two sites in $\left|\psi_1\right\rangle$
 have the same physical properties
 with the second (first) two sites in $\left|\psi_2\right\rangle$.
 Consequently, for the SB phase,
 the two degenerate groundstates can be written as
\begin{subequations}
 \begin{eqnarray}
 \left|\psi_1\right\rangle
  &=& \prod_i\frac{1}{\sqrt{2}}
 \left( \left|\uparrow_{i,u}\downarrow_{i,d}\right\rangle-\left|\downarrow_{i,u}\uparrow_{i,d}\right \rangle \right)
 \left|\uparrow_{i,r}\uparrow_{i+1,l}\right\rangle,
 \label{SB1}
\\
 \left|\psi_2\right\rangle
  &=& \prod_i\frac{1}{\sqrt{2}}
 \left|\uparrow_{i,u}\uparrow_{i,d}\right\rangle
 \left( \left|\uparrow_{i,r}\downarrow_{i+1,l}\right\rangle-\left|\downarrow_{i,r}\uparrow_{i+1,l}\right \rangle \right).
 \label{SB2}
 \end{eqnarray}
\end{subequations}
 These two groundstates $|\psi_1\rangle$ and $|\psi_2\rangle$ are orthogonal to each other, i.e., $\langle\psi_1|\psi_2\rangle=0$.

%%%%%%%%%%%%%%%%%%%%%%%fig. 8%%%%%%%%%%%%%%%%%%%%%%%%%%%%%%%%%%%%%%%%%%%%%%%%%%%%%%%%%
%%%%%%%%%%%%%%%%%%%%%%%%%%%%%%%%%%%%%%%%%%%%%%%%%%%%%%%%%%%%%%%%%%%%%%%%%%%%%%%%%%%%%

%%%%%%%%%%%%%%%%%%%%%%%%%%%%%%%%fig. 9%%%%%%%%%%%%%%%%%%%%%%%%%%%%%%%%%%%%%%%%%%%%%%%%
%%%%%%%%%%%%%%%%%%%%%%%%%%%%%%%%%%%%%%%%%%%%%%%%%%%%%%%%%%%%%%%%%%%%%%%%%%%%%%%%%%%%%

 For the SB phase,
 the explicit forms of the two degenerate groundstates in Eqs.~(\ref{SB1}) and (\ref{SB2})
 can be represented pictorially in Fig. \ref{SBmg}(b).
 One can see that the two degenerate groundstates
 are one-plaquette translational invariant and
 also have a global $\mathrm{U}(1)$-rotational symmetry on the $x$-$y$ plane.
 Our Hamiltonian has the one-plaquette translational and
 the $\mathrm{U}(1)$-rotational symmetries
 of the two groundstates.
 Within the Landau's spontaneous symmetry breaking picture,
 the occurrence of the doubly degenerate groundstates for the SB phase could not be understood
 because no responsible symmetry for the degenerate groundstates exist in the system Hamiltonian.
 One can also notice that under the vertical-to-horizontal site exchange transformation,
 one groundstate becomes the other groundstate
 but the system Hamiltonian is not invariant.
 This fact implies that if the two degenerate groundstates are originated from
 a spontaneous symmetry breaking, the vertical-to-horizontal site exchange symmetry
 might play a significant role to understand an emergent symmetry that is responsible
 for the two groundstates in the SB phase.
 Similar to the MAFP and MFP phases, also,
 each of the two degenerate groundstates have the emergent local
 $\mathrm{SU}(2)$ symmetry that cannot be explained within the
 Hamiltonian symmetry in the SB phase.

%%%%%%%%%%%%%%%%%%%%%%%%%
%%%%%%%%%%%%%%%%%%%%%%%%%%%%%%%%%%%%fig. 10%%%%%%%%%%%%%%%%%%%%%%%%%%%%%%%%%%%%%%%%%%%%
 \begin{figure}
 \begin{center}
  \begin{overpic}[width=0.4\textwidth]{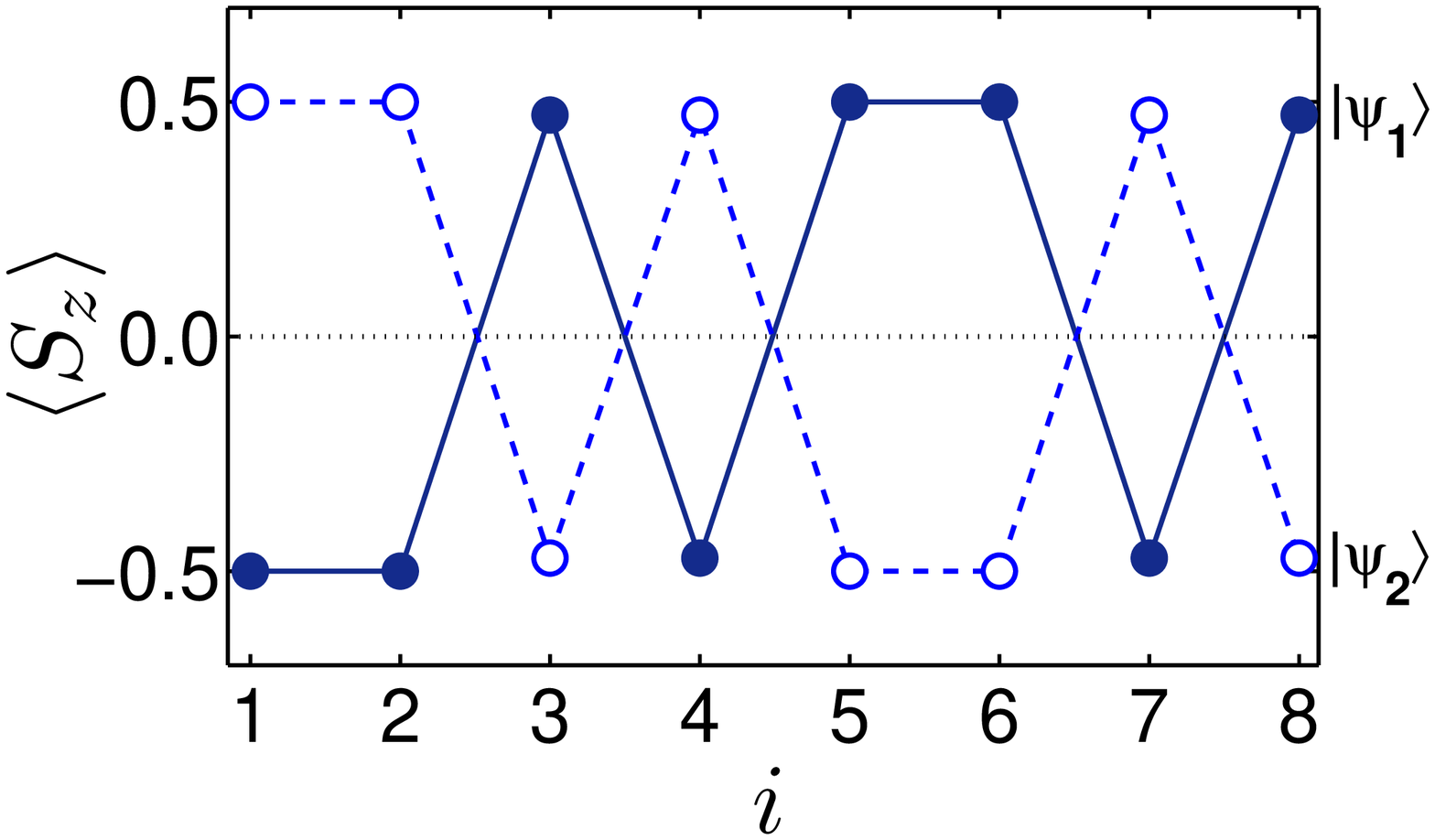}
     \put(1,56){(a)}
     \end{overpic}
  \begin{overpic}[width=0.35\textwidth]{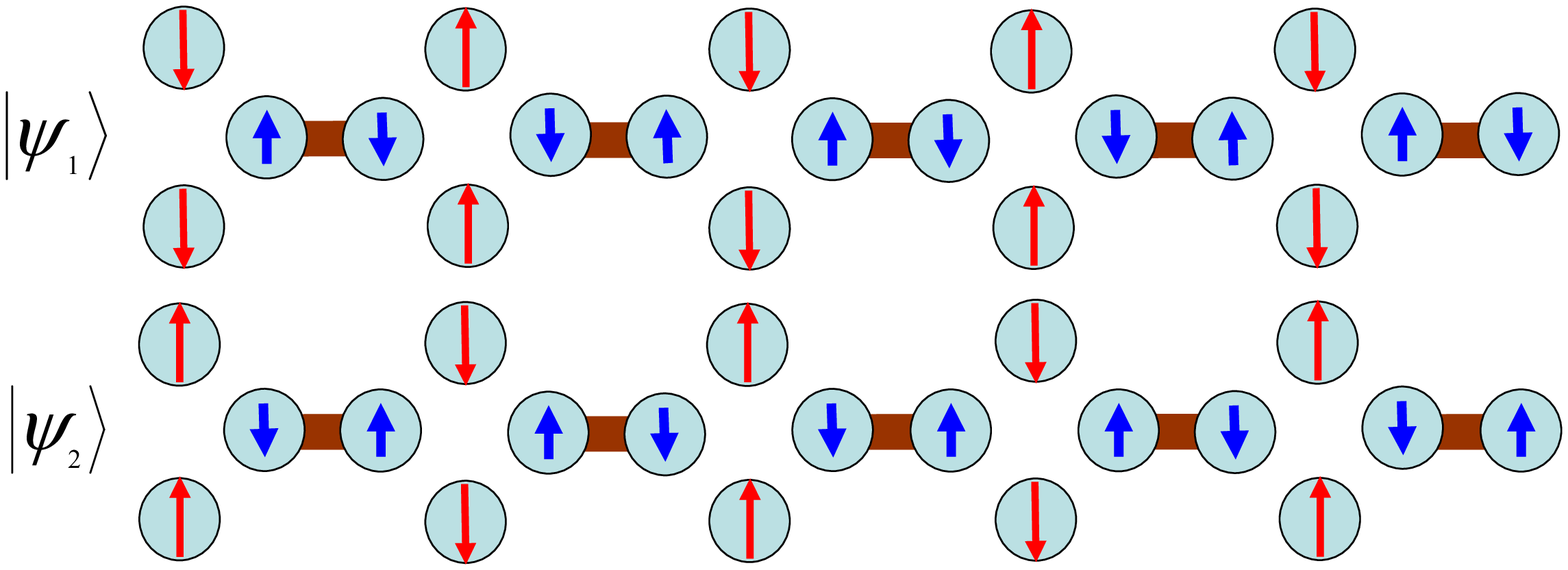}
     \put(-6,35){(b)}
  \end{overpic}
   \end{center}
\caption{(color online)
 (a) Local magnetization $\langle S_z\rangle$ at the lattice site $i$
 for $J'_z=1.43J$ and $h=0$ in the SAFP phase.
 (b) Pictorial representation of the groundstates with spin configuration in the SAFP phase.
   Note that each plaquette has an anti-ferromagnetic configuration
  of local magnetizations in the SAFP phase.}
 \label{MAFmg}
\end{figure}
%%%%%%%%%%%%%%%%%%%%%%%%%%%%%%%%%%%%%%%%%%%%%%%%%%%%%%%%%%%%%%%%%%%%%%%%%%%%%%%%%%%%%

\subsection{Staggered anti-ferromagnetic plaquette phase}
 Two degenerate groundstates are detected in the SAFP phase, as denoted as $|\psi_1\rangle$ and $| \psi_2\rangle$.
 From the two groundstates,
 we plot the local magnetizations
 $\langle \psi_n |  S_z | \psi_n\rangle$ at the lattice site $i$
 in Fig.~\ref{MAFmg}(a).
 In the SAFP phase, Fig.~\ref{MAFmg}(a) shows that for $|\psi_1\rangle$,
 the local magnetizations has a two plaquette (eight-site) periodic structure where
 (i) the first two sites, i.e., $A$ and $B$, have a minimum magnetization $\langle S_z \rangle =-1/2$,
 (ii) the second two sites, i.e., $C$ and $D$,
 have $\langle S^z_C\rangle=-\langle S^z_D\rangle >0$,
 (iii) the third two sites, i.e., $E$ and $F$, have a maximum magnetization $\langle S_z \rangle = 1/2$,
 and
 (iv) the fourth two sites, i.e., $G$ and $H$,
  $\langle S^z_G\rangle=-\langle S^z_H\rangle <0$.
 For other values of system parameters,
 the characteristic behaviors of the local magnetizations are not changed
 and only the values of the local magnetizations at the sites $C$, $D$, $G$, and $H$
 are determined by $J'_z/J$.
 In this SAFP phase, also, the local magnetizations has
 an anti-ferromagnetic configuration in each plaquette.
 Note that for one-plaquette (four-site) shift,
 the local magentizations from $|\psi_2\rangle$ are equal to ones from $|\psi_1\rangle$.
 This implies that one groundstate becomes the other groundstate under one-plaquette shift transformation.

 Actually, from the calculations of two-point spin correlations for $|\psi_1\rangle$ in this SAFP phase,
 we have found that
 for the first two sites $i = A$ and $B$ with the minimum magnetizations
 (the third two sites $i = E$ and $F$ with the maximum magnetizations),
 the two-point spin correlations between $A/B$ ($E/F$) and any site $j$ in the system
 satisfy the conditions in Eqs.~(\ref{eq:6a}) and ~(\ref{eq:6b}), which
 imply that each of the two sites is in a fully polarized state, i.e.,
 $\left|\psi_{AB}\right\rangle=\left|\downarrow_A\right\rangle\left|\downarrow_B\right\rangle$
 and $\left|\psi_{EF}\right\rangle=\left|\uparrow_E\right\rangle\left|\uparrow_F\right\rangle$.
 For the second two sites $i = C$ and $D$ (the fourth two sites $i = G$ and $H$),
 the properties of the two-point spin correlations have been found to be
 the same with those of the second two sites $i = C$ and $D$ (the fourth two site $i = G$ and $H$)
 in the MAFP phase, respectively.
 This means that in the SAFP phase,
 the spin state for the second (fourth) two sites
 has a same form of the spin state for the second (fourth) two sites in the MAFP phase,
 i.e.,
 $\left|\psi_{CD}\right\rangle = a \left|\uparrow_C\downarrow_D\right\rangle
  - |b| \left|\downarrow_C\uparrow_D \right\rangle$
 and
 $\left|\psi_{GH}\right\rangle = |b|\left|\uparrow_G\downarrow_H\right\rangle
  - a \left|\downarrow_G\uparrow_H \right\rangle$,
  where $a$ and $b$ are numerical coefficients depending on $J'_z/J$ with
  $|a|^2+|b|^2=1$.
 Similar to the MAFP phase,
 we find that in this SAFP phase,
 the physical properties from $|\psi_2\rangle$
 are equal to ones from $|\psi_1\rangle$
 under one-plaquette shift transformation.
 Consequently, for the SAFP phase, we obtain
 the groundstates $\left|\psi_1\right\rangle$ and $\left|\psi_2\right\rangle$ as
\begin{subequations}
 \begin{eqnarray}
 \left|\psi_1\right\rangle
 \!\! &=& \!\! \prod_i \left|\Downarrow_{2i,ud} \right\rangle
 |\varphi_{2i,r;2i+1,l}\rangle \left|\Uparrow_{2i+1,ud} \right\rangle
 |\phi_{2i+1,r;2i+2,l}\rangle,
 \label{MAFW1}
\\
\left|\psi_2\right\rangle
 \!\! &=& \!\!  \prod_i\left|\Uparrow_{2i,ud} \right\rangle
 |\phi_{2i,r;2i+1,l}\rangle \left|\Downarrow_{2i+1,ud} \right\rangle
 |\varphi_{2i+1,r;2i+2,l}\rangle,
 \label{MAFW2}
 \end{eqnarray}
\end{subequations}
 where
 $\left|\Downarrow_{2i,ud} \right\rangle \equiv \left|\downarrow_{2i,u}\downarrow_{2i,d}\right\rangle$
 and
 $\left|\Uparrow_{2i,ud} \right\rangle \equiv \left|\uparrow_{2i,u}\uparrow_{2i,d}\right\rangle$.

 In the SAFP phase, both of the two degenerate ground states in Eqs.~(\ref{MAFW1}) and ~(\ref{MAFW2})
 are global $U(1)$-rotational invariant.
 Since the system Hamiltonian has the same global rotational symmetry with the two groundstates,
 the global rotational symmetry has nothing to do with the occurrence of the two groundstates.
 However, both of the two degenerate ground states in Eqs.~(\ref{MAFW1}) and ~(\ref{MAFW2})
 for the SAFP phase can be transformed from one to the other
 under one-plaquette translational or
 spin-flip transformations.
 Then, the plaquette translational symmetry
 and the spin-flip symmetry breakings seem to result in the two degenerate groundstates.
 However, note that for $h =0$,
 the system Hamiltonian has both the one-plaquette translational symmetry and the
 spin-flip symmetry but for $h \neq 0$, it has only the
 one-plaquette
 translational symmetry.
 Hence, for $h = 0$ in the SAFP phase,
 the two-fold degenerate groundstates can be understood
 from the breaking of both the plaquette translational symmetry and the spin-flip symmetry.
 For $h \neq 0$,
 since the system Hamiltonian does not have the spin-flip symmetry,
 more symmetry than the symmetry that the Hamiltonian possesses
 should be involved to be broken for the two degenerate groundstates.
 This situation cannot be understood within the Landau's spontaneous symmetry breaking picture.
 Although, in order to understand why the two groundstates exist in the SAFP phase for $h \neq 0$,
 one may possibly consider the spin-flip symmetry as an emergent symmetry
 which can be broken together with a plaquette-translational symmetry,
 still a question why such an emergent symmetry should be broken
 together with the plaquette-translational symmetry breaking must be left to be answered.
%

%%%%%%%%%%%%%%%%%%%%%%%%%%%%%%%%%%%fig. 11%%%%%%%%%%%%%%%%%%%%%%%%%%%%%%%%%%%%%%%%%%%%%
 \begin{figure}
 \begin{center}
  \begin{overpic}[width=0.4\textwidth]{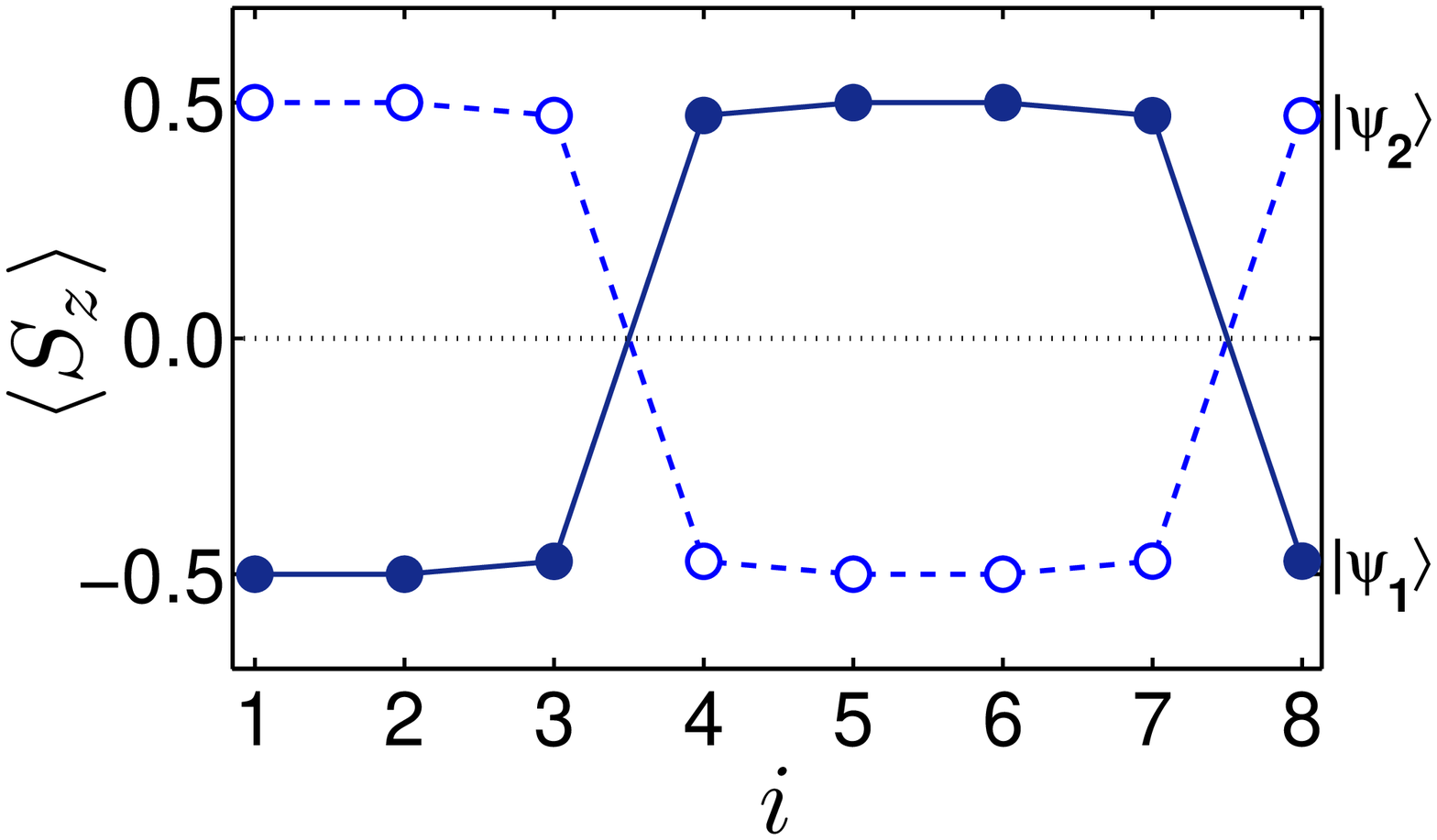}
     \put(1,53){(a)}
  \end{overpic}
    \begin{overpic}[width=0.35\textwidth]{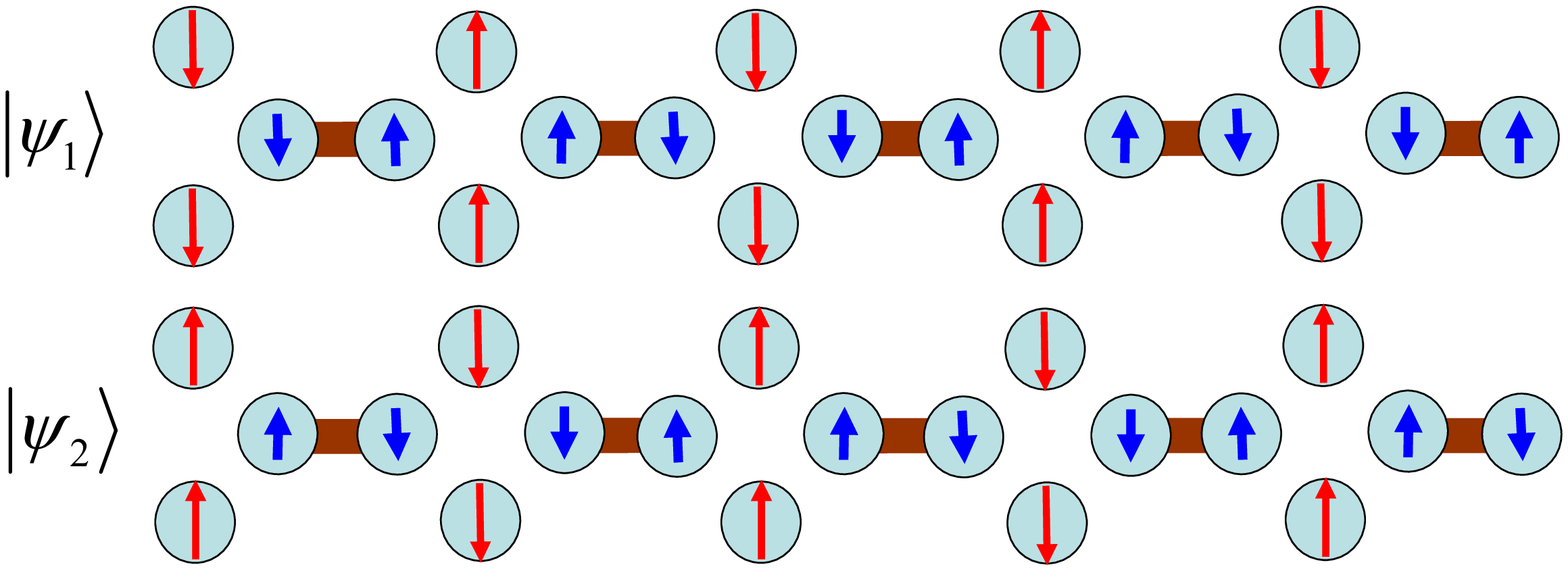}
     \put(-6,35){(b)}
  \end{overpic}
 \end{center}
\caption{(color online)
 (a) Local magnetization $\langle S_z\rangle$ at the lattice site $i$
 for $J'_z=-1.44J$ and $h=0$ in the SFP phase.
 (b) Pictorial representation of the groundstates with spin configuration in the SFP phase.
  In contrast to the SAFP phase,
  note that each plaquette has a ferromagnetic configuration
  of local magnetizations in the SFP phase.}
 \label{SFPmg}
\end{figure}
%%%%%%%%%%%%%%%%%%%%%%%%%%%%%%%%%%%%%%%%%%%%%%%%%%%%%%%%%%%%%%%%%%%%%%%%%%%%%%%%%%%%%

%%%%%%%%%%%%%%%%%%%%%%%%%%%%%%%%%%%%%%%%%%%%%%%%%%%%%%%%%%%%%%%%%%%%%%%%%%%%%%%%%%%%%
\subsection{Staggered ferromagnetic plaquette phase}
 Similar to the SAFP phase,
 there are two degenerate groundstates detected in the SFP phase.
 The wavefunctions can be denoted as $|\psi_1\rangle$ and $| \psi_2\rangle$.
 From the two groundstates,
 we plot the local magnetizations
 $\langle \psi_n |  S_z | \psi_n\rangle$ at the lattice site $i$
 in Fig.~\ref{SFPmg}(a).
 In the SFP phase, Fig.~\ref{SFPmg}(a) shows that for $|\psi_1\rangle$,
 the local magnetizations has a two-plaquette (eight-site) periodic structure where
 (i) the first two sites, i.e., $A$ and $B$, have a minimum magnetization $\langle S_z \rangle =-1/2$,
 (ii) the second two sites, i.e., $C$ and $D$,
 have $\langle S^z_C\rangle=-\langle S^z_D\rangle < 0$,
 (iii) the third two sites, i.e., $E$ and $F$, have a maximum magnetization $\langle S_z \rangle = 1/2$,
 and
 (iv) the fourth two sites, i.e., $G$ and $H$,
  $\langle S^z_G\rangle=-\langle S^z_H\rangle > 0$.
 For other values of system parameters,
 the characteristic behaviors of the local magnetizations are not changed
 and only the values of the local magnetizations at the sites $C$, $D$, $G$, and $H$
 are determined by $J'_z/J$.
 Comparing to the SAFP phase, this SFP phase shows
 a ferromagnetic configuration of the local magnetizations in each plaquette.
 Also, for one-plaquette shift,
 the local magentizations from $|\psi_2\rangle$ are equal to ones from $|\psi_1\rangle$.
 Then, under one-plaquette transformation,
 one groundstate becomes the other groundstate under one-plaquette shift transformation.

 Similar to the case of the SAFP phase,
 from the calculations of two-point spin correlations for $|\psi_1\rangle$ in this SFP phase,
 we have found that
 for the first two sites $i = A$ and $B$ with the minimum magnetizations
 (the third two sites $i = E$ and $F$ with the maximum magnetizations),
 the two-point spin correlations between $A/B$ ($E/F$) and any site $j$ in the system
 satisfy the conditions in Eqs.~(\ref{eq:6a}) and ~(\ref{eq:6b}), which
 imply that each of the two sites is in a fully polarized state, i.e.,
 $\left|\psi_{AB}\right\rangle=\left|\downarrow_A\right\rangle\left|\downarrow_B\right\rangle$
 and $\left|\psi_{EF}\right\rangle=\left|\uparrow_E\right\rangle\left|\uparrow_F\right\rangle$.
 However,
 for the second two sites $i = C$ and $D$ (the fourth two sites $i = G$ and $H$),
 the properties of the two-point spin correlations have been found to be
 the same with those of the fourth two site $i = G$ and $H$ (the second two sites $i = C$ and $D$)
 in the SAFP phase, respectively.
 This means that in this SFP phase,
 the spin state for the second (fourth) two sites
 has a same form of the spin state for the fourth (second) two sites in the SAFP phase,
 i.e.,
 $\left|\psi_{CD}\right\rangle = |b|\left|\uparrow_C\downarrow_D\right\rangle
  - a \left|\downarrow_C\uparrow_D \right\rangle$,
 and
 $\left|\psi_{GH}\right\rangle = a \left|\uparrow_G\downarrow_H\right\rangle
  - |b| \left|\downarrow_G\uparrow_H \right\rangle$
  where $a$ and $b$ are numerical coefficients depending on $J'_z/J$ with
  $|a|^2+|b|^2=1$.
 Similar to the SAFP phase,
 we find that in this SFP phase,
 the physical properties from $|\psi_2\rangle$
 are equal to ones from $|\psi_1\rangle$
 under one-plaquette shift transformation.
 Consequently, for the SFP phase, we obtain
 the groundstates $\left|\psi_1\right\rangle$ and $\left|\psi_2\right\rangle$ as
\begin{subequations}
 \begin{eqnarray}
 \left|\psi_1\right\rangle
 \!\! &=& \!\! \prod_i \left|\Downarrow_{2i,ud} \right\rangle
  |\varphi_{2i,r;2i+1,l}\rangle
  \left|\Uparrow_{2i+1,ud} \right\rangle
 |\phi_{2i+1,r;2i+2,l}\rangle,
 \label{SFPW1}
\\
\left|\psi_2\right\rangle
 \!\! &=& \!\!  \prod_i\left|\Uparrow_{2i,ud} \right\rangle
 |\phi_{2i,r;2i+1,l}\rangle
 \left|\Downarrow_{2i+1,ud} \right\rangle
 |\varphi_{2i+1,r;2i+2,l}\rangle,
 \label{SFPW2}
 \end{eqnarray}
\end{subequations}

 The groundstates in Eqs.~(\ref{SFPW1}) and ~(\ref{SFPW2}) in the SFP phase
 has a very similar form with the groundstates in Eqs.~(\ref{MAFW1}) and ~(\ref{MAFW2})
 in the SAFP phase.
 Then, the groundstates in the two phases have
 a very similar symmetry each other.
 For this SFP phase,
 one groundstate becomes the other groundstate for one-plaquette translation or spin-flip operation,
 which implies that
 the plaquette translational and the spin-flip symmetry breakings
 are responsible for the two degenerate groundstates.
 However, similar to the SAFP phase,
 the Hamiltonian does not have the spin-flip symmetry $h \neq 0$
 and then more symmetry than the symmetry that the Hamiltonian possesses
 should be involved to be broken for the two degenerate groundstates.
 Thus, the occurrence of the two-fold degenerate groundstates in the SFP phase
 is not fully understood within the Landau's spontaneous symmetry breaking picture.
%
%

%%%%%%%%%%%%%%%%%%%%%%%%%%%%%%%%%%%fig. 12%%%%%%%%%%%%%%%%%%%%%%%%%%%%%%%%%%%%%%%%%%%%%
 \begin{figure}
 \begin{center}
  \begin{overpic}[width=0.4\textwidth]{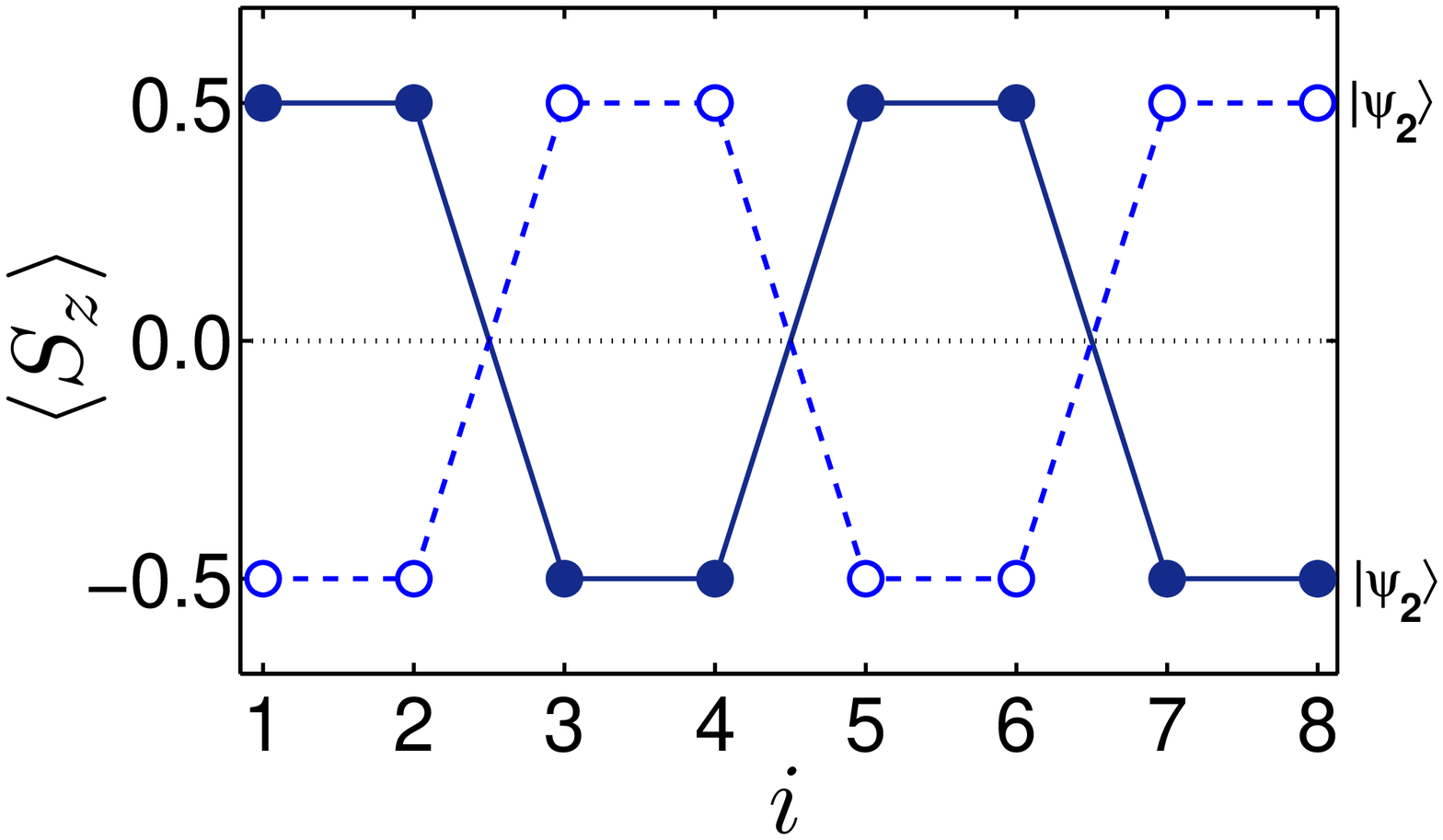}
     \put(1,54){(a)}
  \end{overpic}
    \begin{overpic}[width=0.35\textwidth]{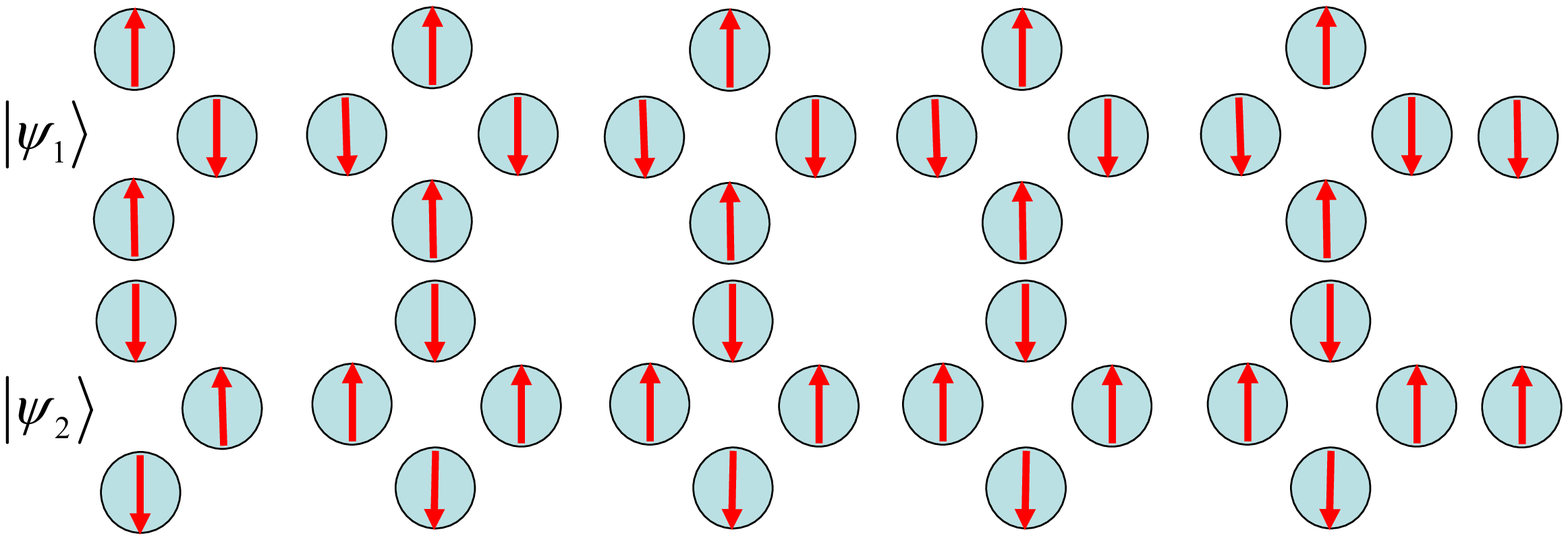}
     \put(-6,35){(b)}
  \end{overpic}
 \end{center}
\caption{(color online)
 (a) Local magnetization $\langle S_z\rangle$ at the lattice site $i$
 for $J'_z=-2|J|$ with $J < 0$ and $h=0$ in the AFP phase.
 (b) Pictorial representation of the groundstates with spin configuration in the AFP phase.
  Note that each plaquette has a ferromagnetic configuration
  of local magnetizations in the AFP phase.}
 \label{AFPmg}
\end{figure}
%%%%%%%%%%%%%%%%%%%%%%%%%%%%%%%%%%%%%%%%%%%%%%%%%%%%%%%%%%%%%%%%%%%%%%%%%%%%%%%%%%%%%

\subsection{Anti-ferromagnetic plaquette phase}
 The AFP phase also have a two-fold degenerate groundstates.
 From the two groundstates,
 we plot the local magnetizations
 $\langle \psi_n |  S_z | \psi_n\rangle$ at the lattice site $i$
 in Fig.~\ref{AFPmg}(a).
 The properties of the local magnetizations
 in Fig.~\ref{AFPmg}(a) can be summarized as follows.
 For $|\psi_1\rangle$ ($|\psi_2\rangle$),
 the local magnetizations has a one-plaquette (four-site) periodic structure where
 (i) the first two sites, i.e., $A$ and $B$, have a maximum (minimum) magnetization
 $\langle S_z \rangle =1/2$ $(-1/2)$,
 (ii) the second two sites, i.e., $C$ and $D$,
  have a minimum (maximum) magnetization $\langle S^z\rangle=-1/2$ $(1/2)$.
 For other values of system parameters,
 the characteristic behaviors of the local magnetizations are not
 changed.
 Comparing to the local magnetizations in plaquettes,
 each plaquette has an anti-ferromagnetic configuration of the local magnetizations.
 Also, for one-plaquette shift,
 the local magentizations from $|\psi_2\rangle$ are equal to ones from $|\psi_1\rangle$.
 Then, under one-plaquette transformation,
 one groundstate becomes the other groundstate under one-plaquette shift transformation.
 From the calculations of two-point spin correlations for the
 groundstates,
 we have confirmed that
 for the first two sites $i = A$ and $B$ with the maximum (minimum) magnetizations
 (the third two sites $i = C$ and $D$ with the minimum (maximum) magnetizations),
 the two-point spin correlations between $A/B$ ($C/D$) and any site $j$ in the system
 satisfy the conditions in Eqs.~(\ref{eq:6a}) and ~(\ref{eq:6b}), which
 imply that each of the two sites is in a fully polarized state.
 Consequently,
 the two groundstate for the AFP phase can be respectively written as
\begin{subequations}
 \begin{eqnarray}
 \left|\psi_1\right\rangle
  &=& \prod_i \left|\uparrow_{i,u} \uparrow_{i,d} \downarrow_{i,l} \downarrow_{i,r}
  \right\rangle,
\\
 \left|\psi_2\right\rangle
  &=& \prod_i \left|\downarrow_{i,u} \downarrow_{i,d} \uparrow_{i,l} \uparrow_{i,r}
  \right\rangle.
 \end{eqnarray}
\end{subequations}
 Figure~\ref{AFPmg}(b) shows the two groundstates in the pictorial representation for the AFP phase.
 By comparing with the symmetry of the Hamiltonian,
 one can notice emergent symmetries for each groundstate.
 For instance, for a lattice rotation of each plaquette,
 each groundstate is invariant but the Hamiltonian is not.
 Also, for a vertical-to-horizontal site exchange
 in each plaquette, e.g, $(A,B) \leftrightarrow (C,D)$, the groundstate is invariant
 but the Hamiltonian is not.
 Hence, each groundstate has such emergent symmetries not belonging to the Hamiltonian symmetry.

 Furthermore, note that
 the two degenerate groundstate are transformed from one to the
 other under the spin-flip, the plaquette-rotational
 (one-site in each plaquette), or the vertical-to-horizontal site-exchange
 transformations.
 However, for $h =0$,
 the system Hamiltonian has only the spin-flip symmetry.
 Even for $h \neq 0$, the system Hamiltonian does not have
 the spin-flip, the plaquette-rotational
 (one-site in each plaquette), and the vertical-to-horizontal site-exchange
 symmetries.
 For $h = 0$ in the AFP phase, hence,
 although the two-fold degenerate groundstates can be understood
 from the breaking of the spin-flip symmetry,
 more symmetries such as the plaquette-rotational
 (one-site in each plaquette) and the vertical-to-horizontal site-exchange
 symmetries should be involved to be broken for the two degenerate groundstates.
 For $h \neq 0$,
 since the system Hamiltonian does not have the spin-flip symmetry,
 no responsible symmetry for the two degenerate groundstates
 exist in the system Hamiltonian.
 This is very similar case to the SB phase.
 Consequently,
 the two degenerate groundstates in the AFP phase
 cannot be understood within the Landau's spontaneous symmetry breaking picture.
 Although, in order to understand why the two groundstates exist in the AFP phase
 for $h \neq 0$,
 one may possibly consider emergent symmetries such as  the spin-flip, the plaquette-rotational
 (one-site in each plaquette), and the vertical-to-horizontal site-exchange
 symmetries that should be broken together,
 still the question why such emergent symmetries should be broken
 together must be an unanswerable.
%

%%%%%%%%%%%%%%%%%%%%%%%%%%%%%%%%%%%fig. 13%%%%%%%%%%%%%%%%%%%%%%%%%%%%%%%%%%%%%%%%%%%%%
 \begin{figure}
 \begin{center}
  \begin{overpic}[width=0.4\textwidth]{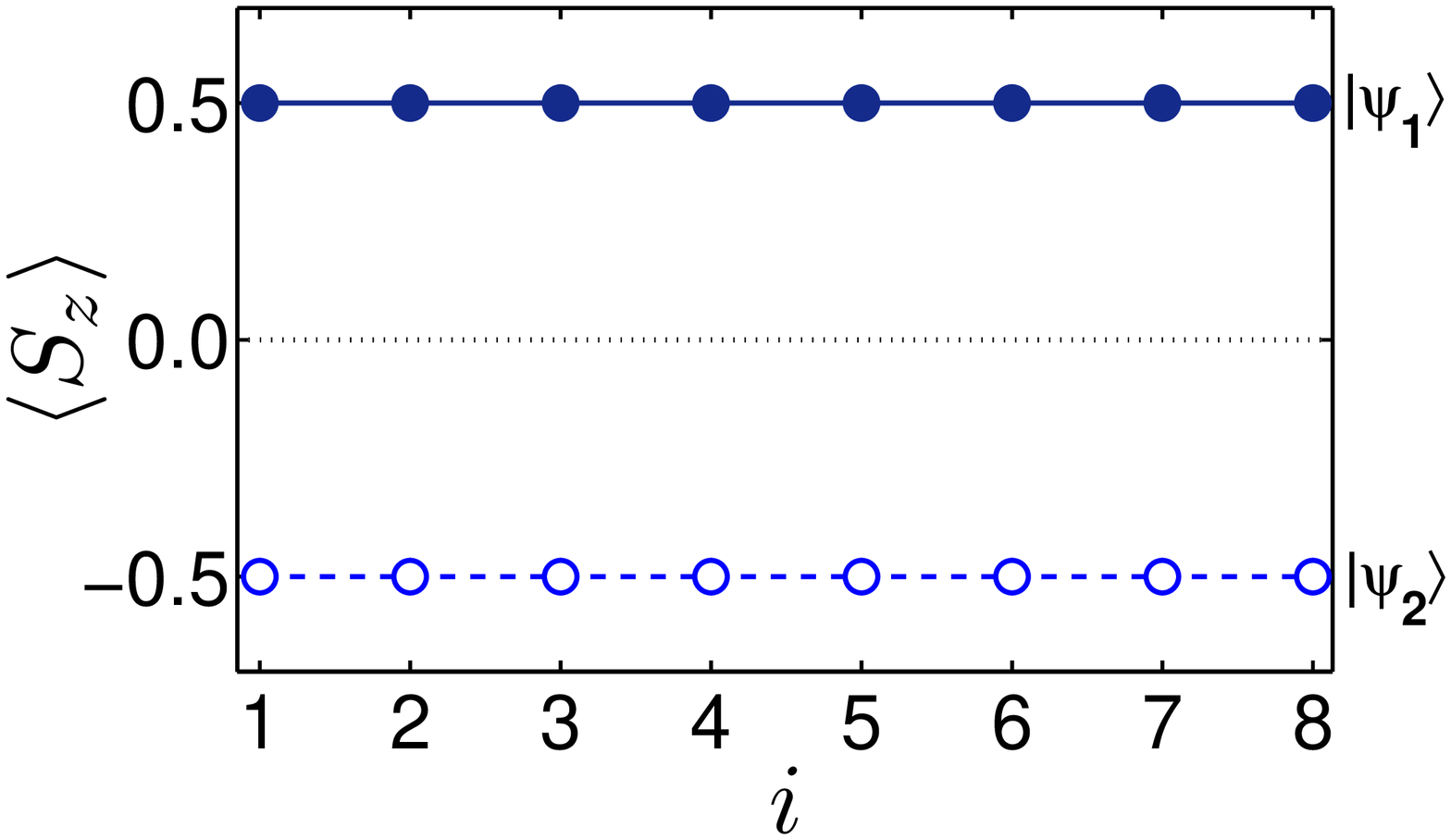}
     \put(1,52){(a)}
  \end{overpic}
    \begin{overpic}[width=0.35\textwidth]{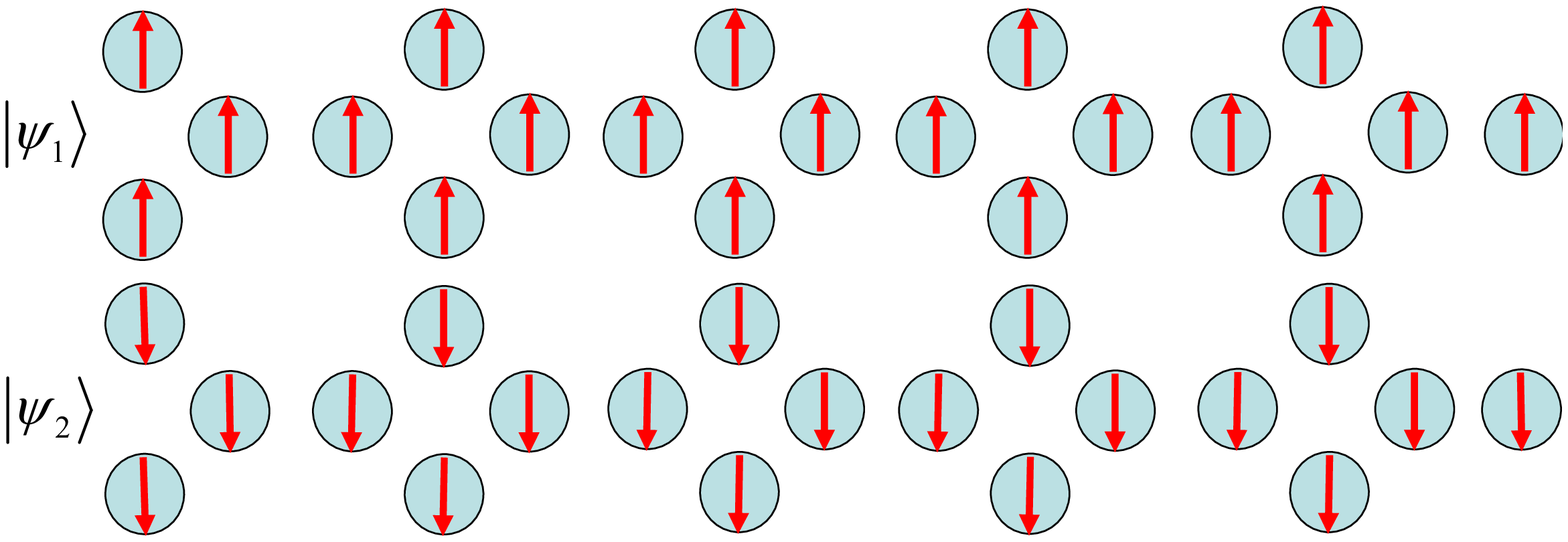}
     \put(-6,35){(b)}
  \end{overpic}
 \end{center}
\caption{(color online)
 (a) Local magnetization $\langle S_z\rangle$ at the lattice site $i$
 for $J'_z=2|J|$ with $J<0$ and $h=0$ in the F phase.
 (b) Pictorial representation of the groundstates with spin configuration in the F phase.
 }
 \label{Fmg}
\end{figure}
%%%%%%%%%%%%%%%%%%%%%%%%%%%%%%%%%%%%%%%%%%%%%%%%%%%%%%%%%%%%%%%%%%%%%%%%%%%%%%%%%%%%%

\subsection{Ferromagnetic phase}
 In the F phase,
 we detect two degenerate groundstates.
 From the two groundstates,
 we plot the local magnetizations
 $\langle \psi_n |  S_z | \psi_n\rangle$ at the lattice site $i$
 in Fig.~\ref{Fmg}(a).
 The local magnetizations
 have $\langle S^z_i \rangle = 1/2$ from $|\psi_1\rangle$
 and $\langle S^z_i \rangle = -1/2$ from $|\psi_2\rangle$.
 Similar to the case of the FP phase,
 it has been observed that
 any site $j$ in the system satisfies
 the conditions in Eqs.~(\ref{eq:6a}) and (\ref{eq:6b})
 in the two-point spin correlations.
 Then,
 the state for the system is in a product state of the spin states of each site.
 According to the values of local magnetizations,
 the two groundstates for the F phase can be respectively written as
\begin{subequations}
 \begin{eqnarray}
 \left|\psi_1\right\rangle
  &=& \prod_i \left|\uparrow_{i,u} \uparrow_{i,d} \uparrow_{i,l} \uparrow_{i,r}
\right\rangle,
\\
 \left|\psi_2\right\rangle
  &=& \prod_i \left|\downarrow_{i,u} \downarrow_{i,d} \downarrow_{i,l} \downarrow_{i,r} \right\rangle.
 \end{eqnarray}
\end{subequations}
 These two groundstates are presented pictorially for the F phase in Fig.~\ref{Fmg}(b).

 Note that this F phase exist for $h =0$.
 One can also easily notice that
 under a spin-flip transformation,
 one groundstate becomes the other groundstate
 and
 the system Hamiltonian is invariant.
 Hence,
 the two degenerate groundstates
 can be understood by the spin-flip symmetry breaking.
 However, by comparing with the symmetry of the Hamiltonian,
 one can notice emergent symmetries for each groundstate.
 Similar to the FP phase,
 each groundstate has more symmetries such as
 the lattice-rotation and the exchange symmetries.
 Consequently, each groundstate in the F phase
 has such emergent symmetries not belonging to the Hamiltonian symmetry.

\section{Discussions on emergent symmetry and spontaneous symmetry breaking}
 In the previous section, for each phase,
 we have discussed how the explicit forms of groundstates can be extracted
 from the characteristic properties of the local magnetizations
 and the two-point spin correlations in our model.
 Emergent symmetries of groundstates have been discussed.
 In this section, as a summary, we will discuss
 how degenerate groundstates can be related to spontaneous symmetry
 breaking in association with the Landau theory.
 What our results can suggest to understand beyond the Landau's
 symmetry breaking mechanism will be also discussed.

\begin{figure}
 \begin{center}
  \begin{overpic}[width=0.3\textwidth]{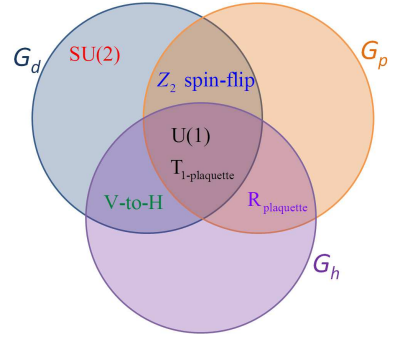}
  \end{overpic}
 \end{center}
\caption{(color online)
 Relevant symmetries to degenerate groundstates in the Hamiltonian of Eq.
 (\ref{Ham}).
 $G_d$, $G_p$, and $G_h$ denote the symmetry groups of the dimer,
 plaquette, and field Hamiltonians, respectively.
 Then, the symmetry group of the Hamiltonian in Eq.
 (\ref{Ham}) is $G = G_d \bigcap G_p \bigcap G_h$ for $h \neq 0$.
 For $h=0$, it becomes $G = G_d \bigcap G_p$.
 `V-to-H', $T_{1-plaquette}$, `$Z_2$ spin-flip', and
 $R_{plaquette}$ indicate the vertical-to-horizontal site-exchange
 symmetry, the one-plaquette translational
 symmetry, the spin-flip symmetry, and the plaquette-rotational
 symmetry, respectively.
 }
 \label{fig14}
\end{figure}

 As we discussed in the introduction,
 if one assumes that two degenerate groundstates $\left|\psi_1\right\rangle$ and $\left|\psi_2\right\rangle$, i.e.,
 $\left|\psi_1\right\rangle \neq \left|\psi_2\right\rangle$,
 are obtained from any method, i.e., numerical or analytical calculations for a system Hamiltonian $H$,
 they satisfy
 $H|\psi_{n}\rangle=E_{gs}|\psi_{n}\rangle$
 and can have a unitary transformation $U$ connecting each other, i.e.,
 $\left|\psi_1\right\rangle = U \left|\psi_2\right\rangle$.
 If the Hamiltonian is invariant under the unitary transformation, i.e.,
 $UHU^{\dagger}=H$ and the unitary transformation $U$
 is related to an element of a subgroup $g$ of the Hamiltonian symmetry group $G$,
 the two degenerate groundstates are originated from a breaking of
 a symmetry consisting of the subgroup $g$ because of
 $\left|\psi_1\right\rangle \neq \left|\psi_2\right\rangle$.
 This case corresponds to the spontaneous symmetry breaking in the Landau theory.
 According to the spontaneous symmetry breaking mechanism in the Landau theory,
 then, the actual key symmetries associated with the two
 degenerate groundstates are presented in Fig. \ref{fig14} for
 the spin-$1/2$ plaquette chain, i.e.,
 one-plaquette translational and spin-flip symmetries for $h=0$
 and one-plaquette translational symmetry for $h \neq 0$.
 As we discussed in Sec.  IV,
 the spontaneous breaking of spin-flip symmetry
 is shown to induce the two degenerate groundsates in the F phase
 and the spontaneous breaking of one-plaquette translational symmetry
 induces the two degenerate groundstates in the MAFP and MFP phases.
 In the cases of SAFP ($h=0$) and SFP ($h=0$) phases,
 the spontaneous breaking of both spin-flip and one-plaquette translational
 symmetries is applicable well to understand the two degenerate groundsates.

%

%%%%%%%%%%%%%%%%%%%%%%%%%%%%%%%%%%%fig. 14%%%%%%%%%%%%%%%%%%%%%%%%%%%%%%%%%%%%%%%%%%%%%
%%%%%%%%%%%%%%%%%%%%%%%%%%%%%%%%%%%%%%%%%%%%%%%%%%%%%%%%%%%%%%%%%%%%%%%%%%%

 %
 However, if the Hamiltonian is not invariant under the unitary transformation $U$,
 i.e., $UHU^{\dagger} \neq H$, the unitary transformation $U$
 is not related to any element of any subgroup $g$ of the Hamiltonian symmetry group $G$.
 Then a symmetry described by the unitary transformation does not belong to the Hamiltonian symmetry.
 In fact,
 such a situation has been observed in the SAFP ($h \neq 0$),
 SFP ($h \neq 0$), AFP ($h=0$), SB, and AFP ($h\neq0$) phases in the spin-$1/2$ plaquette chain system.
 Based on the relations between the unitary transformations and the Hamiltonian symmetry,
 to be more precise,
 we can categorize as for a given system parameter,
 (i) of all unitary transformations connecting degenerate
 groundstates, some are relevant to Hamiltonian symmetry and some are
 not relevant to Hamiltonian symmetry, and
 (ii) all unitary transformations connecting degenerate groundstates
  are not relevant to Hamiltonian symmetry.

 The first case (i) corresponds to the SAFP ($h \neq 0$),
 SFP ($h \neq 0$), and AFP ($h = 0$) phases.
 In the SAFP ($h \neq 0$) and SFP ($h \neq 0$) phases, that is,
 the two degenerate groundstates can transform
 from one groundstate to the other groundstate under
 the one-plaquette translational or the spin-flip transformations,
 but the Hamiltonian is invariant only under the one-plaquette translational
 transformation.
 Also, in the AFP ($h = 0$) phase,
 the two degenerate groundstates can transform
 from one groundstate to the other groundstate under
 the spin-flip, the plaquette-rotational, or the vertical-to-horizonal site-exchange transformations,
 but the Hamiltonian is invariant only under the spin-flip transformation.
 Then, these examples show that more symmetries than Hamiltonian symmetry
 can be broken for a spontaneous symmetry breaking.
 Such situations cannot be explained fully by the spontaneous symmetry
 breaking mechanism in the Landau theory.
 For a complete explanation of occurring such degenerate groundstates
 in a view of spontaneous symmetry breaking,
 if such more symmetries than Hamiltonian symmetry
 are considered as emergent symmetries,
 they can be called an {\it entailed-emergent symmetry}
 that is broken together with a broken-symmetry connecting
 two degenerate groundstates within the Hamiltonian symmetry.

 The second case (ii) corresponds
 to the SB and AFP ($h\neq0$) phases in the spin-$1/2$ plaquette chain system.
 In the AFP ($h \neq 0$) phase, that is,
 the two degenerate groundstates can transform
 from one groundstate to the other groundstate under
 the spin-flip, the plaquette-rotational, or the vertical-to-horizonal site-exchange transformations,
 but the Hamiltonian is not invariant.
 In contrast to the first case (i), moreover, no symmetry in the Hamiltonian symmetry
 is responsible for the two degenerate groundstates.
 In the SB phase,
 also,
 the vertical-to-horizonal site-exchange transformation connects one
 groundstate to the other groundstate,
 but the Hamiltonian is not invariant under the transformation
 and the Hamiltonian symmetry does not have any symmetry responsible
 for the two degenerate groundstates within the spontaneous symmetry breaking
 mechanism in the Landau theory.
 In order to explain such situations for a view of spontaneous symmetry
 breaking, if one can introduce an emergent symmetry that is
 responsible for degenerate groundstates,
 those cases can be called {\it spontaneous emergent symmetry breaking}.
 In this sense,
 such emergent symmetries can be called a {\it to-be-broken emergent
 symmetry} in order to distinguish from emergent symmetries
 that occur in degenerate groundstates, because
 for instance,
 each of the two degenerate groundstates in the F phase
 has emergent symmetries such as the vertical-to-horizontal
 site-exchange symmetry and the plaquette-rotational symmetry.

 However, although introducing such emergent symmetries
 associated with
 degenerate groundstates
 could explain occurring such degenerate groundstates
 in an extended view of the spontaneous symmetry breaking mechanism,
 the raised question, i.e.,
 how such emergent symmetries associated with
 degenerate groundstates in a view of spontaneous symmetry breaking
 are related to Hamiltonian symmetry, is still left to be answered.
 To answer on the question,
 let us consider the Hamiltonian symmetry group
 in the spin-$1/2$ plaquette chain model in Eq. (\ref{Ham}).
 In Fig. \ref{fig14},
 we draw a schematic diagram to show the key symmetries responsible
 for degenerate groundstates in the symmetry groups of
 the dimer, the plaquette, and the field Hamiltonians.
 $G_d$, $G_p$, and $G_h$ indicate the symmetry groups of the dimer,
 plaquette, and field Hamiltonians, respectively.
 The symmetry group of the Hamiltonian in Eq.
 (\ref{Ham}) can be presented as a common subgroup
 of the symmetry groups $G_d$, $G_p$ and $G_h$, i.e.,
 $G = G_d \bigcap G_p \bigcap G_h$ for $h \neq 0$
  and $G = G_d \bigcap G_p$ for $h=0$.
 For the spontaneous symmetry breakings in the F, MAFP, MFP, SAFP ($h=0$) and SFP ($h=0$) phases,
 straightforwardly, the broken-symmetries are
 the one-plaquette translational or/and the spin-flip symmetries
 belonging to the Hamiltonian symmetry group $G$.
 One may also easily notice that
 all so-called emergent symmetries responsible for degenerate groundstates in the
 spin-$1/2$ plaquette chain system belong to the common subgroups
 of pair of the symmetry groups $G_d$, $G_p$ and $G_h$, i.e.,
 $\tilde G - G$
 with $\tilde G = (G_d \bigcap G_p) \bigcup (G_p \bigcap G_h) \bigcup (G_h \bigcap G_d)$.
 For instance, in the SB phase, the vertical-to-horizontal
 site-exchange symmetry belongs to the common group $G_d \bigcap
 G_p - G$.
 Furthermore, compared to the other phases,
 it should be noted that
 the two degenerate groundsates in the SB phase have
 a local $\mathrm{SU}(2)$ symmetry from the singlet states in Eqs.~(\ref{SB1}) and
 (\ref{SB2}).
 Similarly, for in the MAFP (MFP)
 phase, a local $\mathrm{SU}(2)$ symmetry also
 appears in the two degenerate groundstates in Eqs.~(\ref{MFIW1}) and ~(\ref{MFIW2})
 (Eqs. (\ref{MFPW1}) and (\ref{MFPW2})), although the occurrence of the two degenerate groundstates
 can be explained by the spontaneous breaking of the one-plaquette
 translational symmetry
 belonging to the Hamiltonian symmetry.
 Such an occurrence of the local $\mathrm{SU}(2)$ symmetry
 cannot be understood within the Hamiltonian symmetry group
 because only the dimer Hamiltonian $H_d$ has the local $\mathrm{SU}(2)$
 symmetry.
 Hence, these facts imply that a symmetry belonging to the symmetry groups
 of the constituent Hamiltonians, i.e., $H_d$, $H_p$ and $H_h$,
 plays a role for degenerate groundstates.
 In other words, for given system parameters,
 all symmetries belonging to the largest symmetry group $G_d \bigcup G_p \bigcup G_h$
 play a significant role for the system to reach
 its lowest energy states, i.e., degenerate groundstates.
 In a system, then, symmetry of degenerate groundstates might be determined
 within a largest common symmetry groups of constituent Hamiltonians, i.e.,
 $\tilde G$, describing the system.
 Consequently,
 {\it degenerate groundstates can be understood by an extended spontaneous symmetry breaking picture
 including spontaneous emergent symmetry breakings,
 and they do not have a broken-symmetry within a largest common symmetry
 of constituent Hamiltonians describing a given system
 but can have more symmetries than the largest common symmetry.}

%

%%%%%%%%%%%%%%%%%%%%%%%%%
\section{Spin structure factor}
 Experimentally, the quantum spin-$1/2$ plaquette chain can be realized in a number of
 magnetic compounds, especially, some real insulating magnetic materials
 such as
  $[(\mathrm{CuL})_2\mathrm{Dy}][\mathrm{Mo(CN)}_8]$~\cite{mater1},
  [$\mathrm{Fe}\mathrm{(H}_2\mathrm{O)}\mathrm{(L)}][\mathrm{Nb(CN)}_8][\mathrm{Fe(L)}$]
  ~\cite{mater2} and $\mathrm{Dy}(\mathrm{NO}_3)(\mathrm{DMSO})_2\mathrm{Cu(opba)}(\mathrm{DMSO})_2$~\cite{mater3}.
 Those materials can be used to confirm an existence of emergent symmetries of groundstates.
 To do this, a way to detect the studies phases in our model system
 is to observe their characteristic magnetic properties,
 i.e., spin structure factors that can be observed by using neutron scattering experiments.
 For instance, for antiferromagnetic transverse-field Ising
 models in the pyrochlore lattice, a pinch point structure predicted
 theoretically~\cite{sq0,sqn} has been
 observed to correspond to a singularity in the spin structure factor of the
 spin-flip channel in the experiment of neutron scattering~\cite{sq1}.
 In this section, thus, we will discuss spin structure factors for the eight phases in our Hamiltonian
 because the local magnetizations are zero in the SD phase.

\begin{figure}
\begin{center}
 \begin{overpic}[width=0.23\textwidth]{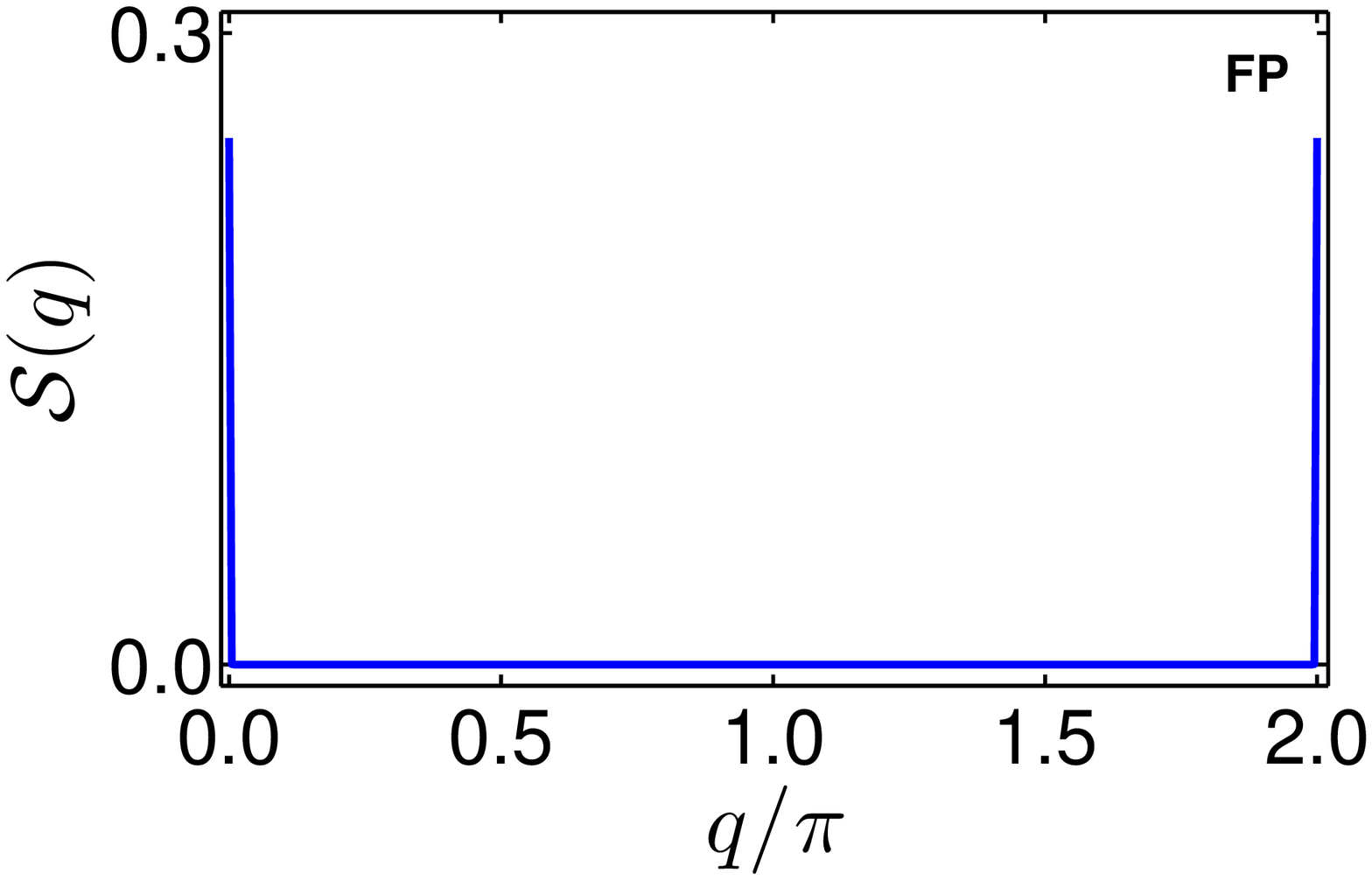}
            \put(-3,55){(a)}
 \end{overpic}
 \begin{overpic}[width=0.23\textwidth]{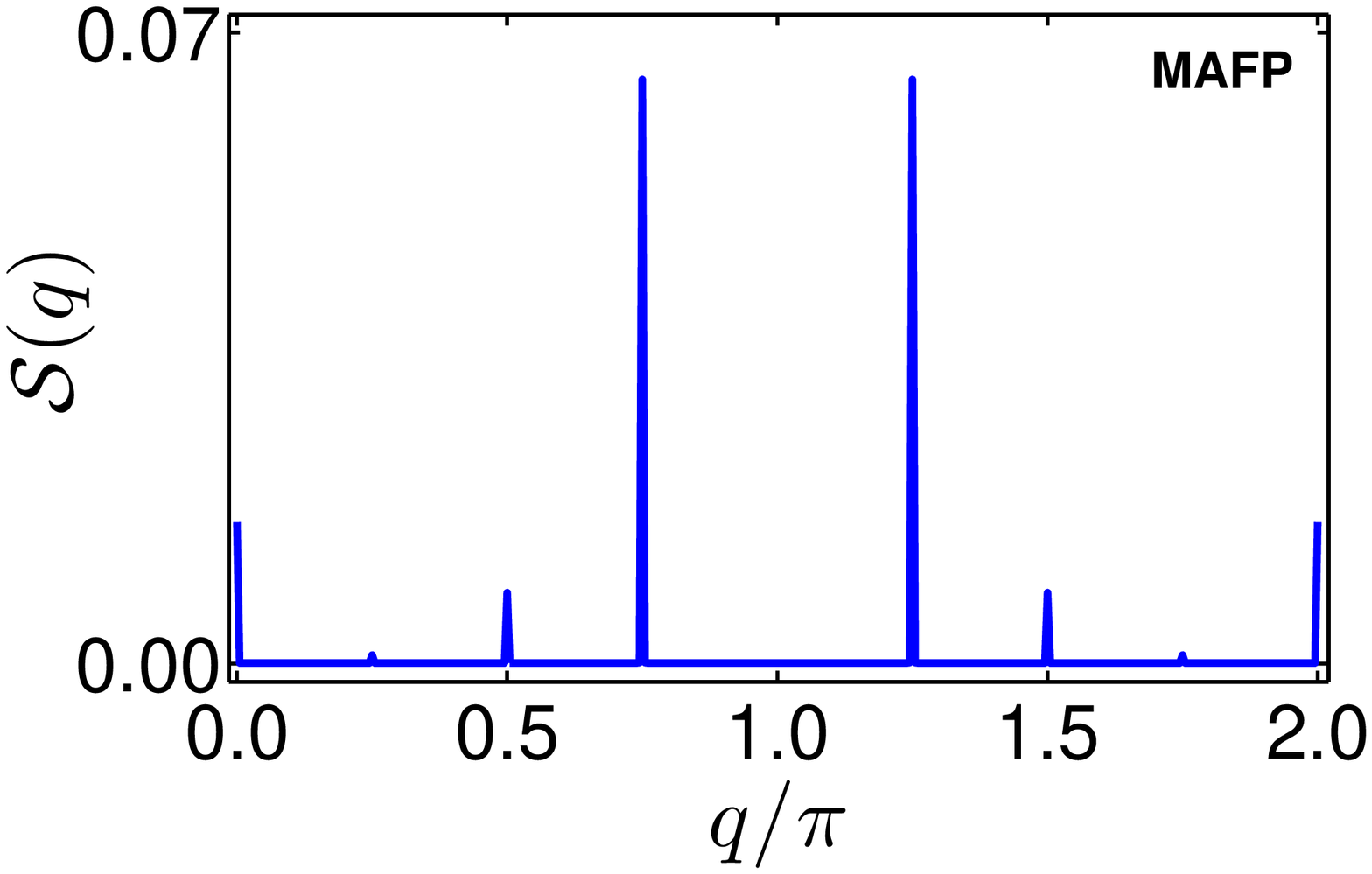}
            \put(-3,55){(b)}
 \end{overpic}
  \begin{overpic}[width=0.23\textwidth]{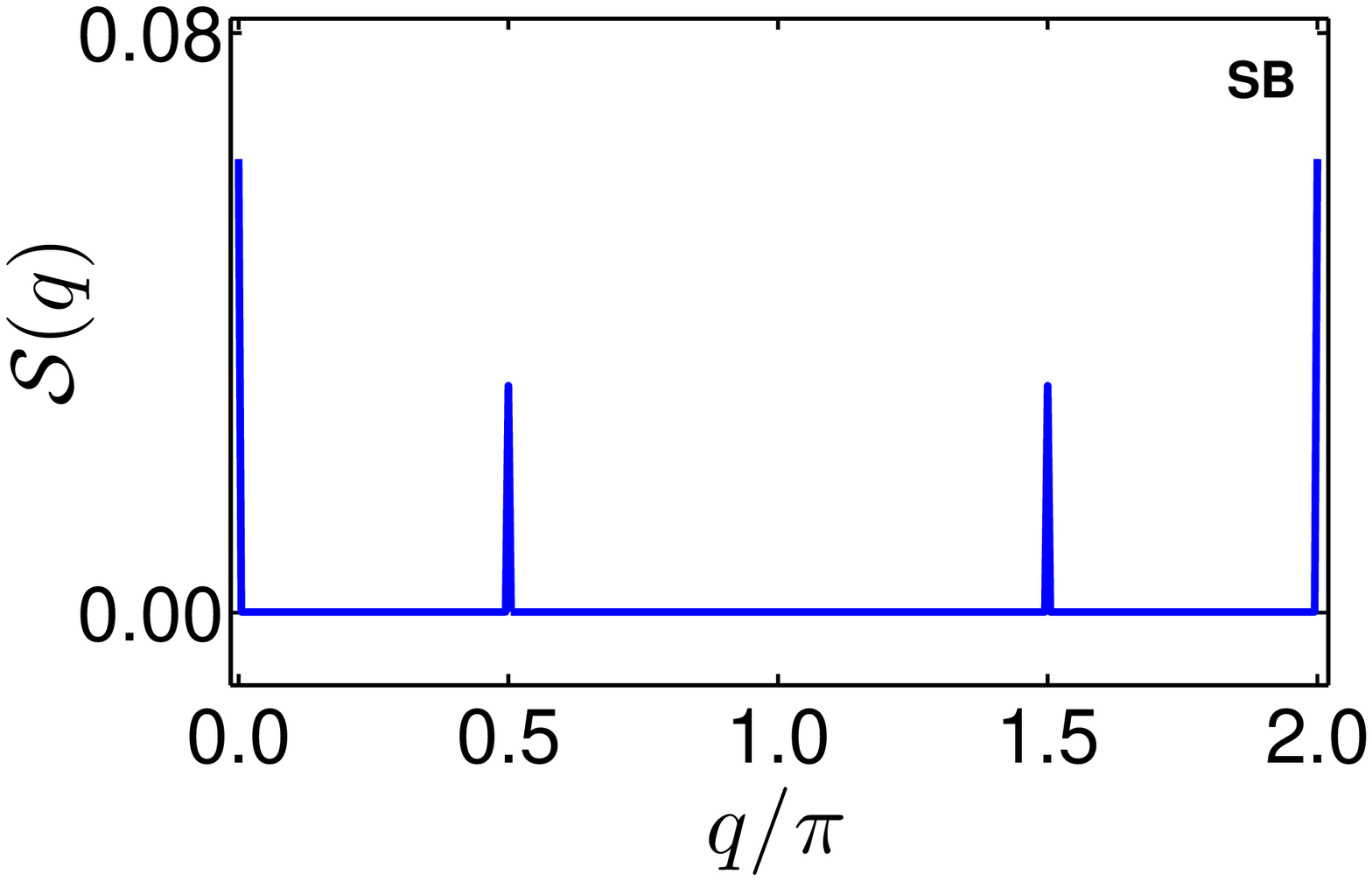}
            \put(-3,55){(c)}
 \end{overpic}
  \begin{overpic}[width=0.23\textwidth]{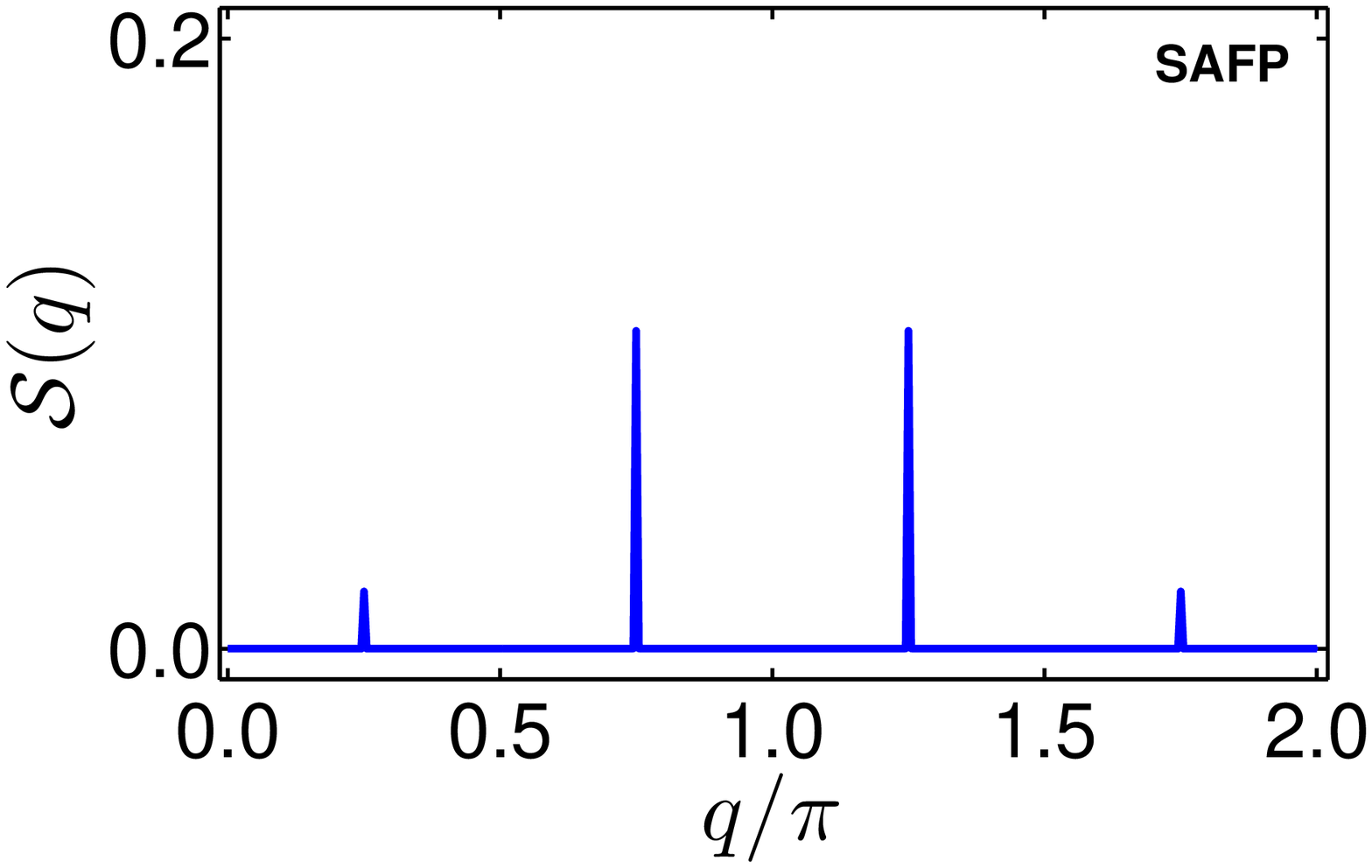}
            \put(-3,55){(d)}
 \end{overpic}
  \begin{overpic}[width=0.23\textwidth]{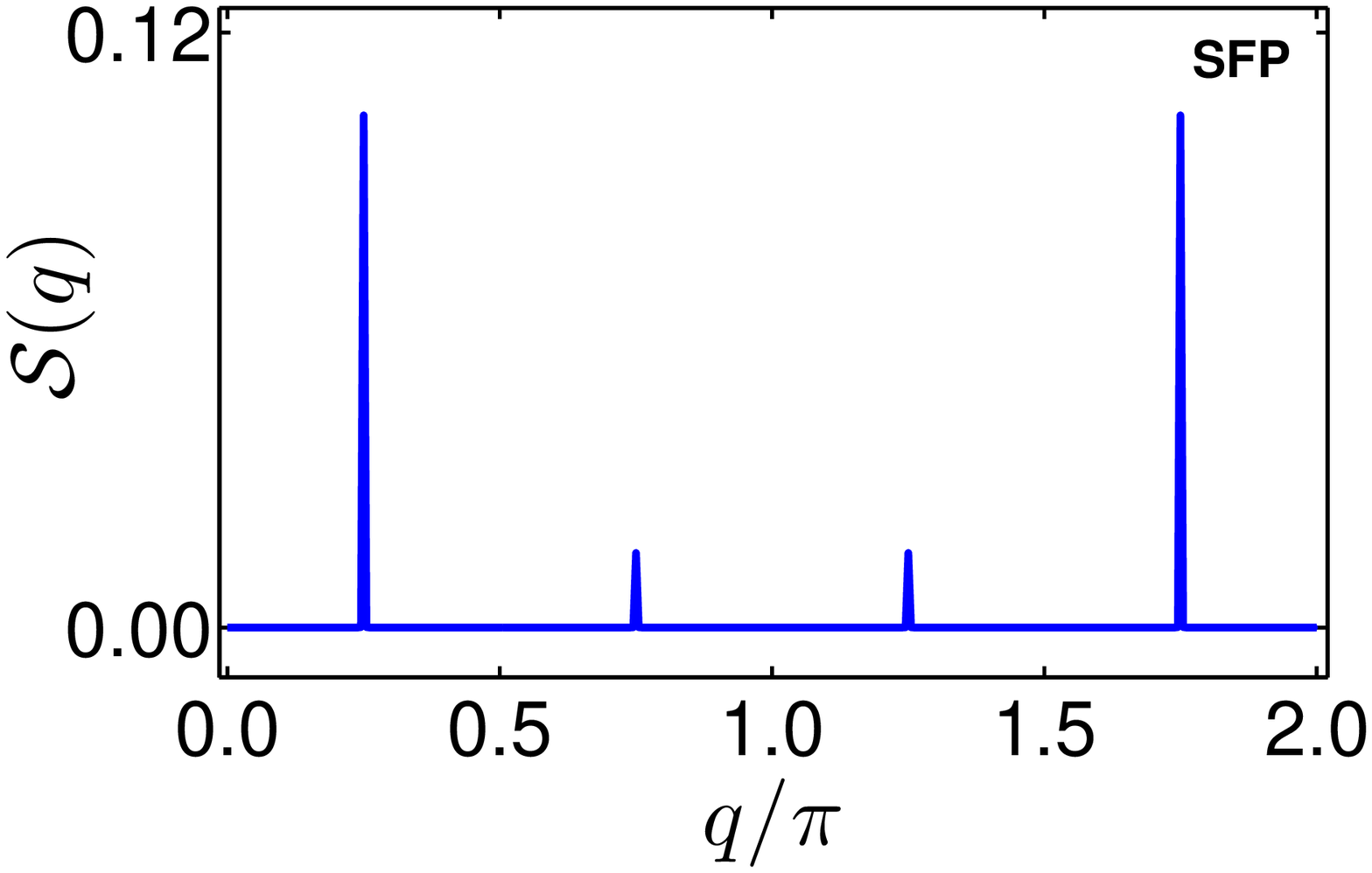}
            \put(-3,55){(e)}
 \end{overpic}
  \begin{overpic}[width=0.23\textwidth]{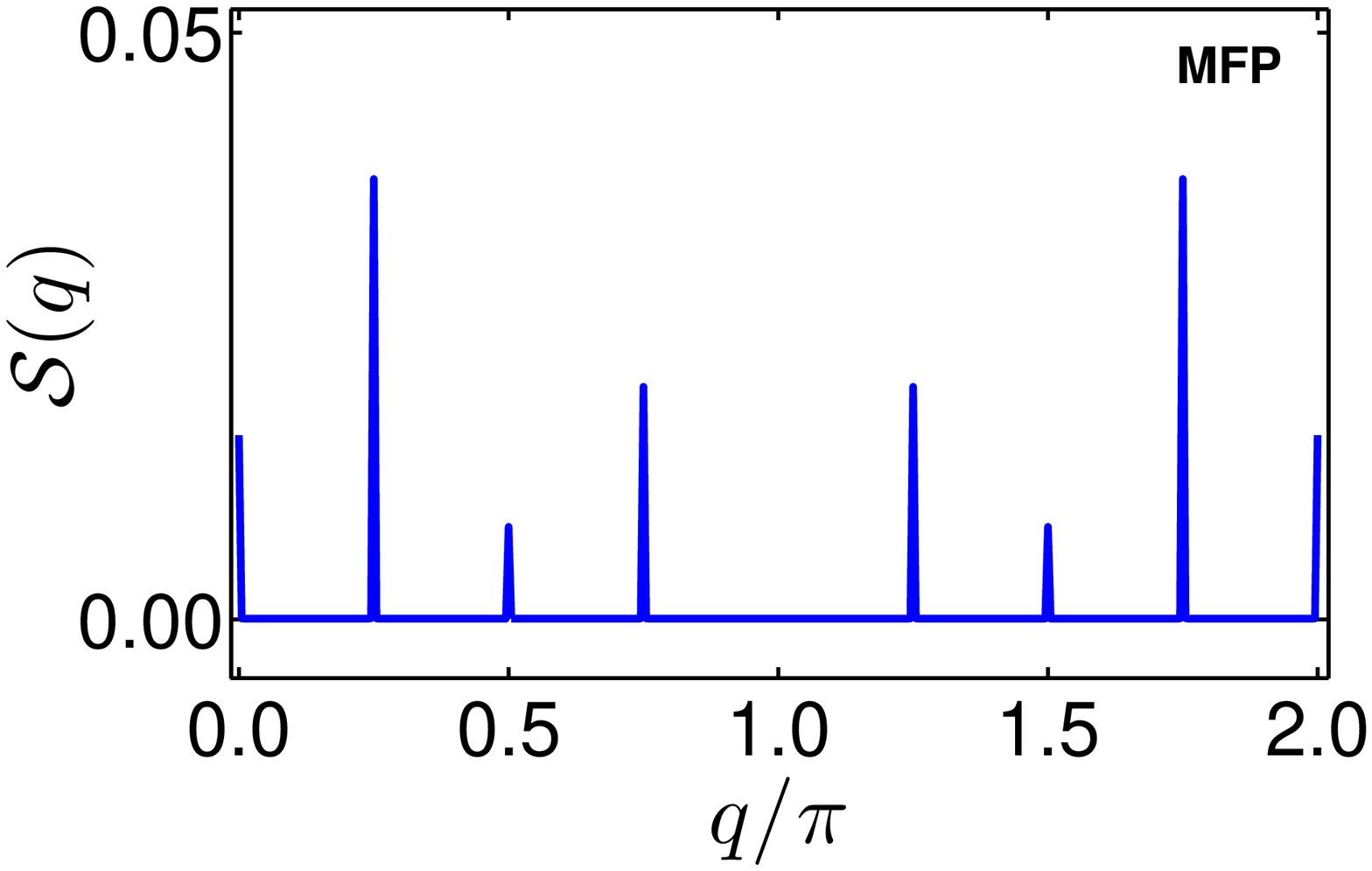}
            \put(-3,55){(f)}
 \end{overpic}
  \begin{overpic}[width=0.23\textwidth]{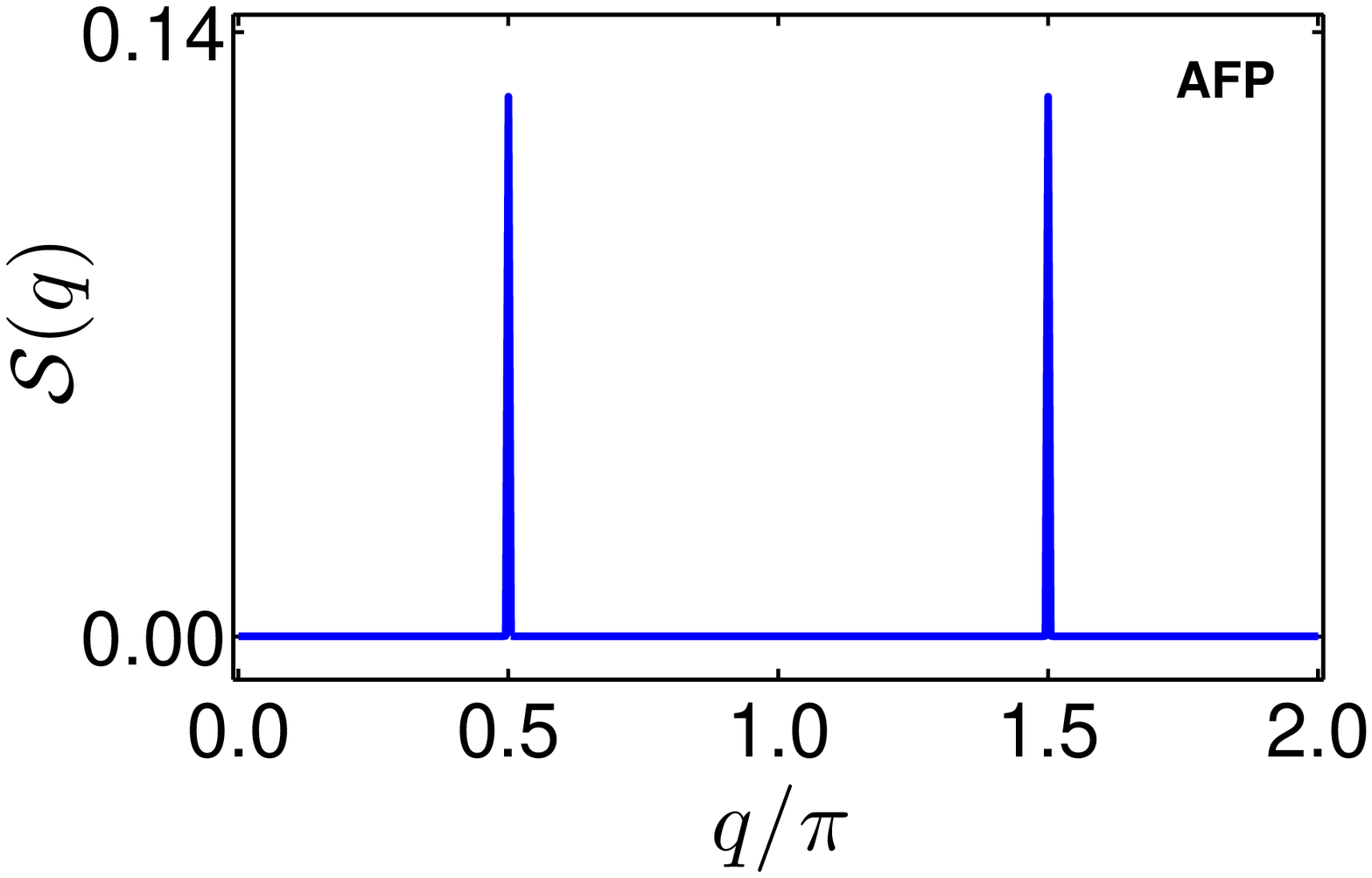}
            \put(-3,55){(g)}
 \end{overpic}
  \begin{overpic}[width=0.23\textwidth]{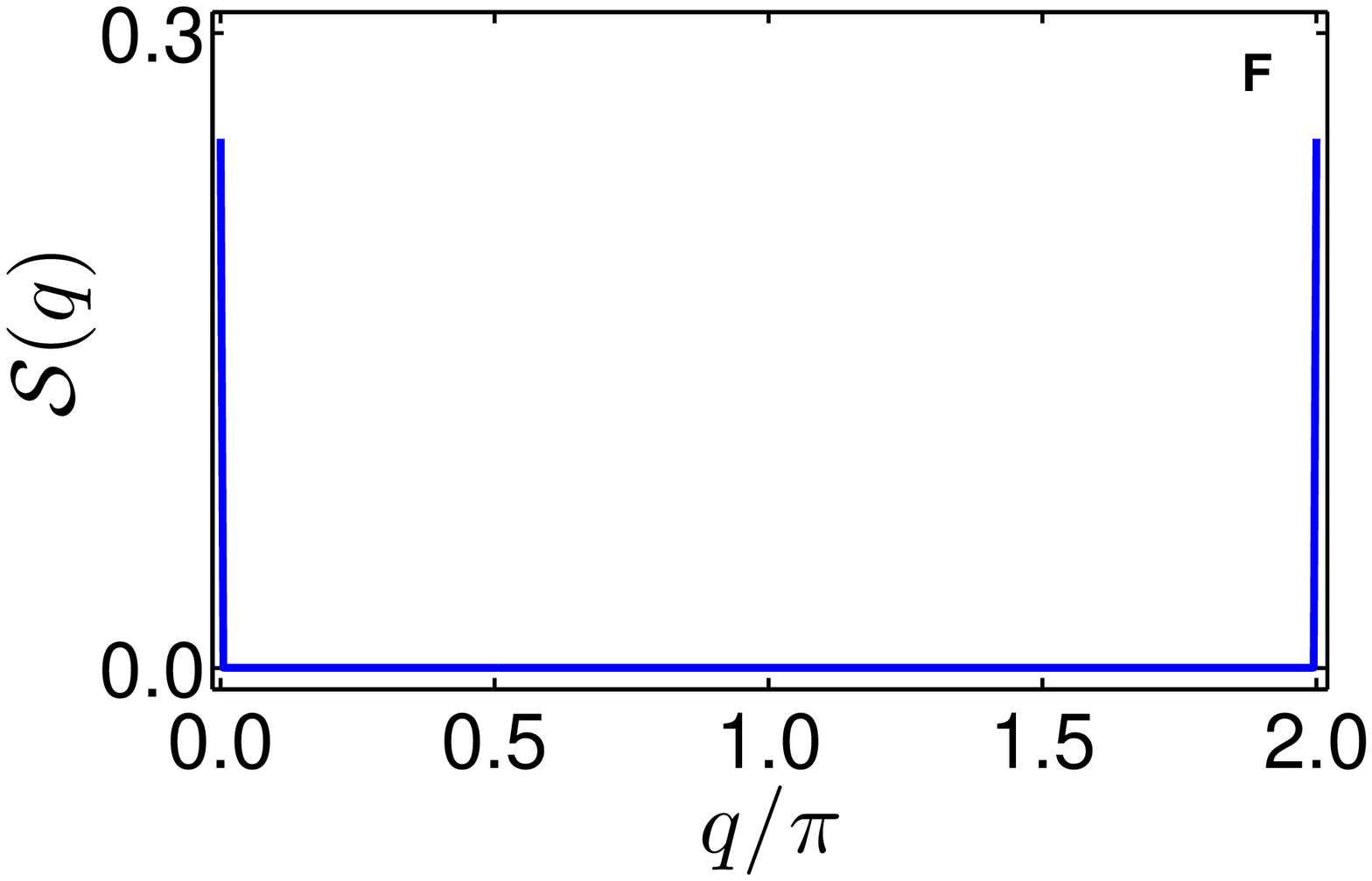}
            \put(-3,55){(h)}
 \end{overpic}
 \end{center}
\caption{(color online) Spin structure factors for
 (a) $J'_z=0.2J$ with $h=1.5J$ in the FP phase,
 (b) $J'_z=2J$ with $h=1.5J$ in the MAFP phase,
 (c) $J'_z=0.8J$ with $h=1.5J$ in the SB phase,
 (d) $J'_z=1.43J$ with $h=0$ in the SAFP phase,
 (e) $J'_z=-1.44J$ with $h=0$ in the SFP phase,
 (f) $J'_z=-1.41J$ with $h=0.3J$ in the MFP phase,
 (g) $J'_z=-2|J|$ with $h=-|J|$ ($J < 0$) in the AFP phase, and
 (h) $J'_z=2|J|$ with $h=0$ ($J < 0$) in the F phase.
  } \label{SF4}
\end{figure}

 For one-dimensional lattice systems with $N$ sites,
 a spin structure factor can be defined by the fourier transformation of spin correlation function.
 For $z$-direction, the spin structure factor can be defined as
\begin{equation}
 {\cal S}(q)=\lim_{N\rightarrow\infty} \frac{1}{N^2}\sum_{m=1}^N\sum_{j=1}^{N}
  {\exp\left[iqr\right]\langle S_{m}^zS_{j}^z\rangle},
 \label{ssf}
\end{equation}
 where $r\equiv j-m$ is the lattice distance, $q \in [0, 2\pi]$, and $m/j =  1, \cdots, N$.
%
%%%
 With the mapped one-dimensional chain structure in our system,
 we calculate spin structure factor ${\cal S}(q)$
 to investigate the characterized behaviors of spin structures in the momentum space $q$.
 Actually, we have calculated ${\cal S}(q)$ numerically and analytically based on
 the two-point spin correlations studied in the previous sections for each phase.
 In Fig.~\ref{SF4}, from numerical calculations,
 the spin structure factor densities ${\cal S}(q)$ as a function of $q$
 with $0\leq q\leq 2\pi$ are plotted for four different phases
 with given parameters.
 It is shown that for each phase, the ${\cal S}(q)$ has a unique peak structure for $0\leq q < 2\pi$:
 (a) for the FP phase with the average magnetization $M_z = 1/2$, there is a peak at $q=0$,
 (b) for the MAFP phase with $M_z = 1/8$, seven peaks at $q = k\pi/4$ with $k = 0, 1, 2, 3, 5, 6$ and $7$,
 (c) for the SB phase with $M_z = 1/4$, three peaks at $q = k\pi/2$ with $k = 0, 1$ and $3$,
 (d) for the SAFP phase with $M_z = 0$, four peaks at $q = k\pi/4$ with $k = 1, 3, 5$ and
 $7$,
 (e) for the SFP phase with $M_z = 0$, four peaks at $q = k\pi/4$ with $k = 1, 3, 5$ and $7$,
 (f) for the MFP phase with $M_z = 1/4$, three peaks at $q = k\pi/2$ with $k = 0, 1$ and
 $3$,
 (g) for the AFP phase with $M_z = 0$, two peaks at $q = k\pi/2$ with $k = 1$ and
 $3$, and
 (h) for the F phase with $M_z = \pm 1/2$, one peak at $q = 0$.
 Obviously, for the SD phase, ${\cal S}(q) =0$
 because the local magnetizations are zero for all lattice sites.
 Our analytic calculation has given the same peak structure in the spin structure factor ${\cal S}(q)$
 and is not presented.
 Such a characteristic peak location in ${\cal S}(q)$ then allows us to distinguish the different phases in our system.
 Consequently, the phases involving an emergent symmetry of groundstates
 can be observed by using a neutron scattering experiment.

%%%%%%%%%%%%%%%%%%%%%%%%%%%%%%%fig. 15%%%%%%%%%%%%%%%%%%%%%%%%%%%%%%%%%%%%%%%%%%%%%%
%%%%%%%%%%%%%%%%%%%%%%%%%%%%%%%%%%%%%%%%%%%%%%%%%%%%%%%%%%%

%%%%%%%%%%%%%%%%%%%%%%%%%
\section{Conclusions}
 We have investigated a relation between degenerate groundstates and spontaneous symmetry breaking
 in the spin-$1/2$ plaquette chain model.
 To do this, the iMPS representation with the iTEBD method
 are employed to calculate groundstates for a given parameter.
 The quantum fidelity has been shown to enable detecting degenerate groundstates for the system paramter space.
 By using the local magnetizations and the two-point spin correlations,
 the explicit forms of groundstates for each phase have been obtained.
 We have found that for whole parameter range, there are nine phases in which
 two phases (i.e., the FP and SD phases) have a single groundstate and seven phases have a
 two-fold degenerate groundstates.

 From unitary transformations connecting
 two degenerate groundstates each other,
 generally there are three types of degenerate groundstates: for a given system parameter,
 (i) all unitary transformations connecting degenerate groundstates are relevant to Hamiltonian symmetry,
 (ii) of all unitary transformations connecting degenerate
 groundstates, some are relevant to Hamiltonian symmetry and some are
 not relevant to Hamiltonian symmetry, and
 (iii) all unitary transformations connecting degenerate groundstates
 are not relevant to Hamiltonian symmetry.
 The first case (i) corresponds to
 the F, MAFP, MFP, SAFP ($h=0$), and SFP ($h=0$) phases,
 which is understood by the Landau's symmetry breaking picture.
 The SAFP ($h \neq 0$),
 SFP ($h \neq 0$), and AFP ($h = 0$) phases belong to the second case
 (ii), which cannot be understood fully by the Landau's symmetry breaking
 picture.
 The third case (iii) are the SB and AFP ($h\neq0$) phases,
 which is beyond the Landau's symmetry breaking
 picture.
 Furthermore,
 the groundstates can have more symmetries than the Hamiltonian
 symmetry.
 For instance,
 the appearances of the local $\mathrm{SU}(2)$ symmetry for the MAFP, MFP, SD, and SB phases
 and of the emergent vertical-to-horizontal site-exchange and  plaquette-rotational
 symmetries for the F phase are not explained by the Landau's symmetry breaking picture.
 Such rich emergent phenomena occurring beyond the Landau theory
 in our model suggest an equal footing theory as an extension of
 the Landau's spontaneous symmetry breaking, i.e.,
 {\it degenerate groundstates are induced by a spontaneous breaking of
 symmetries belonging to a largest common symmetry of continent Hamiltonians
 describing a given system
 but can have more symmetries than the largest common symmetry.}

 Finally, the characteristic properties of the spin structure factors
 in the different phases have been discussed.
 It is shown that the spin structures have the unique peak structures that can be distinguished
 from one another. Such distinguishable peak structures can be observed by using
 a neutron scattering experiment,
 which can be an experimental evidence for the extension of
 the Landau's spontaneous symmetry breaking theory.

%%%%%%%%%%%%%%%%%%%%%%%%%%%%%%%%

\acknowledgements
 M.H. thanks Hai-Tao Wang for useful discussions on the iMPS calculation.
 This work is supported by the National Natural Science Foundation of
 China (Grant No:11104362).

\appendix

\section{\label{app:MAFP1} Spin states for $A,B,E$ and $F$ sites in the MAFP phase}
\begin{figure}
 \begin{center}
  \begin{overpic}[width=0.4\textwidth]{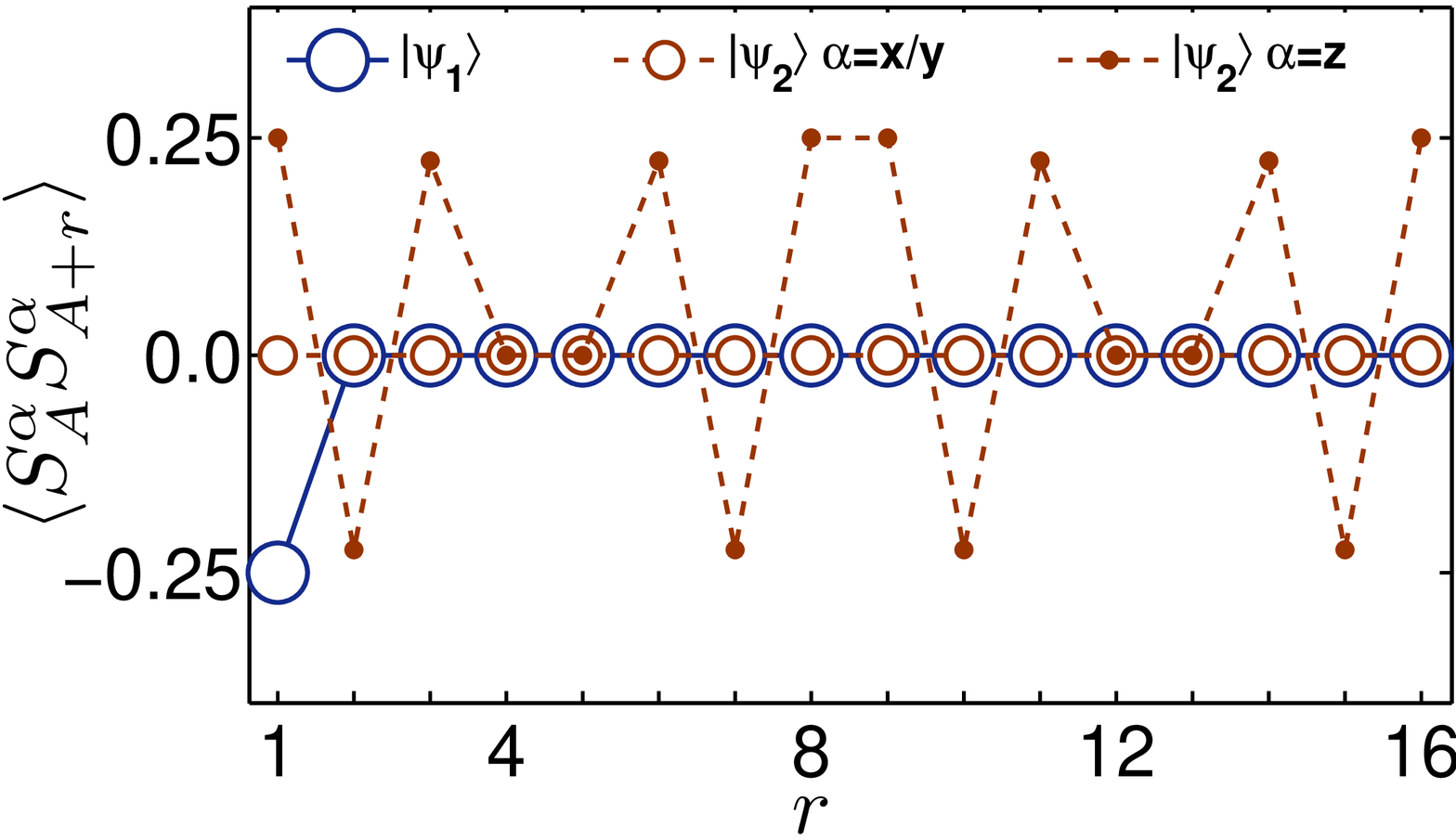}
            \put(1,52){(a)}
  \end{overpic}
  \begin{overpic}[width=0.4\textwidth]{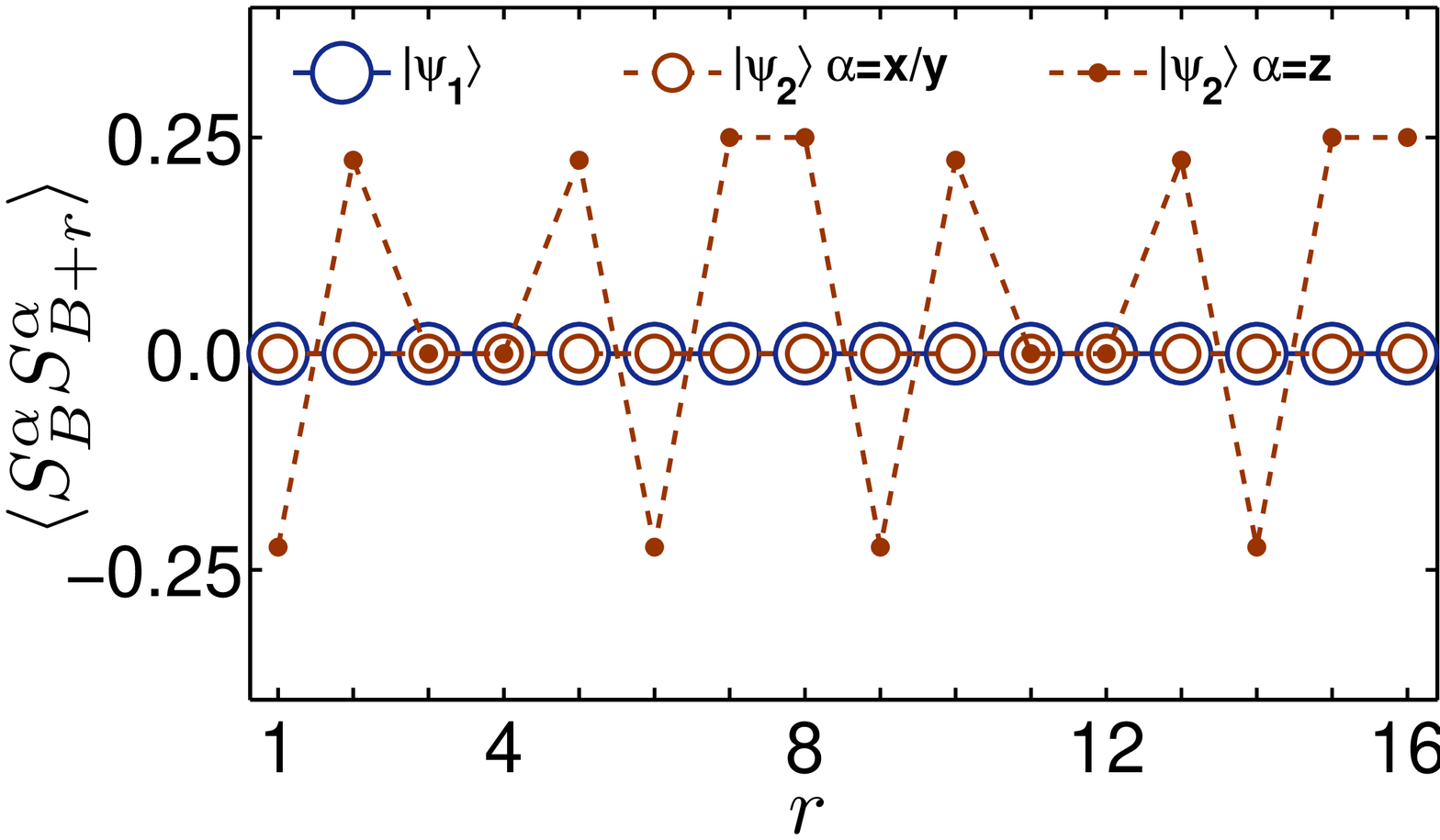}
            \put(1,52){(b)}
  \end{overpic}
  \begin{overpic}[width=0.4\textwidth]{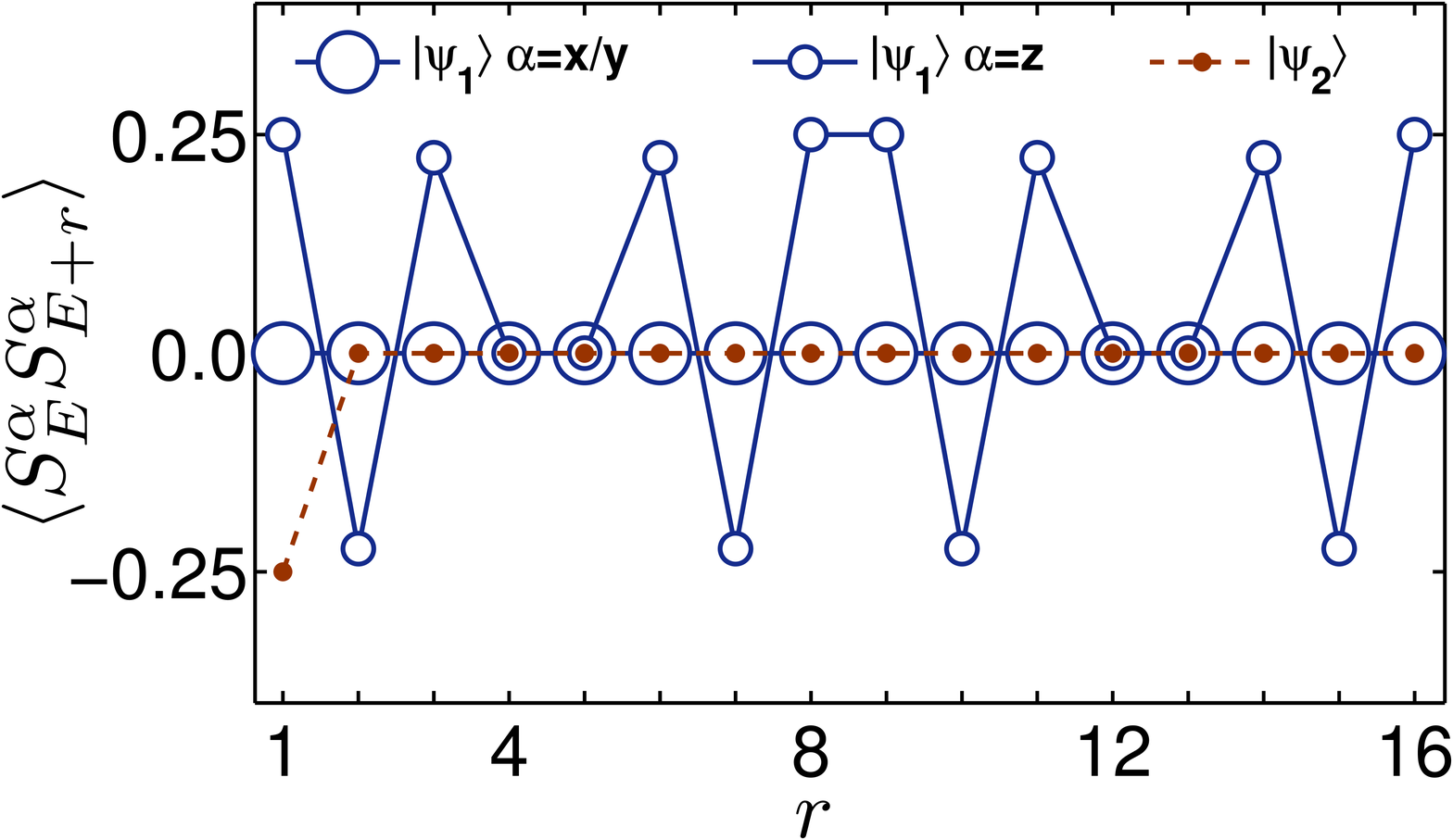}
            \put(1,52){(c)}
  \end{overpic}
  \begin{overpic}[width=0.4\textwidth]{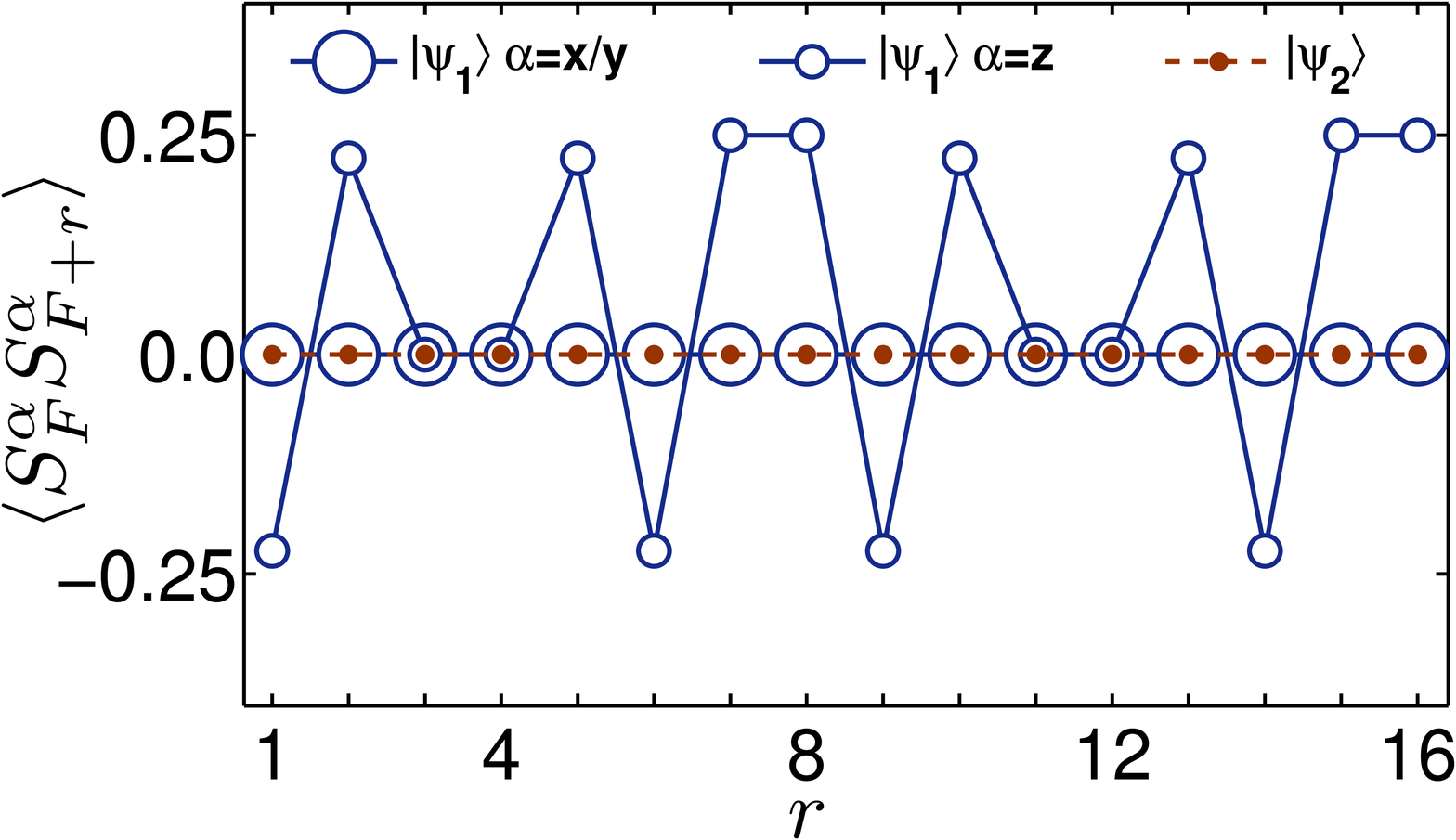}
            \put(1,52){(d)}
  \end{overpic}
\end{center}
\caption{(color online)
 Two-point spin correlations $\langle S^{\alpha}_{\beta} S^{\alpha}_{\beta + r}\rangle$
 as a function of lattice distance between two sites
 $\beta$ and $\beta + r$ for $J'_z=2J$ and $h=1.5J$ in the MAFP phase.
 Here, $\beta \in \{A,B,E,F\}$. }
 \label{MFIcorrab}
\end{figure}

 In the Appendix, we will discuss about the two-point spin
 correlations in order to extract an explicit form of degenerate
 groundstates in the MAFP phase.
 From the first two sites $i = A$ and $B$ in Figs. \ref{MFIcorrab}(a) and \ref{MFIcorrab}(b),
 the two-point spin correlations
 are shown that
 (i)
 $\langle S^{\alpha}_A S^{\alpha}_{A+1} \rangle = -1/4$ and
 $\langle S^{\alpha}_A S^{\alpha}_{A+r} \rangle = 0$ for $r > 1 $,
 (ii)
 $\langle S^{\alpha}_B S^{\alpha}_{B+r} \rangle = 0$.
 Also, we have observed that
 $\langle S^{\alpha}_{A/B}S^{\alpha'}_j\rangle = 0$ for $\alpha \neq \alpha'$
 (not displayed).
 For the third two sites $i = E$ and $F$, in Figs.~\ref{MFIcorrab}(c) and \ref{MFIcorrab}(d),
 the two-point spin correlations
 are shown that
 (i)
 $\langle S^{z}_{E/F} S^{z}_{j} \rangle =\langle S^{z}_{E/F} \rangle\langle S^{z}_{j}\rangle$
 and (ii)
 $\langle S^{x/y}_{E/F} S^{x/y}_{j} \rangle = 0$.
 Also, we have observed that
 $\langle S^{\alpha}_{E/F}S^{\alpha'}_j\rangle = 0$ for $\alpha \neq \alpha'$
 (not displayed).
 As we discussed in the main text,
 such properties of the two-point spin correlations lead to
 the explicit form of the spin states as
 $|\psi_{AB}\rangle=\left(
 \left|\uparrow_A\right\rangle \left|\downarrow_{B}\right\rangle
 - \left|\downarrow_A\right\rangle \left|\uparrow_{B}\right\rangle \right)/\sqrt{2}$
 for the first two sites $A$ and $B$, and
 $\left|\psi_{EF}\right\rangle=\left|\uparrow_E\right\rangle\left|\uparrow_F\right\rangle$
 for the third two sites $E$ and $F$.
%

%%%%%%%%%%%%%%%%%%%%%%%%fig. 16%%%%%%%%%%%%%%%%%%%%%%%%%%%%%%%%%%%%%%%%%%%%%%%%%%%%%%%%
%%%%%%%%%%%%%%%%%%%%%%%%%%%%%%%%%%%%%%%%%%%%%%%%%%%%%%%%%%%%%%%%%%%%%%%%%%%%%%%%%%%%%

\section{\label{app:MAFP2} Spin states for $C,D,G$ and $H$ sites in the MAFP phase}

%%%%%%%%%%%%%%%%%%%%%%%%%%%%%fig. 17%%%%%%%%%%%%%%%%%%%%%%%%%%%%%%%%%%%%%%%%%%%%%%%%%%%
 \begin{figure}
 \begin{center}
    \begin{overpic}[width=0.4\textwidth]{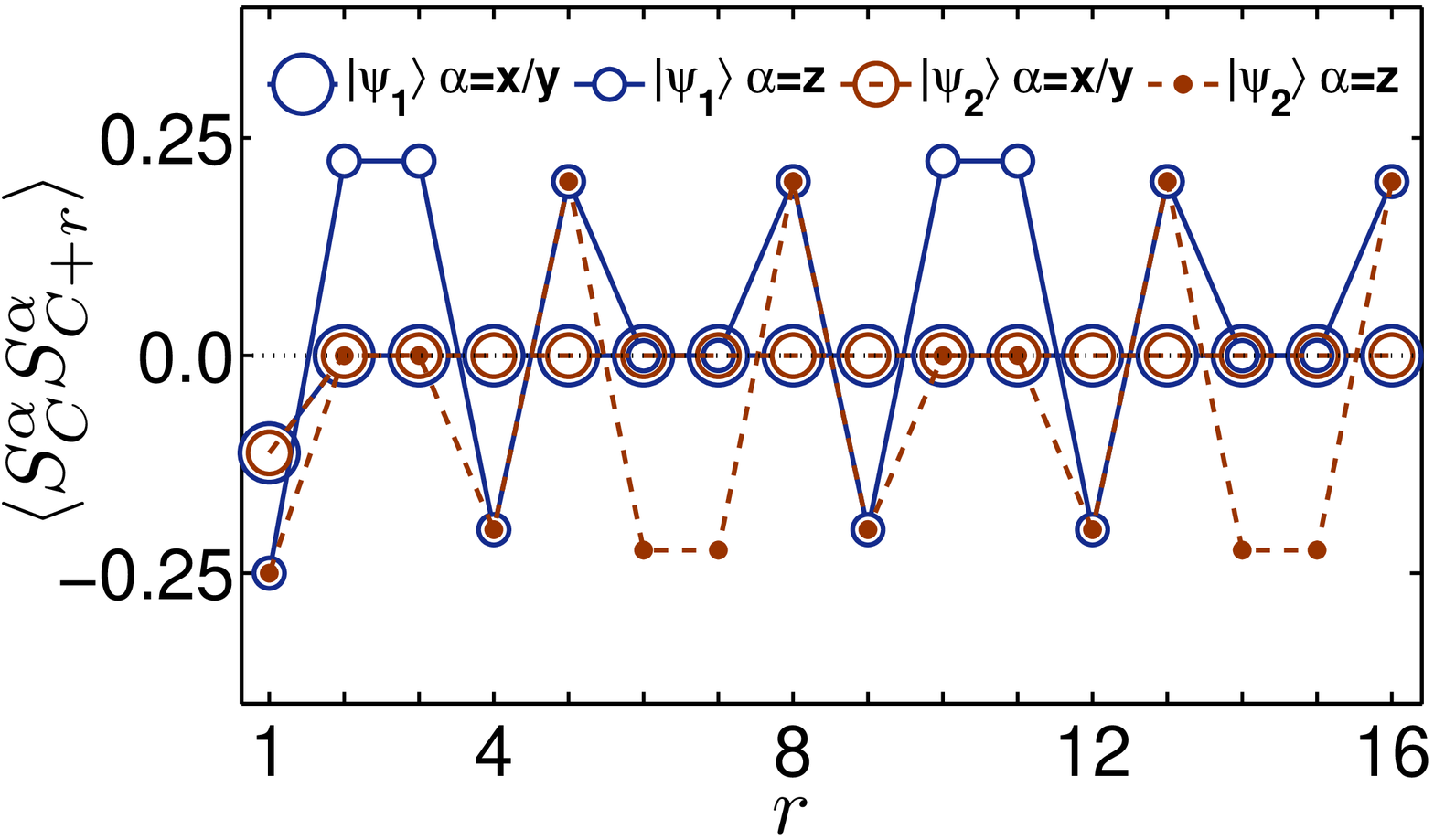}
            \put(1,52){(a)}
  \end{overpic}
  \begin{overpic}[width=0.4\textwidth]{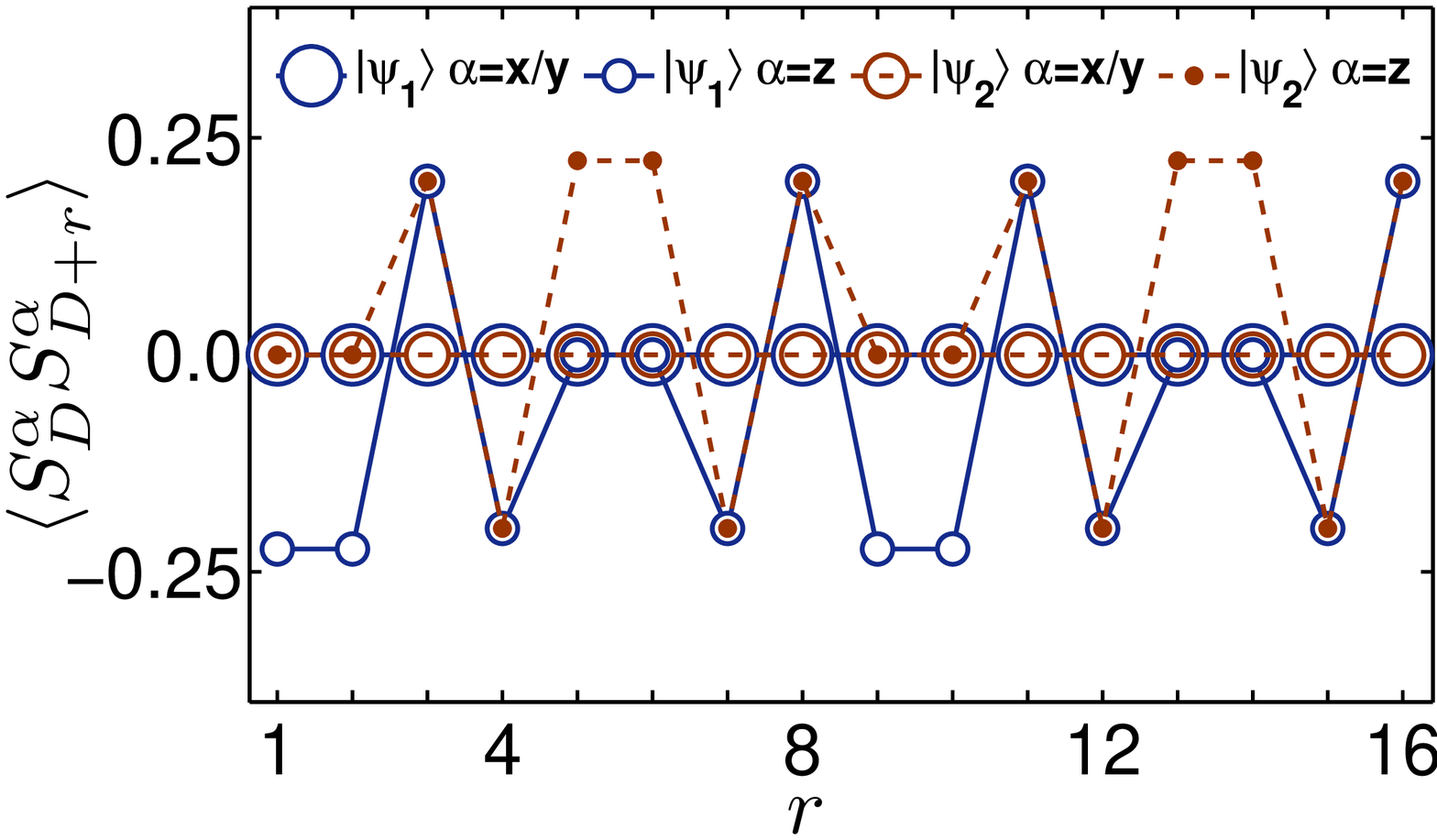}
            \put(1,52){(b)}
  \end{overpic}
 \end{center}
\caption{(color online)
 Two-point spin correlations $\langle S^{\alpha}_{C/D} S^{\alpha}_{C/D + r}\rangle$
 as a function of lattice distance between two sites
 $C/D$ and $C/D + r$ for $J'_z=2J$ and $h=1.5J$ in the MAFP phase.
 }
 \label{MFIcorrcd}
\end{figure}
%%%%%%%%%%%%%%%%%%%%%%%%%%%%%%%%%%%%%%%%%%%%%%%%%%%%%%%%%%%%%%%%%%%%%%%%%%%%%%%%%%%%%

%
 For the second two sites $i = C$ and $D$ in Fig. \ref{MFIcorrcd},
 the properties of the two-point spin correlations for $J'_z=2J$ and $h=1.5J$
 are summarized as follows:
 (i)
 $\langle S^{x/y}_C S^{x/y}_{C+1} \rangle = -0.1118$
 and $\langle S^{x/y}_C S^{x/y}_{C+r} \rangle = 0$ for $r > 1$.
 (ii)
 $\langle S^{z}_C S^{z}_{C+r} \rangle$ has an aperiodic structure, i.e.,
 except for the first period, $\langle S^{z}_{C} S^{z}_{C+r}\rangle$
 has an eight-site periodicity.
 The only difference between the properties of the first period and the other periods
 is $\langle S^{z}_C S^{z}_{C+1} \rangle = -1/4$ and
 $\langle S^{z}_{C} S^{z}_{C+8m+1} \rangle = -0.2
 = \langle S^{z}_{C} \rangle \langle S^{z}_{C+8m+1} \rangle$ with $m = 1,2,\cdots$.
 (iii) $\langle S^{\alpha}_C S^{\alpha'}_{C+r} \rangle = 0$ for $\alpha \neq \alpha'$.
 (iv)
 $\langle S^{x/y}_D S^{x/y}_{D+r} \rangle = 0$.
 (v)
 $\langle S^{z}_D S^{z}_{D+r} \rangle$ has an eight-site periodic structure.
 Also, we have observed that $\langle S^{z}_D S^{z}_{D+r} \rangle
 = \langle S^{z}_D\rangle \langle S^{z}_{D+r} \rangle$.
 (vi) For $\alpha \neq \alpha'$, $\langle S^{\alpha}_D S^{\alpha'}_{D+r} \rangle = 0$.
 From the summary, it should be noted that
 spin correlation,
 $\langle S^{\alpha}_C S^{\alpha}_{C+r} \rangle = \langle S^{\alpha}_C\rangle \langle S^{\alpha}_{C+r} \rangle$
 for $r > 1$ and
 $\langle S^{\alpha}_D S^{\alpha}_{D+r} \rangle = \langle S^{\alpha}_D\rangle \langle S^{\alpha}_{D+r} \rangle$
 for $r \geq 1$.
 These properties of the two-point spin correlations at $C$ and $D$
 satisfy the dimerzation conditions in Eqs. (\ref{eq:8a}), (\ref{eq:8b}), and (\ref{eq:8c}).
 However,
 $\langle S^{x/y}_C S^{x/y}_{C+1} \rangle \neq -1/4$ even though
 $\langle S^{z}_C S^{z}_{C+1} \rangle = -1/4$,
 which implies that the spin state for the two sites
 is not in a singlet state.
 Furthermore, for different system parameters in the MAFP phase, the values of
 $\langle S^{x/y}_C S^{x/y}_{C+1}\rangle$, $\langle S^{z}_C\rangle$, and
 $\langle S^{z}_{C+1}\rangle$ change even though
 $\langle S^{z}_C S^{z}_{C+1} \rangle = -1/4$ does not change.
 Also, from the numerical calculation, the values of
 $\langle S^{x/y}_C S^{x/y}_{C+1}\rangle$, $\langle S^{z}_C\rangle$, and
 $\langle S^{z}_{C+1}\rangle$ are observed to depend on only $J'/J$.
 Actually, for any parameter in the MAFP phase,
 the $\langle S^{z}_C S^{z}_{C+1} \rangle = -1/4$
 implies that the state for the two sites
 is a linear combination of two possible spin states, i.e., $\left|\uparrow_C\downarrow_D\right\rangle$
 and $\left|\downarrow_C\uparrow_D\right\rangle$,
 and then it can be written as
 $\left|\psi_{CD}\right\rangle = a(J'_z/J) \left|\uparrow_C\downarrow_D\right\rangle
 + b(J'_z/J) \left|\downarrow_C\uparrow_D \right\rangle $ with $|a(J'_z/J)|^2+|b(J'_z/J)|^2=1$.
 In terms of the numerical coefficients $a$ and $b$,
 the local magnetizations and the two-point spin correlation are expressed as
 $\langle S^{z}_C\rangle=-\langle S^{z}_D\rangle=\left(|a(J'_z/J)|^2-|b(J'_z/J)|^2 \right)/2$
 and
 $\langle S^{x}_C S^{x}_D\rangle=\langle S^{y}_C S^{y}_D\rangle
 =|a(J'_z/J)| |b(J'_z/J)]| \cos\theta /2$ with a relative phase $\theta$ between $a$ and $b$,
 respectively,
 from the expression of $\left|\psi_{CD}\right\rangle$.
 Numerically, the relations of $\langle S^{z}_C\rangle=-\langle S^{z}_D\rangle$ and
 $\langle S^{x}_C S^{x}_D\rangle=\langle S^{y}_C S^{y}_D\rangle$
 are manifested from Figs. ~\ref{MFImg} and ~\ref{MFIcorrcd}.
 Also, for all the system parameters in the MAFP phase,
 we have numerically found that
 $\langle S^{x}_C S^{x}_D\rangle=\langle S^{y}_C S^{y}_D\rangle = -
 (1+2 \langle S^{z}_C\rangle)^{1/2} (1+2 \langle S^{z}_D\rangle )^{1/2} /4$,
 which implies that the relative phase $\theta$ does not depend on the system parameters
 and $\theta = \pi$.
 Then, one has a freedom to set that $a$ is a positive real number
 and $b$ can be written as $-|b|$.
 Consequently, a best expression for the spin state of the two sites $C$ and $D$
 can be
 $\left|\psi_{CD}\right\rangle = a(J'_z/J) \left|\uparrow_C\downarrow_D\right\rangle
 - |b(J'_z/J)| \left|\downarrow_C\uparrow_D \right\rangle $.

%

%%%%%%%%%%%%%%%%%%%%%%%%%%%%%%fig. 18%%%%%%%%%%%%%%%%%%%%%%%%%%%%%%%%%%%%%%%%%%%%%%%%%%
 \begin{figure}
 \begin{center}
  \begin{overpic}[width=0.4\textwidth]{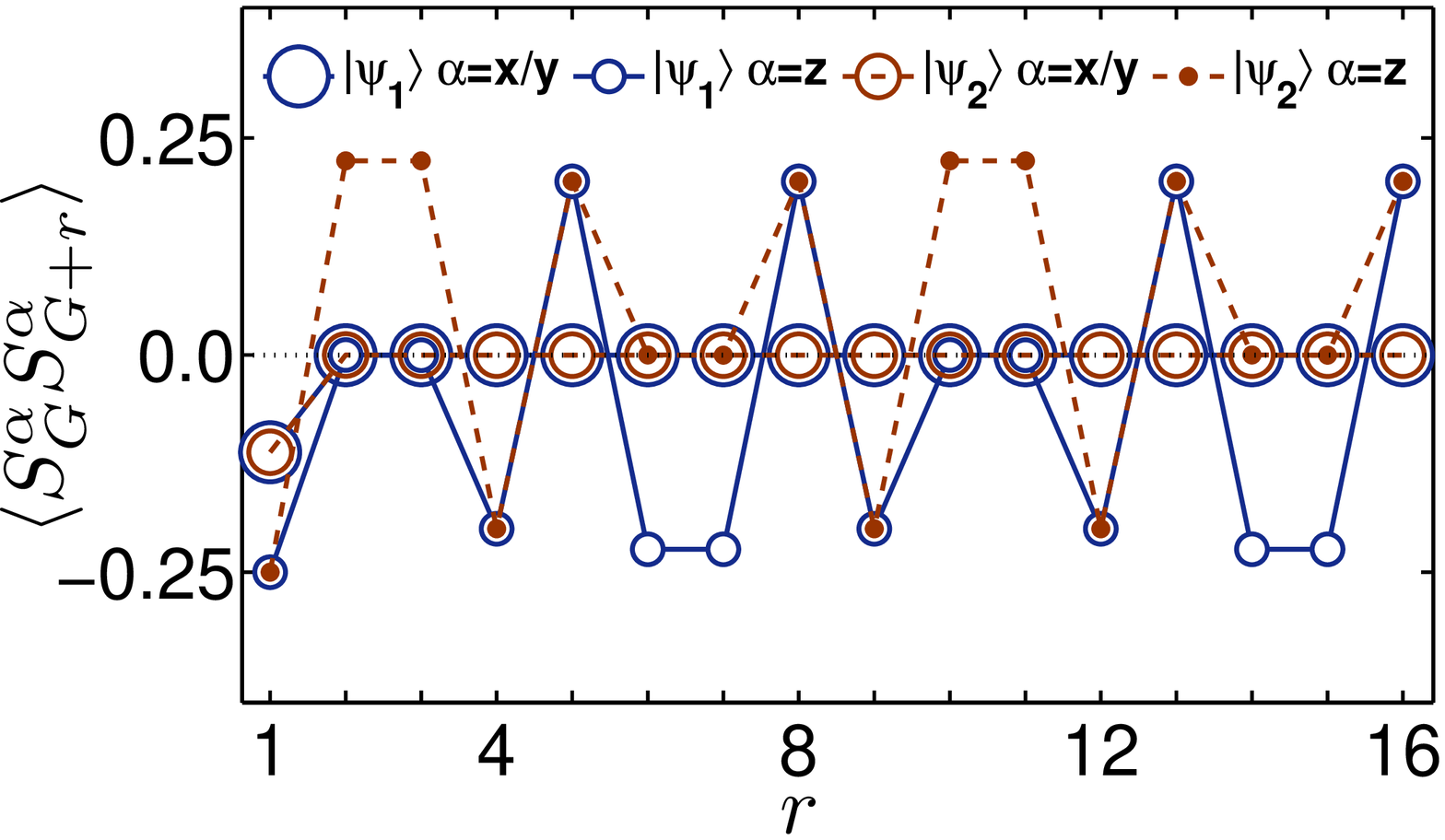}
            \put(1,52){(a)}
  \end{overpic}
  \begin{overpic}[width=0.4\textwidth]{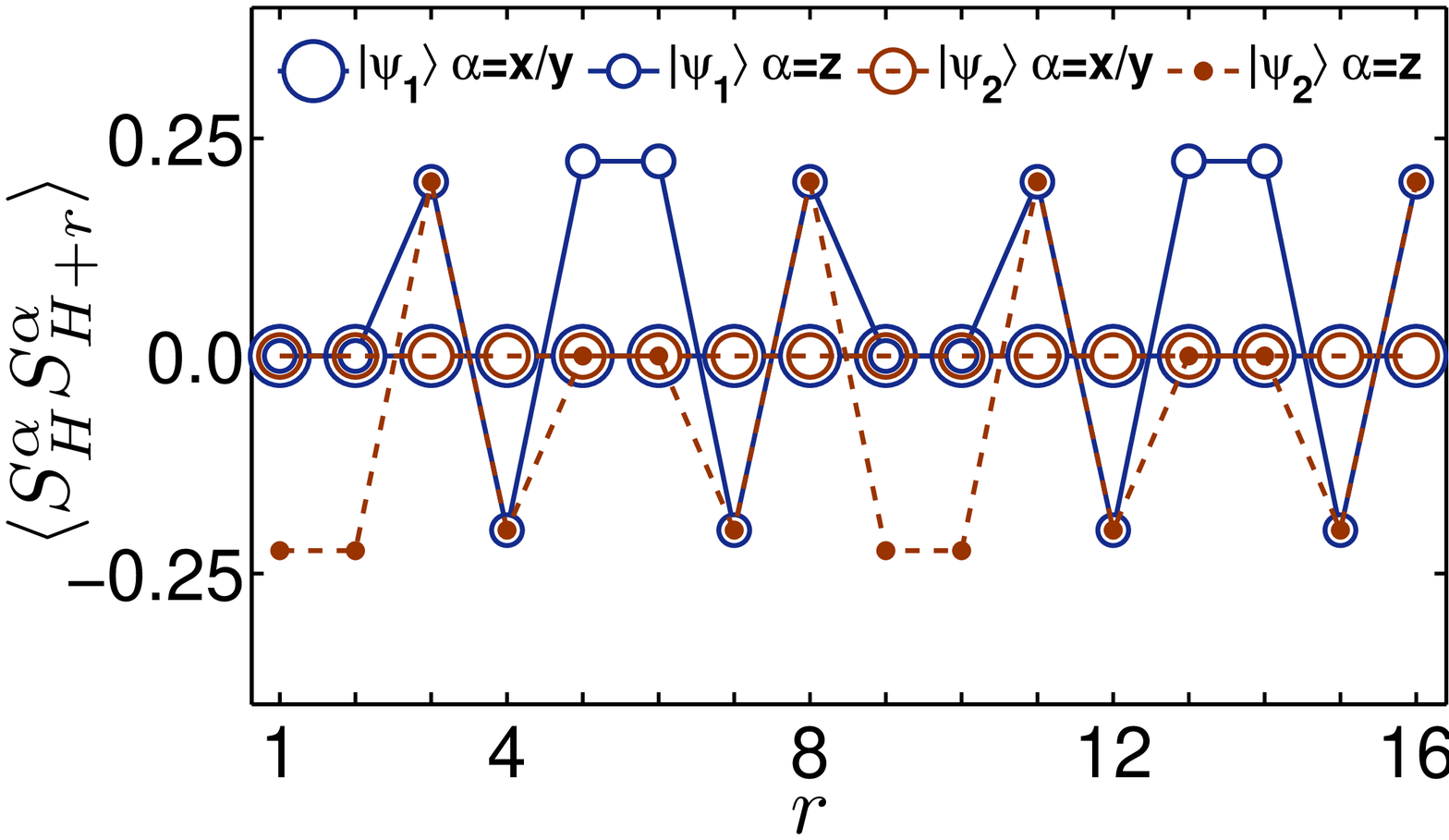}
            \put(1,52){(b)}
  \end{overpic}
 \end{center}
\caption{(color online)
 Two-point spin correlations $\langle S^{\alpha}_{G/H} S^{\alpha}_{G/H + r}\rangle$
 as a function of lattice distance between two sites
 $G/H$ and $G/H + r$ for $J'_z=2J$ and $h=1.5J$ in the MAFP phase.
 }
 \label{MFIcorrgh}
\end{figure}
%%%%%%%%%%%%%%%%%%%%%%%%%%%%%%%%%%%%%%%%%%%%%%%%%%%%%%%%%%%%%%%%%%%%%%%%%%%%%%%%%%%%%

 For the fourth two sites $i = G$ and $H$ in Fig. ~\ref{MFIcorrgh},
 the properties of the two-point spin correlations for $J'_z=2J$ and $h=1.5J$
 are summarized as follows:
 (i)
 $\langle S^{x/y}_G S^{x/y}_{G+1} \rangle = -0.1118$
 and $\langle S^{x/y}_G S^{x/y}_{G+r} \rangle = 0$ for $r > 1$.
 (ii)
 $\langle S^{z}_G S^{z}_{G+r} \rangle$ has an aperiodic structure, i.e.,
 except for the first period, $\langle S^{z}_{G} S^{z}_{G+r}\rangle$
 has an eight-site periodicity.
 The only difference between the properties of the first period and the other periods
 is $\langle S^{z}_G S^{z}_{G+1} \rangle = -1/4$ and
 $\langle S^{z}_{G} S^{z}_{G+8m+1} \rangle = -0.2
 = \langle S^{z}_{G} \rangle \langle S^{z}_{G+8m+1} \rangle$ with $m = 1,2,\cdots$.
 (iii) $\langle S^{\alpha}_G S^{\alpha'}_{G+r} \rangle = 0$ for $\alpha \neq \alpha'$.
 (iv)
 $\langle S^{x/y}_H S^{x/y}_{H+r} \rangle = 0$.
 (v)
 $\langle S^{z}_H S^{z}_{H+r} \rangle$ has an eight-site periodic structure.
 Also, we have observed that $\langle S^{z}_H S^{z}_{H+r} \rangle
 = \langle S^{z}_H\rangle \langle S^{z}_{H+r} \rangle$.
 (vi) For $\alpha \neq \alpha'$, $\langle S^{\alpha}_H S^{\alpha'}_{H+r} \rangle = 0$.
 From the summary, it should be noted that
 spin correlation,
 $\langle S^{\alpha}_G S^{\alpha}_{G+r} \rangle = \langle S^{\alpha}_G\rangle \langle S^{\alpha}_{G+r} \rangle$
 for $r > 1$ and
 $\langle S^{\alpha}_H S^{\alpha}_{H+r} \rangle = \langle S^{\alpha}_H\rangle \langle S^{\alpha}_{H+r} \rangle$
 for $r \geq 1$.
 These properties of the two-point spin correlations at $G$ and $H$
 satisfy the dimerization conditions in Eqs. (\ref{eq:8a}), (\ref{eq:8b}), and (\ref{eq:8c}).
 However,
 $\langle S^{x/y}_G S^{x/y}_{H+1} \rangle \neq -1/4$ even though
 $\langle S^{z}_G S^{z}_{H+1} \rangle = -1/4$,
 which implies that the spin state for the two sites
 is not in a singlet state.
 Furthermore, for different system parameters in the MAFP phase, the values of
 $\langle S^{x/y}_G S^{x/y}_{G+1}\rangle$, $\langle S^{z}_G\rangle$, and
 $\langle S^{z}_{G+1}\rangle$ change even though
 $\langle S^{z}_G S^{z}_{G+1} \rangle = -1/4$ does not change.
 Similar to the pair state of the two sites $C$ and $D$,
 the invariant $\langle S^{z}_G S^{z}_{G+1} \rangle = -1/4$ of the two sites $G$ and $H$ for any parameter in the MAFP phase
 allows the state of the MAFP phase as a linear combination of two possible spin states, i.e.,
 $\left|\psi_{GH}\right\rangle = a'(J'_z/J) \left|\uparrow_G\downarrow_H\right\rangle
 + b'(J'_z/J) \left|\downarrow_G\uparrow_H \right\rangle $ with $|a'(J'_z/J)|^2+|b'(J'_z/J)|^2=1$,
 where the numerical coefficients have been confirmed to be independent on the external magnetic field.
 In terms of the numerical coefficients $a'$ and $b'$,
 the local magnetizations and the two-point spin correlation are expressed as
 $\langle S^{z}_G\rangle=-\langle S^{z}_H\rangle=\left(|a'(J'_z/J)|^2-|b'(J'_z/J)|^2 \right)/2$
 and
 $\langle S^{x}_G S^{x}_H\rangle=\langle S^{y}_G S^{y}_H\rangle
 =|a'(J'_z/J)| |b'(J'_z/J)]| \cos\theta' /2$ with a relative phase $\theta'$ between $a'$ and $b'$,
 respectively,
 from the expression of $\left|\psi_{GH}\right\rangle$.
 Numerically, the relations of $\langle S^{z}_G\rangle=-\langle S^{z}_H\rangle$ and
 $\langle S^{x}_G S^{x}_H\rangle=\langle S^{y}_G S^{y}_H\rangle$
 are manifested from Figs.~\ref{MFImg}, and \ref{MFIcorrgh}.
 Also, for all the system parameters in the MAFP phase,
 we have numerically found that
 $\langle S^{x}_G S^{x}_H\rangle=\langle S^{y}_G S^{y}_H\rangle = -
 (1+2 \langle S^{z}_G\rangle)^{1/2} (1+2 \langle S^{z}_H\rangle )^{1/2} /4$,
 which implies that the relative phase $\theta'$ does not depend on the system parameters
 and then $\theta' = \pi$.
 Then, one has a freedom to set that $a'$ is a positive real number
 and $b'$ can be written as $-|b'|$.
 As a result,
 a best expression for the spin state of the two sites $G$ and $H$
 can be
 $\left|\psi_{GH}\right\rangle = a'(J'_z/J) \left|\uparrow_G\downarrow_H\right\rangle
 - |b'(J'_z/J)| \left|\downarrow_G\uparrow_H \right\rangle $.
 Interestingly, the $\left|\psi_{GH}\right\rangle$ has a similar form with the $\left|\psi_{CD}\right\rangle$.
 Comparing with the local magnetizations and the two-point spin correlations in the sites in Figs.~\ref{MFImg}, ~\ref{MFIcorrcd} and ~\ref{MFIcorrgh},
 it should be noted that
 $\langle S^z_G\rangle=-\langle S^z_C\rangle$, $\langle S^z_H\rangle=-\langle S^z_D\rangle$,
 and  $\langle S^{x}_G S^{x}_H\rangle=\langle S^{x}_C S^{x}_D\rangle$,
 which give the relations between the coefficients, i.e., $a^2-|b|^2=-(a'^2-|b'|^2)$ and $a|b|=a'|b'|$.
 With $a^2+|b|^2=1$ and $a'^2+|b'|^2=1$,
 one can obtained the relations $a'=|b|$ and $|b'|=a$.
 In terms of the coefficients $a$ and $b$,
 the spin state of the two sites $G$ and $H$ can be expressed as
 $\left|\psi_{GH}\right\rangle = |b(J'_z/J)|\left|\uparrow_G\downarrow_H\right\rangle
 - a(J'_z/J) \left|\downarrow_G\uparrow_H \right\rangle $.
%

%%%%%%%%%%%%%%%%%%%%%%%%%%%%%%%%%%%%%%%%%%%%%%%%%%%%%%%%%%%%
%%%%%%%%%%%%%%%%%%%%%%%%%%%%%%%%%%%%%%%%%%%%%%%%%%%%%%%%%%%%

\end{document}